\documentclass[prd, twocolumn, superscriptaddress, nofootinbib, amsmath, amssymb, aps]{revtex4-2}
\usepackage{color} 
\usepackage{bm} 
\usepackage{hyperref} 

\usepackage{float}

\usepackage{graphicx}
\usepackage{grffile}
\usepackage{epstopdf} 
\usepackage{amsthm}
\usepackage{subcaption} 
\usepackage[UTF8]{ctex} 

\usepackage{placeins}

\newtheorem{theorem}{Theorem}[section]

\begin{document}

\title{Stability Boundary of Neutron-Dark Matter Mixed Stars}

\author{Xiao-Ding Zhou (周小丁)}
\email{18038288465@shu.edu.cn}
\affiliation{Department of Physics and Institute for Quantum Science and Technology, Shanghai University, Shanghai 200444, China}

\author{Tian-Shun Chen (陈天舜)}
\email{deltachen302@gmail.com}
\affiliation{Department of Physics and Institute for Quantum Science and Technology, Shanghai University, Shanghai 200444, China}

\author{Si-Man Wu (吴思熳)}
\email{2045152271@qq.com}
\affiliation{Department of Physics and Institute for Quantum Science and Technology, Shanghai University, Shanghai 200444, China}

\author{Kilar Zhang (张弘)}
\email{kilar@shu.edu.cn; Corresponding Author}
\affiliation{Department of Physics and Institute for Quantum Science and Technology, Shanghai University, Shanghai 200444, China}
\affiliation{Shanghai Key Lab for Astrophysics, Shanghai 200234, China}
\affiliation{Shanghai Key Laboratory of High Temperature Superconductors, Shanghai 200444, China}

\begin{abstract}
Regarding the stability of two-fluid star models, we rigorously prove the equivalence between the emergence  of a zero-frequency radial oscillation mode and the static critical-curve criterion for mixed stars, after briefly reviewing the hybrid star case. This establishes a sufficient-condition relation between two independently developed stability criteria. Although this connection has often been implicitly assumed in previous studies of mixed stars, it has rarely been demonstrated explicitly. Our derivation can be extended to general multi-fluid systems. As an illustrative example, we consider dark matter-admixed neutron star models and show that their stability boundary differs from that of single-fluid stars. In this case, stable configurations form a surface in the three-dimensional parameter space spanned by central pressure, mass, and radius, giving rise to a class of stable mixed stars. This class includes “twin stars” with identical masses and radii but distinct internal compositions and structures. These results provide a useful framework for interpreting compact star observations and for constraining dark matter properties through astrophysical measurements.
\end{abstract}

\maketitle

\section{Introduction}
\label{sec:intro}
Evaluating the stability of compact stars is an important issue in astrophysics, especially in the multi-messenger era marked by the gravitational wave observation of neutron stars (NS) \cite{PhysRevLett.119.161101}. For single fluid models, the stability criterion has two equivalent formulations, one based on the sign of fundamental radial oscillation frequencies \cite{Chandrasekhar:1964}, the other based on the extrema in mass-radius ($M$-$R$) curves (known as BTM criterion \cite{bardeen1966} by Bardeen, Thorne, and Meltzer). The former is time consuming but also gives non-zero oscillation modes, and works for multi-fluid case. The latter is intuitive and convenient, but its extension to multi-fluid systems is not guaranteed.

There are two main scenarios for multi-fluid stellar structures, hybrid stars \cite{MISHRA_1993,Ghosh_1995,KHADKIKAR_1995,Di_Clemente_2020} where the interactions among different fluids are strong, and mixed stars \cite{HENRIQUES1990511,Zhang_2025,panotopoulos2019compactstarschallengeview,Rezaei_2018,Wang_2019,Mukhopadhyay_2016,Xiang_2014,Leung_2011,Leung_2012, zhang2022gw170817} where only gravitational interactions are present. The hybrid star case is relatively simple, and its stability rule is found in \cite{Di_Clemente_2020}.
As a byproduct in another work \cite{zhang2022gw170817}, other collaborators and one of the current authors have conjectured necessary conditions from BTM criteria for mixed stars, which needs further modifications. In fact, for mixed star case, there are two known independent  approaches for determining stability. One approach is to compute the radial oscillation frequencies \cite{Chandrasekhar:1964, Kain:2020oho} as mentioned before, and the other is to locate critical curves in certain parameter space \cite{HENRIQUES1990511}. The equivalence of these two criteria was suggested in \cite{HENRIQUES1990511} and applied with specific equations of state (EoS) in \cite{Kain:2021poc}. In this paper, we give a rigorous proof (and more mathematically complete in \cite{Chen2025Equivalence}) that the critical curve criterion is equivalent to the existence of zero-frequency radial oscillation modes, and further explore its application to dark matter (DM) -- nuclear matter (NM) mixed stars.

NS, with their extremely high compactness, serve as natural laboratories for testing theoretical physics. As relativistic objects, they should be described within the framework of general relativity. Their macroscopic properties (such as mass and radius) are determined by the Tolman-Oppenheimer-Volkoff (TOV) equations \cite{PhysRev.55.364,oppenheimer1939massive}, which describe gravitational equilibrium, and by the equation of state (EoS), which governs microscopic nuclear interactions \cite{panotopoulos2017dark}. In our discussion, these objects are treated  at zero temperature as as is standard in this context. Consequently, the pressure is treated as depending solely on the energy density.

Evidence for the existence of DM comes from multiple sources, including galactic rotation curves, large-scale structure formation, and gravitational lensing \cite{Leung_2011}. Precise cosmological observations and measurements further indicate that DM constitutes approximately 25\% of the total energy budget of the universe \cite{kolb2018early}.

At particle physics level, one of the DM candidates is the weakly interacting massive particle (WIMP), such as the neutralino in supersymmetric theories, whose thermal production mechanism can naturally explain the observed DM abundance \cite{steffen2009dark,guver2014capture}. Beyond the traditional WIMP paradigm, DM may possess other properties. For example, asymmetric DM models, in which the DM particle is not self-conjugate \cite{PhysRevD.88.095004,PhysRevD.84.101302}, provide another promising theoretical framework. 

Additionally, DM could be bosonic particles with self-interactions, which would affect its dynamical evolution within NS \cite{guver2014capture}.
This so-called
self-interacting dark matter (SIDM) extends the standard cold DM model by incorporating self-interactions, which can effectively alleviate small-scale structure issues \cite{Tulin_2018,van_den_Aarssen_2012}. A key constraint on this model requires the ratio of its self-scattering cross-section to the mass to lie within a narrow range \cite{Tulin_2018,PhysRevLett.110.111301,Feng_2010}. SIDM provides a viable theoretical pathway for forming compact stellar structures \cite{zhang2023dark}.

DM detection strategies are broadly classified into direct and indirect detections. Direct detection experiments (e.g., DAMA, CoGeNT, CDMS, XENON \cite{PhysRevLett.105.131302,PhysRevLett.106.131302,aalseth2011results,bernabei2008first}) aim to detect signals from DM-nucleus scattering within terrestrial detectors \cite{Leung_2011}. Indirect detection, on the other hand, is achieved by studying the effects of DM on astrophysical objects. For instance, the extremely strong gravitational field and high density of NS interiors may make them possible DM trappers.   Subsequent physical changes—such as heating, accumulating or collapse—in NS can then be used to constrain DM properties \cite{guver2014capture}. To be specific, the mechanisms by which DM may be captured into NS  have been extensively studied \cite{PhysRevD.40.3221,PhysRevD.77.023006,PhysRevD.77.043515,PhysRevD.81.123521,PhysRevD.82.063531,PhysRevLett.115.111301,Cermeno_Perez_Garcia_Silk_2017}. Should the captured DM be self-annihilating, its subsequent annihilation can significantly heat the stellar interior, leading to observable increases in surface temperature or luminosity \cite{PhysRevD.77.023006,PhysRevD.77.043515,PhysRevD.81.123521}. In contrast, non-annihilating DM would gradually accumulate within the star, potentially giving rise to stable mixed configurations \cite{gresham2019asymmetric}, or alternatively, collapsing into a black hole once the accumulated DM reaches its self-gravitating limit \cite{guver2014capture}. Beyond capture, an alternative scenario is that the collapse leading to NS formation could be triggered within a pre-existing DM halo \cite{PhysRevD.97.123007,PhysRevD.99.063015,nelson2019dark}; that is, the mixing of DM and NM may occur during the formation of the compact object itself.

There has been extensive research on the influence of DM on the properties of compact stars, see, e.g. \cite{Leung_2011,bramante2025darkmattercompactstars,Alford_2005,issifu2025rotatingprotoneutronstarsadmixed}. It is necessary to discuss the conditions for the stable existence of mixed stars composed of these components; relevant studies include Refs.~\cite{zhang2022gw170817, zhang2020constraint, Kain:2021poc,pitz2024generatingultracompactneutronstars,kumar2025stabilityanalysistwofluidneutron,Ivanytskyi_2020,Gresham_2019}. Our study focuses primarily on mixed stars consisting of NM and SIDM or fermionic DM, and we systematically investigate their stability and macroscopic properties through different methods.

This paper is organized as follows. In Section \ref{sec:Stellar Model Setup}, we establish the theoretical framework for multi-fluid stellar models, presenting the general-relativistic equilibrium equations and outlining the microphysical inputs, namely EoS, used in our numerical examples. Section \ref{sec:Stability Criteria} contains the central theoretical result of this work: a rigorous proof, based on a variational principle, of the formal equivalence between the existence of zero‑frequency radial oscillation mode and the static critical curve condition. Section \ref{sec:Results} provides numerical validation of our theorem, showing that the stability boundaries computed via direct radial oscillation analysis precisely match the theoretically derived critical curves for various EoS combinations. Finally, Section \ref{sec:Conclusion} summarizes our main results and discusses their astrophysical implications. For completeness, we provide background review on the static stability criteria and detailed framework of the standard radial pulsation equations in Appendices \ref{app:static_criteria} and \ref{app:pulsation_derivation}, respectively.

\section{Star Model Setup}
\label{sec:Stellar Model Setup}
To model the coexistence of DM and NM in compact astrophysical objects and their corresponding structure, we begin with a basic description of their possible internal composition, followed by the TOV equations for spherical equilibrium structure within the framework of general relativity. The formulation is completed by specifying the EoS that relates pressure to energy density.

\subsection{Possible Mixing Models}
\begin{figure}
\centering
\includegraphics[width=\linewidth]{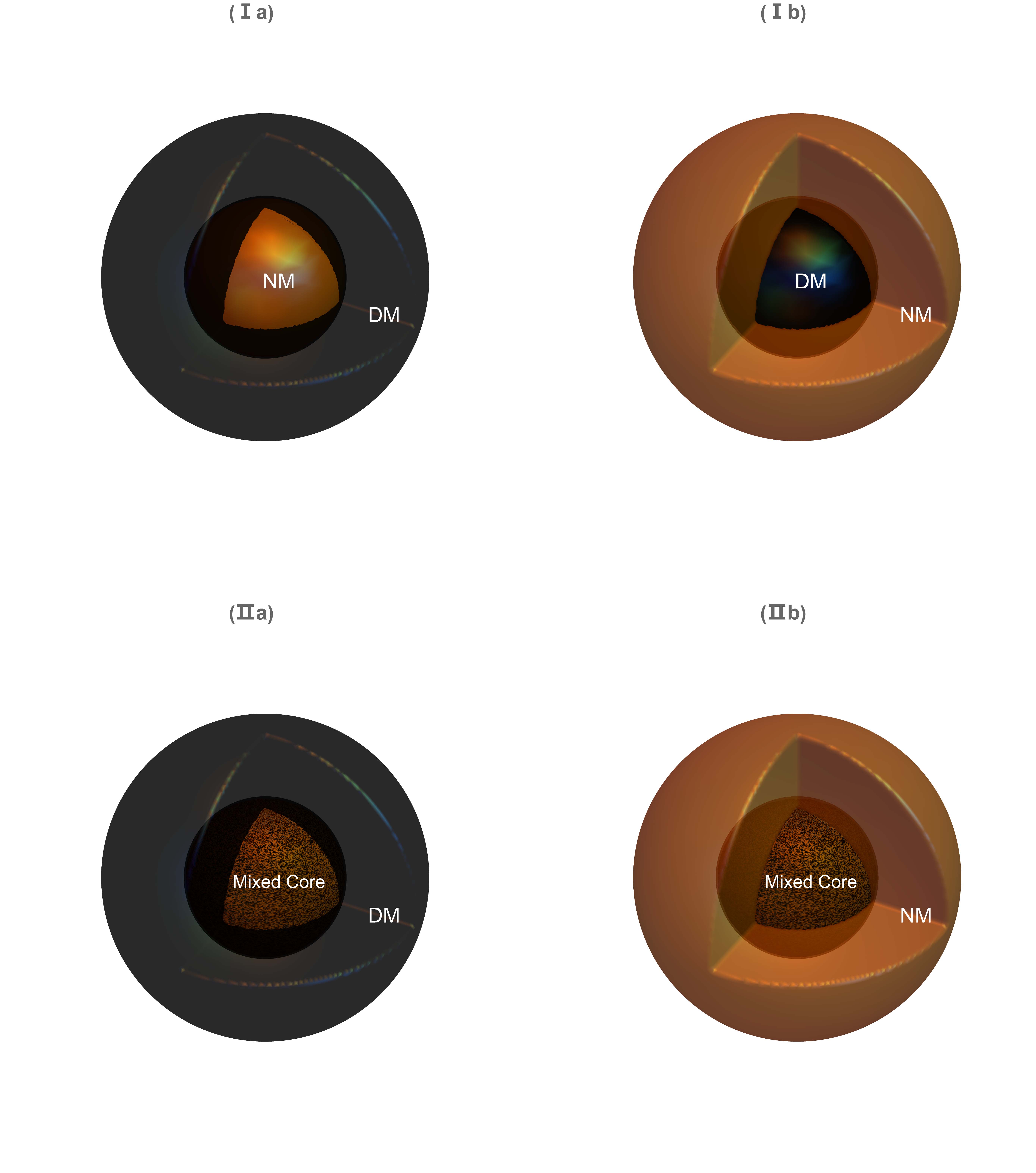} 
\caption{Schematic diagram: two different possible structures: hybrid stars (I) and mixed stars (II).}
\label{fig:Structure settings}
\end{figure}

From the perspective of geometric configuration, two-fluid star models can be preliminarily classified into two categories. Hybrid star: a NM (DM) core enveloped by a DM (NM) crust; mixed star: a mixed core enveloped by a single-component shell \cite{zhang2022gw170817,zhang2020constraint}\footnote{In these two references, ``Hybrid'' includes both hybrid and mixed cases in this paper.}, as shown in Fig.\ref{fig:Structure settings}.

The formation mechanisms of the layered and mixed-core configurations are fundamentally different in the perspective of microscopic interactions. The emergence of a layered structure requires the existence of non-gravitational interactions \cite{nelson2019dark} between DM and NM, such that the total energy of the two-component system in a phase-separated state is lower than that of the mixed state, whereas a mixed core arises when only gravitational interactions are present, lacking the driving force for a sharp interface.

This structural difference profoundly influences their macroscopic manifestations. In $M$-$R$ relations, layered (Hybrid) models depends on a single parameter variation, with a clear evolutionary path of a one-dimensional curve. In contrast, mixed-core models exhibit more complex parameter dependence corresponding to the central parameters of both DM and NM, causing their $M$-$R$ relation to topologically form a two-dimensional continuous surface.

\begin{figure}
\centering
\includegraphics[width=\linewidth]{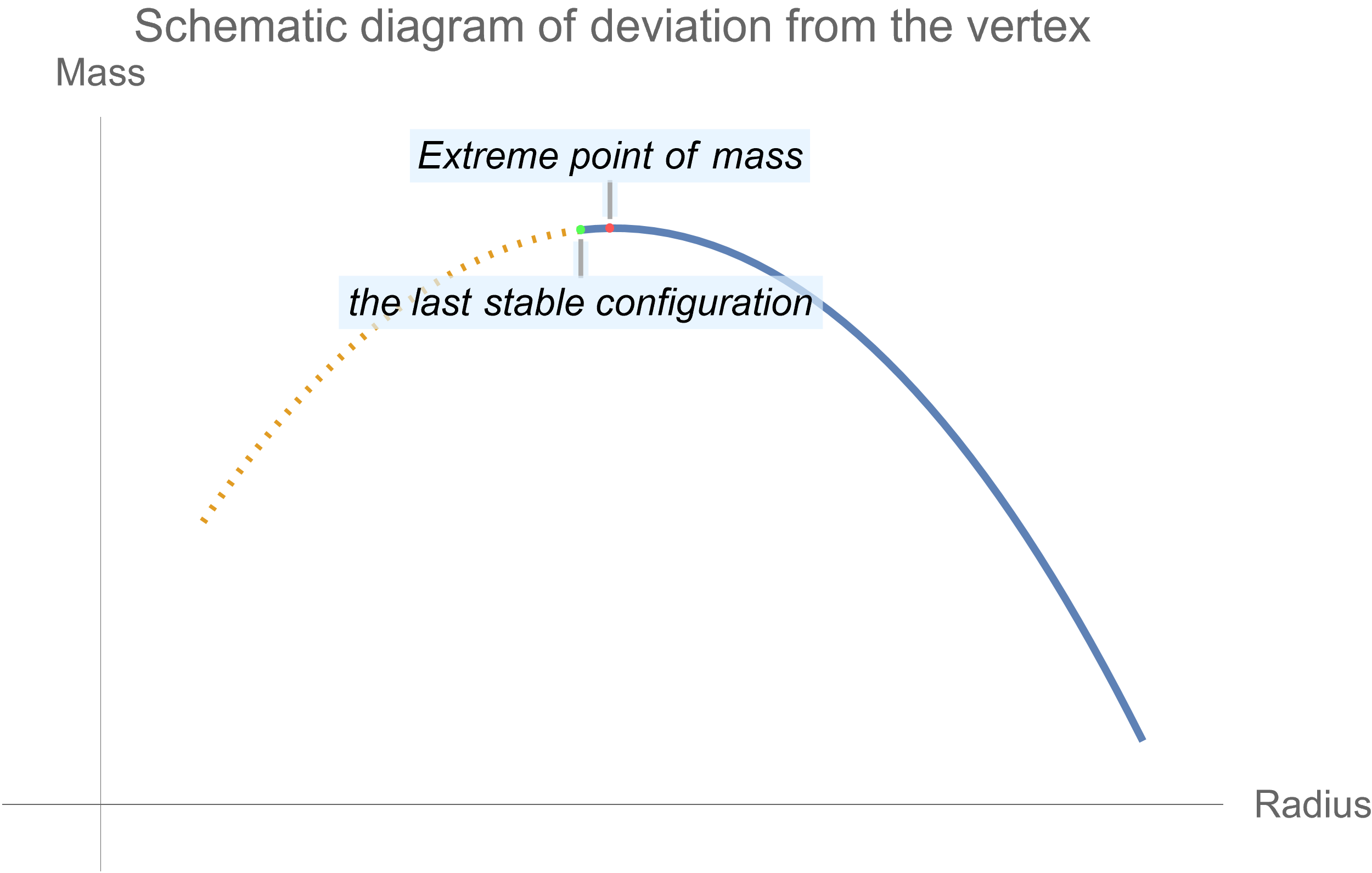}
\caption{Schematic diagram: the stability turning point of the hybrid star deviates from the mass vertex.}
\label{fig:SchematicOfDeviation}
\end{figure}

Regarding stability, both types can have their stable regions rigorously determined via radial perturbation analysis. However, as the schematic diagram Fig.\ref{fig:SchematicOfDeviation} shows, unlike pure NS, the stability boundary characterization for layered stars exhibits a slight offset from the traditional BTM criteria \cite{Bardeen:1966} in the $M$-$R$ relation, according to the radial mode results calculated in \cite{Di_Clemente_2020}. For mixed-core models, the stability analysis is more complex, and their manifestation in macroscopic quantities must satisfy a generalization of the BTM criteria, which will be discussed in the next sections.
In the following, we focus on this mixed star case, which is the scenario II in Fig.\ref{fig:Structure settings}.

\subsection{Equilibrium Configuration}
The commonly used model for describing compact star equilibrium configuration is provided by the TOV formalism. This framework comprises a set of differential equations for a static, non-rotating star modeled as an ideal fluid in spherical symmetry.

We consider a spherically symmetric spacetime with Schwarzschild-like coordinates $(t, r, \theta, \phi)$. The line element is
\begin{equation}
ds^2=-e^{2\phi \left( r \right)}c^2dt^2+e^{2\nu(r)} dr^2+r^2 d\varOmega ^2,
\end{equation}
where $d\Omega ^2=d\theta ^2+\sin ^2\theta d\phi ^2$, and the metric functions satisfy $e^{2\nu(r)} = \left(1 - \frac{2G m(r)}{r c^{2}}\right)^{-1}$.
and the energy-momentum tensor for an ideal fluid
\begin{equation}
T^{\mu \nu}=\left( \rho \left( 1+\zeta /c^2 \right) +p/c^2 \right) u^{\mu}u^{\nu}+pg^{\mu \nu},
\end{equation}
with $p$ the fluid pressure, $\zeta$ the specific energy density, $\rho$ the rest mass density, and  $u^{\mu}$ the four-velocity vector. The TOV equations can be derived by substituting them into the Einstein equation
\begin{equation}
G^{\mu \nu}=\frac{8\pi G}{c^4}T^{\mu \nu},
\end{equation}
together with the energy-momentum conservation
\begin{equation}
\nabla T^{\mu \nu}=0.
\end{equation}
The explicit TOV equations are given as follows:
\begin{equation}
\begin{aligned}
	\frac{dp}{dr}=&-G\left( \rho \left( r \right) \left( 1+\zeta \left( r \right) /c^2 \right) +p\left( r \right) /c^2 \right) \\ &\frac{m\left( r \right) +4\pi r^3p\left( r \right) /c^2}{r\left( r-2Gm\left( r \right) /c^2 \right)},\\
	\frac{dm}{dr}=&4\pi r^2\rho \left( r \right) \left( 1+\zeta \left( r \right) /c^2 \right) ,\\
	\frac{d\phi}{dr}=&\frac{m\left( r \right) +4\pi r^3p\left( r \right) /c^2}{r\left( r-2Gm\left( r \right) /c^2 \right)},
\end{aligned}
\label{eq:full_tov}
\end{equation}
where $m\left( r \right)$ is the gravitational mass within radius $r$.

Defining the total energy density as $\varepsilon = \rho c^2 + \rho\zeta$ and adopting the natural unit system, we simplify \eqref{eq:full_tov} to
\begin{equation}
\begin{aligned}
	\frac{dp}{dr}&=-\left( \varepsilon \left( r \right) +p\left( r \right) \right) \frac{m\left( r \right) +4\pi r^3p\left( r \right)}{r\left( r-2m\left( r \right) \right)},\\
	\frac{dm}{dr}&=4\pi r^2\varepsilon \left( r \right) ,\\
	\frac{d\phi}{dr}&=\frac{m\left( r \right) +4\pi r^3p\left( r \right)}{r\left( r-2m\left( r \right) \right)}.
\end{aligned}
\label{eq:tov}
\end{equation}
We adopt the equations \eqref{eq:tov} in our subsequent single-fluid numerical calculations.

The boundary conditions are prescribed as follows: at the star center, the mass within an infinitesimal radius must approach zero; at the surface (where \( r = R \), with \( R \) denoting the star radius), the metric potential \(\phi\) assumes its vacuum form, \(\phi = \frac{1}{2} \ln \left(1 - 2GM / (R c^2)\right)\), which matches the Schwarzschild exterior solution, where \( M \equiv m(R) \) represents the total mass enclosed within the star radius. The central pressure \( p_c \equiv p(0) \), on the other hand, serves as a free parameter that may be specified independently.

While the above TOV equations typically models the interior of a compact star as an ideal fluid composed of a single species of matter, the physical scenario under consideration here is more complex: a DM/NM mixed star. Generally, models with multiple matter components can be constructed using either a single-fluid approximation \cite{shahrbaf2024observational}\cite{shahrbaf2025probingstrangedarkmatter} (where the system is still treated as one effective fluid, but with an EoS incorporating contributions from multiple components) or a multi-fluid formalism (where distinct matter components are described by separate EoS). In the case of a weakly coupled multi-fluid model with only gravitational interactions—meaning the energy-momentum tensors of the individual components are decoupled—it is possible to derive a multi-fluid generalization of the TOV equation \cite{Mukhopadhyay_2017,Rezaei_2018}:
\begin{equation}
\begin{aligned}\label{tov_multi}
	\frac{dp_I}{dr} &= -\left( \varepsilon_I(r) + p_I(r) \right) \frac{m_t(r) + 4\pi r^3 p_t(r)}{r\left( r - 2m_t(r) \right)}, \\
	\frac{dm_I}{dr} &= 4\pi r^2 \varepsilon_I(r), \\
	\frac{d\phi}{dr} &= \frac{m_t(r) + 4\pi r^3 p_t(r)}{r^2 \left(1 - \frac{2m_t(r)}{r}\right)},
\end{aligned}
\end{equation}
where the subscript $I$ indicates the components of the substance. $m_{t}=\sum_I{m_I}$, $p_{t}=\sum_I{p_I}$ and $\varepsilon _{t}=\sum_I{\varepsilon_I}$ are the sums of respective quantities from contributions of each fluid component. When $p_I$ reaches zero, the corresponding radius and mass for that component are $R_I$ and $M_I$. When $p_{t}$ reaches zero, we have the total star radius $R_t$, and the total star mass  $M=\sum_I{M_I}$.

In addition to the nature units, we further apply astrophysical units for later convenience.
The units of mass, pressure, energy density and radius are $M_{\odot}$, $p_{\odot}=\varepsilon_{\odot} = 1/ M_{\odot} ^2$, and $\mathrm{km}$, respectively, where $M_{\odot}$ denotes the solar mass.

\subsection{Equation of State}
\label{subsec:EoS}
There are many existing EoS models of NS and quark stars, such as FPS \cite{Pandharipande1989}, APR (ALF) \cite{akmal1998equation}, WFF \cite{wiringa1988equation}, MPA1 \cite{muther1987nuclear}, MSL \cite{mueller1996relativistic}, H4, \cite{PhysRevD.73.024021} and the BSK family of models \cite{goriely2009skyrme}.

To check the stability, we can deal with all kinds of EoS, either numerical or analytical.  In our project, we take one numerical EoS and three analytical EoS for example. 

For NS EoS, we first consider the famous SLy4 \cite{CHABANAT1998231,CHABANAT1997710,Douchin2001}, which survives from the GW170817 \cite{PhysRevLett.119.161101} and NICER \cite{Choudhury_2024} observation constraints.
It is a Skyrme-type effective nuclear interaction parametrization, tailored for describing asymmetric NM. Its parameters are derived by fitting to microscopic NM calculations and experimental data of doubly magic nuclei, achieving self-consistency between nuclear and NS physics. However it has many adjustable parameters and doesn't have an analytic form.

The second EoS is a double-polytropic model \cite{Li_2024} motivated by AdS/QCD and the Witten-Sakai-Sugimoto framework \cite{Sakai_2005,witten1984cosmic}, where an instanton-gas structure is introduced in the D-brane background and the corresponding free-energy variational equations are solved. This analytic EoS should be regarded here as a simplified reference model rather than a fully calibrated microphysical description. For the particular parameter choice used in this work, it supports a maximum mass of about $1.85M_{\odot}$, and therefore is not intended to represent a complete phenomenological fit to the most restrictive massive-pulsar constraints. Nevertheless, the double-polytropic form is flexible: with suitable choices of parameters, it can be made compatible with current observational constraints and the causality condition. In the present study, we use this model mainly because its closed analytic form makes the numerical comparison transparent and easily reproducible. The focus of this work is the proof and numerical verification of the equivalence between the dynamical and static stability criteria. The EoS for double-polytropic model is:
\begin{equation}
\varepsilon =0.140\mathcal{A} ^{0.571}p^{0.429}+3.896\mathcal{A} ^{-0.335}p^{1.335},\label{eq:double-polytropic EoS}
\end{equation}
where $\mathcal{A} =1.8\times 10^{-5}\times \ell ^{-7}$, and $\ell$ (brane asymptotic separation) is the only adjustable parameter. We refer this EoS as Holographic EoS later.

For the DM component in mixed stars, the EoS is determined by the chosen DM model. Studies have employed free Fermi gas DM combined with the SLy4 NS model to investigate mixed star stability \cite{Kain:2021poc}. In this work, we also include this EoS for comparison. The EoS for such free Fermionic DM is given by \cite{shapiro1983black,glendenning2000compact}:
\begin{equation}
    \begin{aligned}
        \varepsilon &=\frac{1}{2\pi ^2}\int_0^{k_F}{dk\,\,k^2\sqrt{k^2+m_{f}^{2}}}\\
        &=\frac{1}{8\pi ^2}\left[ k_F\sqrt{{k_F}^2+m_{f}^{2}}\left( 2{k_F}^2+m_{f}^{2} \right) \right. \\
        &\quad \left. -m_{f}^{4}\ln \left( \frac{k_F+\sqrt{{k_F}^2+m_{f}^{2}}}{m_f} \right) \right], \\
        p&=\frac{1}{6\pi ^2}\int_0^{k_F}{dk\frac{k^4}{\sqrt{k^2+m_{f}^{2}}}}\\
        &=\frac{1}{24\pi ^2}\left[ k_F\sqrt{{k_F}^2+m_{f}^{2}}\left( 2{k_F}^2-3m_{f}^{2} \right) \right. \\
        &\quad \left. +3m_{f}^{4}\ln \left( \frac{k_F+\sqrt{{k_F}^2+m_{f}^{2}}}{m_f} \right) \right],
    \end{aligned}\label{eq:EoS_free fermion}
\end{equation}where $m_{f}$ is the fermion mass and $k_F$ is the Fermi
momentum. Various interactions can be incorporated into this free Fermi gas model to construct more sophisticated EoS.

For Bosonic DM, a stable mixed star model cannot be formed with free bosons due to the absence of degeneracy pressure. Given the weak interaction between DM and NM, it is natural to consider self-interacting bosonic DM to construct the DM EoS that balances the self-gravitational effect. In this project, we utilize a scalar field theory with the potential $V\left( \phi \right) =\frac{{m_b}^2}{2}\left| \phi \right|^2+\frac{\lambda _4}{4}\left| \phi \right|^4$, as proposed in \cite{colpi1986boson}, where $m_b$ is the boson mass and $\lambda_4$ is the coupling. In the isotropic limit, this theory yields the following EoS \cite{zhang2023dark}:
\begin{equation}
\varepsilon =3p+B\sqrt{\varepsilon},\label{eq:self-interaction EoS}
\end{equation}
where the free parameter is $B=\frac{0.08}{\sqrt{\lambda _4}}\left( \frac{m}{\mathrm{GeV}} \right) ^2$.

\section{Stability Criteria}
\label{sec:Stability Criteria}

An equilibrium configuration of a compact star, determined by solving the Tolman-Oppenheimer-Volkoff (TOV) equations, can be either stable or unstable against perturbations. The stability can be assessed through two main approaches: a \textit{dynamical method}, which involves solving the equations for radial oscillations \cite{Chandrasekhar:1964}, and a \textit{static method}, which analyzes the properties of sequences of equilibrium solutions.

While the dynamical method provides a definitive answer, it is often computationally intensive. For single-fluid stars, an efficient static alternative exists in the form of the well-known Bardeen-Thorne-Meltzer (BTM) criterion \cite{Bardeen:1966}. For multi-fluid stars, where the BTM criterion fails, a generalization known as the ``critical curve criterion" has been developed by Henriques et al. \cite{HENRIQUES1990511} and utilized in subsequent works on dark matter admixed stars \cite{Kain:2021poc}. A detailed description of these static methods is provided in Appendix~\ref{app:static_criteria}. For completeness and formula consistency, the standard radial pulsation equations (the dynamical method) which was first derived by Kain in \cite{Kain:2020oho} is outlined in Appendix~\ref{app:pulsation_derivation}.

While the ``critical curve criterion'' has been numerically verified in several contexts such as\cite{Leung_2012,Kain:2021poc}, a formal proof founded upon first principles such as variational methods and covariant conservation laws has been lacking.
In this section, we present the core contribution of our work: a rigorous demonstration of the formal equivalence between the existence of zero-frequency oscillation mode and the static critical curve criterion, and explain the connection with the dynamical stability boundary.

\subsection{Equivalence between Zero‑Mode Condition and Critical Curve}
\label{subsec:equivalence}

\subsubsection*{Preliminary Remarks}

Before proceeding to the equivalence proof, several important considerations must be emphasized:

\begin{enumerate}
\item \textbf{Stability vs. Zero-mode Condition}: The fundamental mode vanishing ($\omega_0^2=0$) represents a stronger condition than the existence of any zero-mode ($\exists \,n\in \mathbb{N} \text{ s.t. } \omega _{n}^{2}=0$). The original critical curve method strictly identifies configurations where \emph{some} mode has zero frequency. Additional verification (which we will discuss in the ending part of the proof section) is required to find the curve exactly corresponds to the fundamental zero mode.

\item \textbf{Mathematical Foundations}: The existence of fundamental modes, the fact that the eigenvalues are real, and the completeness of mode expansions have been rigorously established by \cite{Caballero_2024}. We adopt these conclusions without further proof.

\item \textbf{Model Assumptions and Mode Continuity}: Our proof of the equivalence between the zero-frequency mode and the critical curve criterion relies on several physical assumptions that should be made explicit. First, we consider stars composed of ideal fluids, for which the energy-momentum tensor takes the perfect-fluid form and no viscous or heat-conduction effects are present. Second, we assume that the fluids are non-dissipative, so that the dynamical equations are time-reversal invariant and no mode damping or growth arises from entropy production. Third, we work with smooth EoS, ensuring that the relevant thermodynamic derivatives are well-defined and that the pulsation equations are regular throughout the star. Fourth, we fix the chemical composition and specific entropy of each fluid, so that the only dynamical variables are the fluid displacements and the metric perturbations; in particular, no additional degrees of freedom (such as composition gradients or thermal diffusion) enter the problem. Under these assumptions, the dynamical equations are time-reversal invariant. Consequently, the eigenvalues $\omega_n^2$ are real and vary continuously as functions of the central pressures in the two-dimensional parameter space \cite{weinberg2013gravitation,Caballero_2024}. This continuity guarantees that the critical curve, defined by the locus where the fundamental mode frequency vanishes, properly separates stable and unstable regions, and that the stability boundary cannot be crossed without passing through a zero-frequency mode.
\end{enumerate}

The equivalence proof proceeds by establishing that the critical curve condition is equivalent to the existence of zero-frequency modes. We begin by reformulating the problem and then proceed through two main theorems.

In practical analysis, we identify the zero-mode locus as the set of points in the two-dimensional parameter space spanned by the central pressures of the two fluids,
\begin{equation}
\boldsymbol{\sigma} = (p_c^{\mathrm{DM}}, p_c^{\mathrm{NM}}),
\label{eq:central-pressure-parameter}
\end{equation}
where the gradient operator is defined as
\begin{equation}
\nabla_{\boldsymbol{\sigma}} \equiv 
\left( 
\frac{\partial}{\partial p_c^{\mathrm{DM}}}, 
\frac{\partial}{\partial p_c^{\mathrm{NM}}} 
\right).
\label{eq:parameter-space-gradient}
\end{equation}
The quantities $M$, $N_{\mathrm{NM}}$, and $N_{\mathrm{DM}}$ are scalar functions on this parameter space. The equivalence we aim to prove can then be formulated as:
\begin{equation}
\label{eq:equivalence_condition}
\begin{aligned}
&(\nabla_{\boldsymbol{\sigma}} M \times \nabla_{\boldsymbol{\sigma}} N_{\mathrm{NM}} = 0) \;\lor\; (\nabla_{\boldsymbol{\sigma}} M \times \nabla_{\boldsymbol{\sigma}} N_{\mathrm{DM}} = 0) \\
&\quad \lor\; (\nabla_{\boldsymbol{\sigma}} N_{\mathrm{NM}} \times \nabla_{\boldsymbol{\sigma}} N_{\mathrm{DM}} = 0) \;\Longleftrightarrow\; \exists n \text{ s.t. } \omega_n^2 = 0,
\end{aligned}
\end{equation}
that is, the existence of a zero mode is equivalent to that any two of the gradients $\nabla_{\boldsymbol{\sigma}}M$, $\nabla_{\boldsymbol{\sigma}}N_{\mathrm{NM}}$, and $\nabla_{\boldsymbol{\sigma}}N_{\mathrm{DM}}$ become parallel.

\subsubsection{Theorem 1: Variational Extremum Principle}

The first theorem establishes a rigorous connection between the hydrostatic equilibrium and the variational properties of the system, forming the foundation of \ref{Theorem:2} and of simplifying the condition to the actual critical curve method. A similar proof of the variational principle for two-fluid systems was previously outlined in \cite{Goldman2013}. In the present work, we provide a self-contained and explicit derivation, introducing a unified auxiliary function $F(r)$ that consolidates the nonlocal integral contributions, and defining a common proportional function $\Gamma(r)=\lambda_I n_I/(p_I+\epsilon_I)$ to eliminate the Lagrange multipliers efficiently. This construction leads directly to the standard two-fluid TOV equations. Moreover, we establish the  necessity and sufficiency of the relation between the variational extremum condition and the hydrostatic equilibrium, providing a complete and transparent foundation for the subsequent stability analysis.

\begin{theorem}
A \textbf{two-fluid isentropic mixed star} configuration satisfies the coupled general-relativistic hydrostatic equilibrium equations (TOV equations) if and only if the total mass $M$ takes its extremal value with respect to all time-independent variations that conserve the particle numbers of both fluids.\label{Theorem:1}
\end{theorem}

\begin{proof}
We provide the rigorous proof for Theorem \ref{Theorem:1}. The star configuration satisfies the equilibrium equations if and only if the quantity \( M \) defined by:
\begin{equation}
M\equiv \sum_{I=1,2}{\int_0^{R_I}{4\pi r^2\varepsilon _I(r)dr}}
\end{equation}
is extremal under all variations of \( \varepsilon_I(r) \) that keep the quantities
\begin{equation}
N_I \equiv \int_0^{R_I}{4\pi r^2 n_I(r) \left[ 1 - \frac{2 m_t(r)}{r} \right]^{-\frac{1}{2}} dr}
\end{equation}
constant for \( I=1,2 \). This is equivalent to the existence of Lagrange multipliers \( \lambda_1 \) and \( \lambda_2 \) such that \( \delta H = \delta M - \lambda_1 \delta N_1 - \lambda_2 \delta N_2 \) is zero for all variations.

For given variations \( \delta \varepsilon_I(r) \), we have:
\begin{equation}
\begin{aligned}
    \delta H\left[ \left\{ \delta \varepsilon _I\left( r \right) \right\} \right]  =& \delta M\left[ \left\{ \delta \varepsilon _I\left( r \right) \right\} \right]  - \lambda_1 \delta N_1\left[ \left\{ \delta \varepsilon _I\left( r \right) \right\} \right] \\& - \lambda_2 \delta N_2\left[ \left\{ \delta \varepsilon _I\left( r \right) \right\} \right] .
\end{aligned}
\end{equation}

Now expand each term as (since all the integrand vanishes above $R_I+\delta R_I$, we write the upper limit of integration as positive infinity)
\begin{equation}
\begin{aligned}
\delta M = &4\pi \sum_{I=1,2} \int_0^{\infty } r^2 \delta \varepsilon_I(r) dr,\\
    \delta N_I = &4\pi \int_0^{\infty } r^2 \left( 1 - \frac{2 m_t(r)}{r} \right)^{-1/2} \delta n_I(r) dr \\&+ 4\pi \int_0^{\infty } r \left( 1 - \frac{2 m_t(r)}{r} \right)^{-3/2} \\&\times n_I(r) \delta m_t(r) dr,
\end{aligned}
\end{equation}
where \( m_t(r) = m_1(r) + m_2(r) \). Thus:
\begin{equation}
\begin{aligned}
    \delta H =& 4\pi \sum_{I=1,2} \int_0^{\infty} r^2 \delta \varepsilon_I(r) dr - 4\pi \sum_{I=1,2} \lambda_I \\&\times \left[ \int_0^{\infty} r^2 \left( 1 - \frac{2 m_t}{r} \right)^{-\frac{1}{2}} \delta n_I(r) dr
    \right.\\&\left.
    + \sum_{J=1,2} \int_0^{\infty} r \left( 1 - \frac{2 m_t}{r} \right)^{-\frac{3}{2}} n_I(r) \delta m_J(r) dr \right].
\end{aligned}
\end{equation}

Using the isentropic conditions:
\begin{equation}
\delta n_I(r) = \frac{n_I(r)}{p_I(r) + \varepsilon_I(r)} \delta \varepsilon_I(r)
\end{equation}
and the mass variation:
\begin{equation}
\delta m_I(r) = 4\pi \int_0^r r'^2 \delta \varepsilon_I(r') dr',
\end{equation}
We do substitution by steps. First, substitute \( \delta n_I \):
\begin{equation}
\begin{aligned}
    \delta H =& 4\pi \sum_{I=1,2} \int_0^{\infty} r^2 \delta \varepsilon_I(r) dr - 4\pi \sum_{I=1,2} \left[ \lambda_I \right.\\&\left. \times \int_0^{\infty} r^2 \left( 1 - \frac{2 m_t}{r} \right)^{-\frac{1}{2}} \frac{n_I(r)}{p_I(r) + \varepsilon_I(r)} \delta  \varepsilon_I(r) dr\right] \\&- 4\pi \sum_{I=1,2} \lambda_I \sum_{J=1,2} \int_0^{\infty} \left[r \left( 1 - \frac{2 m_t}{r} \right)^{-\frac{3}{2}} \right.\\&\left.n_I(r) \delta m_J(r) dr\right].
\end{aligned}
\end{equation}

Now substitute \( \delta m_J(r) \) into the last term:
\begin{equation}
\begin{aligned}
    &- 4\pi \sum_{I=1,2} \lambda_I \sum_{J=1,2} \int_0^{\infty} r \left( 1 - \frac{2 m_t}{r} \right)^{-\frac{3}{2}} \\&
    n_I(r) 
    \left[ 4\pi \int_0^r r'^2 \delta \varepsilon_J(r') dr' \right] dr
\end{aligned}
\end{equation}

Defining
\begin{equation}
\begin{aligned}
    D(r) = \sum_{K=1,2} \lambda_K \int_{r}^{\infty} dr' \, r' \left( 1 - \frac{2 m_t(r')}{r'} \right)^{-\frac{3}{2}} n_K(r')
\end{aligned}
\end{equation} and exchanging the order of integration in the double integral, the last term becomes:
\begin{equation}
 - 16\pi^2 \sum_{J=1,2} \int_0^{\infty} dr \, r^2 \delta \varepsilon_J(r) D(r)
\end{equation}

Now combine all terms in \( \delta H \):
\widetext
\begin{equation}
\begin{aligned}
    \delta H =& 4\pi \sum_{I=1,2} \int_0^{\infty} r^2 \delta \varepsilon_I(r) dr - 4\pi \sum_{I=1,2} \lambda_I \int_0^{\infty} r^2 \left( 1 - \frac{2 m_t(r)}{r} \right)^{-\frac{1}{2}} \frac{n_I(r)}{p_I(r) + \varepsilon_I(r)} \delta \varepsilon_I(r) dr \\&- 16\pi^2 \sum_{J=1,2} \int_0^{\infty} dr \, r^2 \delta \varepsilon_J(r) D(r) \\
    =& 4\pi \sum_{I=1,2} \int_0^{\infty} r^2 \delta \varepsilon_I(r) \left[ 1 - \lambda_I \left( 1 - \frac{2 m_t(r)}{r} \right)^{-\frac{1}{2}} \frac{n_I(r)}{p_I(r) + \varepsilon_I(r)} - 4\pi D(r) \right] dr \\
    =& 4\pi \sum_{I=1,2} \int_0^{\infty} r^2 \delta \varepsilon_I(r) \left[ 1 - \lambda_I \left( 1 - \frac{2 m_t(r)}{r} \right)^{-\frac{1}{2}} \frac{n_I(r)}{p_I(r) + \varepsilon_I(r)} - 4\pi \sum_{J=1,2} \lambda_J \int_{r}^{\infty} r' 
    \right.\\&\left.
    \left( 1 - \frac{2 m_t(r')}{r'} \right)^{-\frac{3}{2}}
    n_J(r') dr' \right] dr.
\end{aligned}
\end{equation}
\endwidetext

For \( \delta H \) to be zero for all variations \( \left[ \left\{ \delta \varepsilon _I\left( r \right) \right\} \right]  \), the integrand must vanish for each fluid component \( I \):
\begin{equation}
\begin{aligned}
    &1 - \lambda_I A(r)^{-1/2} \frac{n_I(r)}{p_I(r) + \varepsilon_I(r)} \\&
    - 4\pi \sum_{J=1,2} \lambda_J \int_{r}^{\infty} r' A(r')^{-3/2} n_J(r') dr' = 0,
\end{aligned}
\end{equation}
where $A(r) = 1 - \frac{2 m_t(r)}{r}$. This is the extremization condition for the two-fluid system.

Define:
\begin{equation}
F(r) = 4\pi \sum_{J=1,2} \lambda_J \int_{r}^{\infty} r' A(r')^{-3/2} n_J(r') dr'.
\end{equation}

Then the condition becomes:
\begin{equation}
\lambda_I A(r)^{-1/2} \frac{n_I(r)}{p_I(r) + \varepsilon_I(r)} + F(r) = 1 \quad \text{for } I=1,2.
\end{equation}

From these two equations, we obtain a key constraint that reflects the effect of Lagrange multipliers in this two-fluid system:
\begin{equation}
\lambda_1 A(r)^{-1/2} \frac{n_1(r)}{p_1(r) + \varepsilon_1(r)} = \lambda_2 A(r)^{-1/2} \frac{n_2(r)}{p_2(r) + \varepsilon_2(r)},
\end{equation}
which implies:
\begin{equation}
\frac{\lambda_1 n_1}{p_1 + \varepsilon_1} = \frac{\lambda_2 n_2}{p_2 + \varepsilon_2}.
\end{equation}
This reflects that the chemical potential between the two fluids satisfies a proportional relationship, which is somewhat different from the case of a single fluid \cite{weinberg2013gravitation}.

Define a common function:
\begin{equation}
\Gamma(r) = \lambda_I \frac{n_I(r)}{p_I(r) + \varepsilon_I(r)} \quad, \text{for } I=1,2.
\end{equation}

Differentiate the condition with respect to $r$:
\begin{equation}
\begin{aligned}
    &\frac{d}{dr} \left[ \lambda_I A(r)^{-1/2} \frac{n_I(r)}{p_I(r) + \varepsilon_I(r)} \right] + F'(r) = 0,
\\& \text{for } I=1,2.
\end{aligned}\label{eq:cond_derivate}
\end{equation}
where:
\begin{equation}
F'(r) = -4\pi \sum_{J=1,2} \lambda_J r A(r)^{-3/2} n_J.
\end{equation}

Take the derivative in \eqref{eq:cond_derivate} and substitute the isentropic condition $n_I' = \frac{n_I \varepsilon_I'}{p_I + \varepsilon_I}$:
\begin{equation}
\begin{aligned}
    &\lambda_I \left[ - n_I A^{-1/2} \frac{ p_I' }{(p_I + \varepsilon_I)^2} 
    \right.\\&\left.
    - \frac{1}{2} n_I A^{-3/2} \frac{ A' }{p_I + \varepsilon_I} \right] - 4\pi \sum_{J=1,2} \lambda_J r A^{-3/2} n_J = 0,
    \\& \text{for } I=1,2.
\end{aligned}
\end{equation}

Multiply by $A^{3/2}$ and $(p_I + \varepsilon_I)^2$, then rearrange to solve for $p_I'$:
\begin{equation}
p_I' = - \frac{1}{2} \frac{ A' }{A} (p_I + \varepsilon_I) - 4\pi \frac{ \sum_{J=1,2} \lambda_J r n_J }{\lambda_I n_I A} (p_I + \varepsilon_I)^2.\label{eq:pprime_I}
\end{equation}

From the constraint definition $\Gamma(r) = \lambda_I \frac{n_I}{p_I + \varepsilon_I}$, we have:
\begin{equation}
\lambda_I n_I = \Gamma(r) (p_I + \varepsilon_I),
\end{equation}
\begin{equation}
\sum_{J=1,2} \lambda_J n_J = \Gamma(r) (p_1 + \varepsilon_1 + p_2 + \varepsilon_2) = \Gamma(r) (p_t + \varepsilon_t),
\end{equation}
Thus:
\begin{equation}
\frac{ \sum_{J=1,2} \lambda_J n_J }{\lambda_I n_I} = \frac{ p_t + \varepsilon_t }{ p_I + \varepsilon_I }.
\end{equation}

Substitute back to \eqref{eq:pprime_I}:
\begin{equation}
\begin{aligned}
    p_I' =& - \frac{1}{2} \frac{ A' }{A} (p_I + \varepsilon_I) - 4\pi r \frac{ p_t + \varepsilon_t }{ A } (p_I + \varepsilon_I)
\\=&- (p_I + \varepsilon_I) \left( \frac{1}{2} \frac{ A' }{A} + 4\pi r \frac{ p_t + \varepsilon_t }{ A } \right).
\end{aligned}
\end{equation}

Now compute $A'$: using $m_t' = 4\pi r^2 \varepsilon_t$ we obtain
\begin{equation}
A' = 2 \left( \frac{m_t}{r^2} - 4\pi r \varepsilon_t \right).
\end{equation}

Substitute into the expression for $p_I'$:
\begin{equation}
\begin{aligned}
    p_I' =& - (p_I + \varepsilon_I) \frac{1}{A} \left( \frac{1}{2} A' + 4\pi r ( p_t + \varepsilon_t ) \right)
\\
=& - (p_I + \varepsilon_I) \frac{1}{A} \left( \frac{ m_t}{r^2} + 4\pi r p_t \right)
\\
=& - (p_I + \varepsilon_I) \frac{m_t + 4\pi r^3 p_t }{ r (r - 2 m_t ) }.
\end{aligned}
\end{equation}
This is exactly the two-fluid equilibrium equation:
\begin{equation}
\begin{aligned}
    &\frac{dp_I}{dr} = - \left( \varepsilon_I(r) + p_I(r) \right) \frac{m_t(r) + 4\pi r^3 p_t(r) }{ r\left( r - 2 m_t(r) \right) }, \\& \text{for } I=1,2.
\end{aligned}
\end{equation}

We have explicitly shown that the condition $\delta H = 0$ leads directly to the TOV equations. Conversely, since every step in the derivation above (substitution, differentiation, algebraic rearrangement) is reversible, if one starts from the TOV equations and chooses the constants $\lambda_1,\lambda_2$ such that $\lambda_1/\lambda_2 = \mu_1(r)/\mu_2(r)$, one can integrate back to show that $\delta H=0$ (here $\mu_I(r) \equiv (p_I+\varepsilon_I)/n_I$ is the local chemical potential of zero temperature fluid $I$). This completes the proof of necessity and sufficiency.
\end{proof}


\subsubsection{Theorem 2: Zero Modes and Parallel Gradients}

This conjecture was initially proposed in Henriques' work \cite{HENRIQUES1990511}. We demonstrate it by using conservation laws to extend a theorem that holds in a single-fluid spherically symmetric static compact star (see \cite{weinberg2013gravitation}) to the case of two fluids. A complete proof of the converse direction, together with a more comprehensive and mathematically rigorous analysis, are presented in a separate work \cite{Chen2025Equivalence}.

\begin{theorem}\label{Theorem:2}
For a two-fluid equilibrium configuration, the three gradients with respect to the central-pressure parameters,
$\nabla_{\boldsymbol{\sigma}} M$, $\nabla_{\boldsymbol{\sigma}} N_{\mathrm{NM}}$, and $\nabla_{\boldsymbol{\sigma}} N_{\mathrm{DM}}$,
are mutually parallel in the central-pressure parameter space if there exists at least a zero mode with respect to radial perturbation.
\end{theorem}

\begin{proof}
If for a certain static equilibrium configuration, there exists $n \in \mathbb{N}$ such that $\omega_{n}^{2} = 0$ under radial perturbation, the corresponding eigenmode is strictly time-independent:
\begin{equation}
\xi_{In}(r,t) = \xi_{In}(r) e^{-i\omega_n t} = \xi_{In}(r).
\end{equation}
This static solution satisfies the linearized field equations and corresponds to an infinitesimal variation between the equilibrium configuration $\boldsymbol{\sigma}$ and a neighboring equilibrium configuration $\boldsymbol{\sigma} + \delta\boldsymbol{\sigma}$, where $\delta\boldsymbol{\sigma} = (\delta p_2^c, \delta p_1^c)$ is the direction in the central-pressure parameter space associated with the displacement field $\xi_{In}(r)$.

The connection between the existence of static solutions and the critical curve relies on conserved quantities. Recalling the derivation of the two-fluid equilibrium and oscillation equations, the definitions of the physical quantities and their perturbations in terms of Lagrangian displacements originate from the Einstein field equations and the covariant conservation of the particle number density for each fluid $\nabla _{\mu}J_{I}^{\mu}=\nabla _{\mu}\left( n_Iu_{I}^{\mu} \right) =0$\label{eq.ni_conserve}, together with their linearizations. For a perturbation solution such as the zero-frequency mode, the linearized particle number conservation directly implies that the first-order variation of the total particle number vanishes:
\begin{equation}
\delta N_I = 0 .
\end{equation}


We now explicitly verify this natural conclusion of particle number conservation:

\begin{itemize}

\item \textbf{Symbol definition: Conserved Charges and Currents}

We define the conserved charges and currents as:
\begin{equation}
    \begin{aligned}
    J_{N_I}^{\mu} \equiv n_I u^{\mu}, \,
    Q_{N_I} \equiv -\int_{\Sigma} J_{N_I}^{\mu} a_{\mu} \sqrt{h} \, d^3 x,
\end{aligned}
\end{equation}
where \(\Sigma\) is a spacelike hypersurface of constant coordinate time \(t\), \(a_{\mu} = (-e^{\phi},0,0,0)\) is the future-directed unit normal vector, \(h\) is the determinant of the induced 3-metric, and \(\sqrt{h}\,d^{3}x = e^{\nu} r^{2}\sin\theta\, dr\,d\theta\,d\phi\).

\item \textbf{Explicit Form of Conserved Charges}

Define \( B = e^{2\phi}, A = e^{-2\nu} \). For \( i = N_I \), the conserved charge is explicitly given by:
\begin{equation}
\begin{aligned}
Q_{N_I} &= -\int_{\Sigma}{J_{N_I}^{\mu}a_{\mu}\sqrt{h}d^3x} \\
&= \int_{\Sigma}{J_{N_I}^{0}\sqrt{B}\sqrt{h}d^3x} \\
&= \int_{\Sigma}{J_{N_I}^{0}\sqrt{-g}d^3x} \\
&= \int_{\Sigma}{n_I u^0 r^2 \frac{\sqrt{B}}{\sqrt{A}} \sin \theta \, dr \, d\phi \, d\theta} \\
&= \int_0^\infty 4\pi r^2 n_I u^0 \frac{\sqrt{B}}{\sqrt{A}} \, dr \\
&= \int_0^\infty 4\pi r^2 n_I \frac{1}{\sqrt{B}} \frac{\sqrt{B}}{\sqrt{A}} \, dr \\
&= \int_0^\infty 4\pi r^2 n_I \left( 1 - \frac{2m_{\text{total}}}{r} \right)^{-\frac{1}{2}} dr \\
&= N_I.
\end{aligned}
\end{equation}

\item \textbf{Covariant Conservation Law}

The covariant conservation law is given by:
\[
\nabla_{\mu} J_{N_I}^{\mu} = 0.
\]

\item \textbf{Proof of Conservation}

Consider two adjacent spacelike hypersurfaces \(\Sigma_1\) and \(\Sigma_2\), which together with a timelike boundary surface form a closed hypersurface. Since no current flows through the boundary, the flux of \( J_{N_I}^{\mu} \) through the boundary vanishes.

The change in the conserved charge from \(\Sigma_1\) to \(\Sigma_2\) is equal to the total flux of \( J_{N_I}^{\mu} \) through the closed hypersurface.

Applying the Stokes' theorem, this flux equals the volume integral of the covariant divergence of \( J_{N_I}^{\mu} \) over the enclosed 4-volume:
\[
\Delta Q_{N_I} = \int \nabla_{\mu} J_{N_I}^{\mu} \sqrt{-g} d^4 x.
\]

By the covariant conservation law \( \nabla_{\mu} J_{N_I}^{\mu} = 0 \), we obtain:
\[
\Delta Q_{N_I} = 0,
\]
which implies \( \Delta N_I = 0 \), thus proving that the covariant conservation law leads to the global conservation of the total particle numbers $N_1$ and $N_2$.
\end{itemize}

Since the zero‑frequency mode is a static solution of the radial perturbation equations, and the equations incorporate the covariant conservation of the particle number density for each fluid, the linearized system directly forces the first‑order variations of the particle numbers to vanish:
\begin{equation}
\delta N_1 = 0,\qquad \delta N_2 = 0 .
\label{eq:deltaN_zero}
\end{equation}

For the mass variation, a similar conserved‑current construction is not available: the Komar energy current built from a timelike Killing vector field does not yield the total gravitational mass $M$ defined in our framework (the two differ by a metric-function factor), and its conservation relies on a global timelike Killing symmetry that is absent once radial perturbations are excited. Nevertheless, for the static zero‑frequency mode we can determine the first‑order behaviour of $M$ through the variational principle. Theorem~\ref{Theorem:1} established that, for an equilibrium configuration, there exist constants $\lambda_1,\lambda_2$ such that for any static variation
$\delta M - \lambda_1\,\delta N_1 - \lambda_2\,\delta N_2 = 0$.
For the zero‑frequency static mode, the conservation $\delta N_1=\delta N_2=0$ reduce the variational relation to $\delta M = 0$ (to the first order).

Based on the above, we can conclude: if an equilibrium configuration possesses a radial normal mode with zero eigenvalue, then this mode enters the tangent space of this configuration point in the equilibrium space, and along this tangent vector the first order variations of the mass and particle number functionals vanish, i.e., $\delta M[\boldsymbol{\sigma}; \delta\boldsymbol{\sigma}]=\delta N[\boldsymbol{\sigma}; \delta\boldsymbol{\sigma}]=0$. Consequently, the directional derivatives of $M$ and $N_I$ along the same vector $\delta\boldsymbol{\sigma}$ in parameter space are zero:
\begin{equation}
\nabla_{\boldsymbol{\sigma}}M\cdot\delta\boldsymbol{\sigma}=0,\quad \nabla_{\boldsymbol{\sigma}}N_1\cdot\delta\boldsymbol{\sigma}=0,\quad \nabla_{\boldsymbol{\sigma}}N_2\cdot\delta\boldsymbol{\sigma}=0.
\end{equation}
This is equivalent to the parallelism of the gradients of $M$ and $N_I$:
\begin{equation}
\nabla M\parallel \nabla N_1\parallel \nabla N_2,
\end{equation}
where the symbol $\parallel$ denotes the parallel relation.
\end{proof}

\textbf{Analysis and Conclusion:}
Since the Lagrange multipliers can be rescaled arbitrarily (for example, dividing $\nabla M = \lambda_1 \nabla N_{NM} + \lambda_2 \nabla N_{DM}$ by $\lambda_1$ or $\lambda_2$), the statement of Theorem~\ref{Theorem:1} remains valid under any permutation of $M$, $N_{NM}$, and $N_{DM}$. This permutation symmetry implies that if any two of the three gradients ($\nabla M$, $\nabla N_{NM}$, and $\nabla N_{DM}$) are parallel (indicating simultaneous extrema), the third must also be parallel to them. Hence, the critical curve can be determined by the parallelism of any two gradients.

So far, we have established that the stability boundary (the locus where the zero fundamental modes arise) constitutes a sufficient condition for the one-dimensional sequence derived via the critical curve method. Nevertheless, to verify whether the resulting curve indeed corresponds to the stability boundary, supplementary tests are necessary. A straightforward approach involves conducting radial oscillation analyses on two sample configurations situated on opposite sides of the curve. Alternatively, leveraging astrophysical insights (such as data from white dwarfs and neutron stars), we can often directly estimate the appropriate range of central parameters, thereby enabling the direct determination of the stability boundary. Hereby we can treat the static method as a reliable and efficient criteria for rigorous stability analysis.

\subsubsection{Physical Interpretation via Energy Considerations}
The stability of an equilibrium configuration, which is an extremum of $M$, is determined by the second-order variation $\delta^2 M$. The connection to dynamical stability is given by the thermodynamic relation \cite{Friedman:1978,Caballero_2024}:
\begin{equation}
\operatorname{sgn}(\delta^{2} M[\boldsymbol{\xi}]) = \operatorname{sgn}(\omega_0^2),
\end{equation}
where $\boldsymbol{\xi}$ is the perturbation mode. The stability conditions are:
\begin{itemize}
\item $\delta^{2} M > 0 \Leftrightarrow \omega_0^2 > 0$: Stable configuration (energy minimum).
\item $\delta^{2} M = 0 \Leftrightarrow \omega_0^2 = 0$: Stability boundary (marginal stability).
\item $\delta^{2} M < 0 \Leftrightarrow \omega_0^2 < 0$: Unstable configuration (saddle point).
\end{itemize}
This framework provides the physical intuition: a stable star resides at the bottom of an energy well. The stability boundary is reached when the well becomes flat in some direction, allowing for a transition to another state without energy cost, which is precisely the correspondence of a zero-frequency fundamental mode.

\section{Results}\label{sec:Results}

\begin{figure*}[!htb] 
\centering
\begin{subfigure}[b]{0.48\textwidth}
\centering
\includegraphics[width=\linewidth]{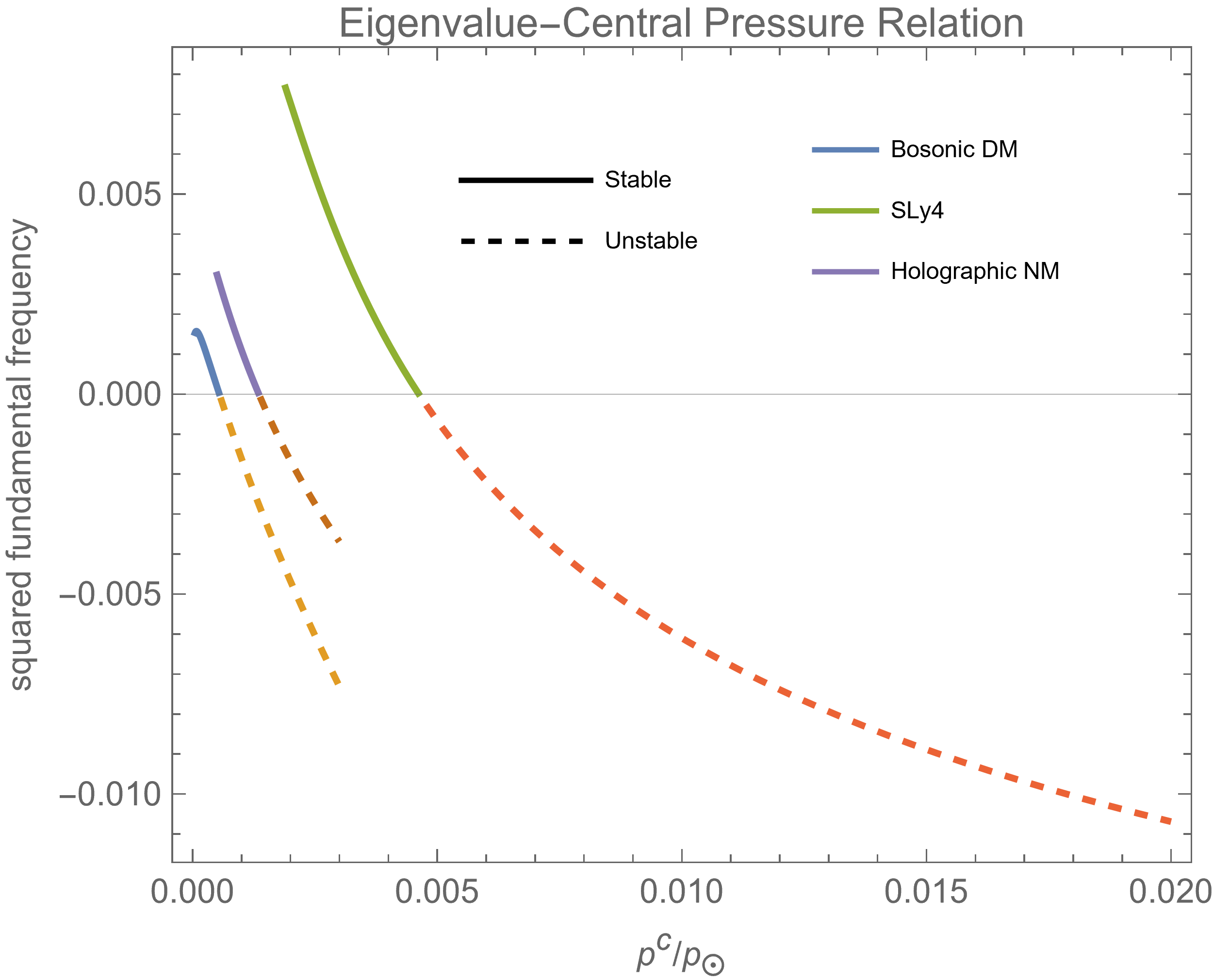} 
\caption{Single fluid eigenvalue curves as functions of central configuration for different EoS.}
\label{fig:single_fluid_eigenvalue_a}
\end{subfigure}
\hfill
\begin{subfigure}[b]{0.48\textwidth}
\centering
\includegraphics[width=\linewidth]{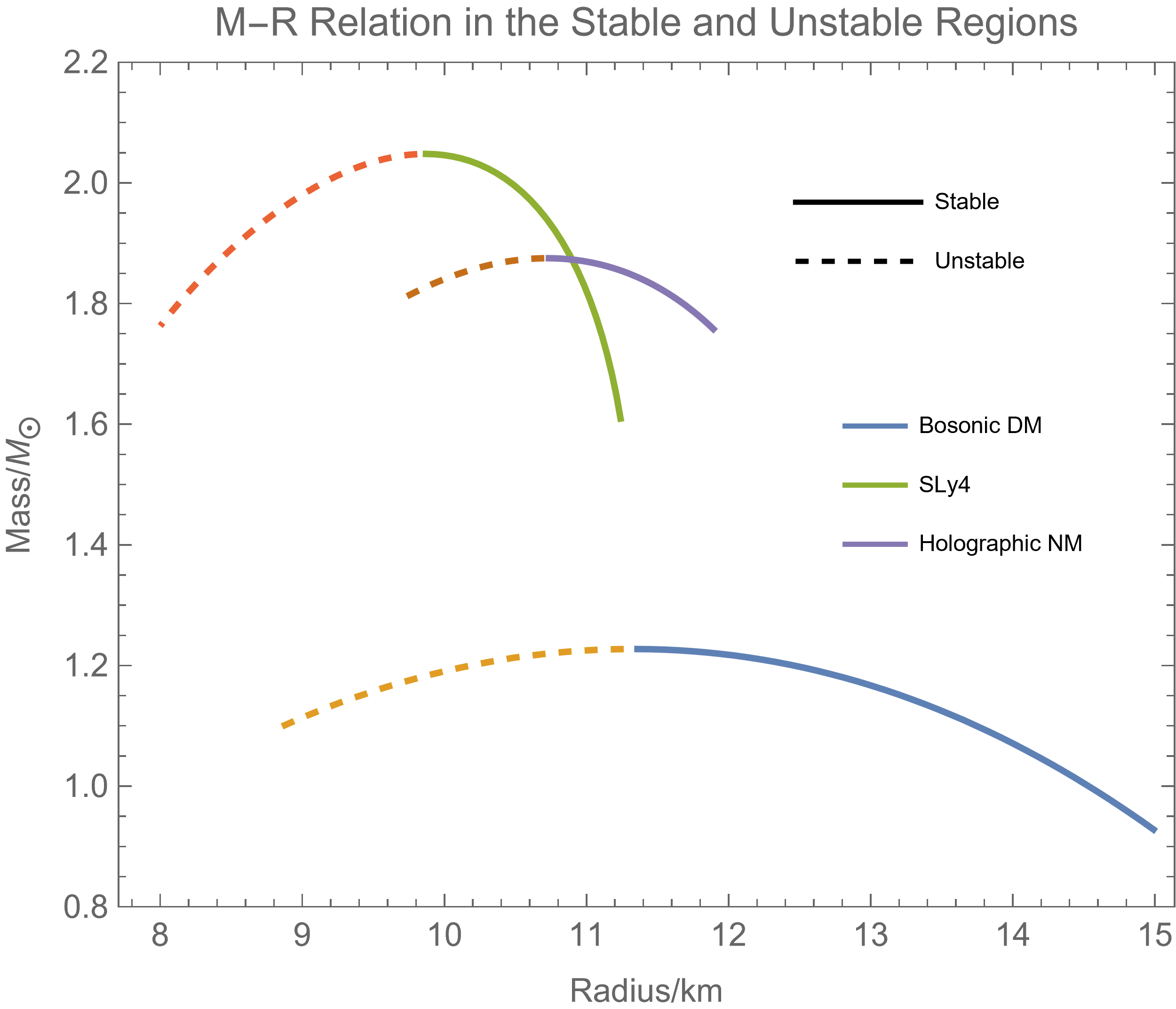} 
\caption{Corresponding $M$-$R$ relation curves.}
\label{fig:single_fluid_eigenvalue_b}
\end{subfigure}
\caption{Single fluid stability analysis. Left panel (Fig.\ref{fig:single_fluid_eigenvalue_a}) shows fundamental eigenvalue curves as functions of central configuration for different EoS: Bosonic DM ($B_4 = 0.1$), SLy4, and Holographic NM ($\ell ^{-7}=10300$) respectively. Right panel (Fig.\ref{fig:single_fluid_eigenvalue_b}) displays the corresponding $M$-$R$ relation curves.}
\label{fig:single_fluid_eigenvalue}
\label{fig.3}
\end{figure*}

\begin{figure*}[!htb] 
\centering
\begin{subfigure}[b]{0.48\textwidth}
\centering
\includegraphics[width=\linewidth]{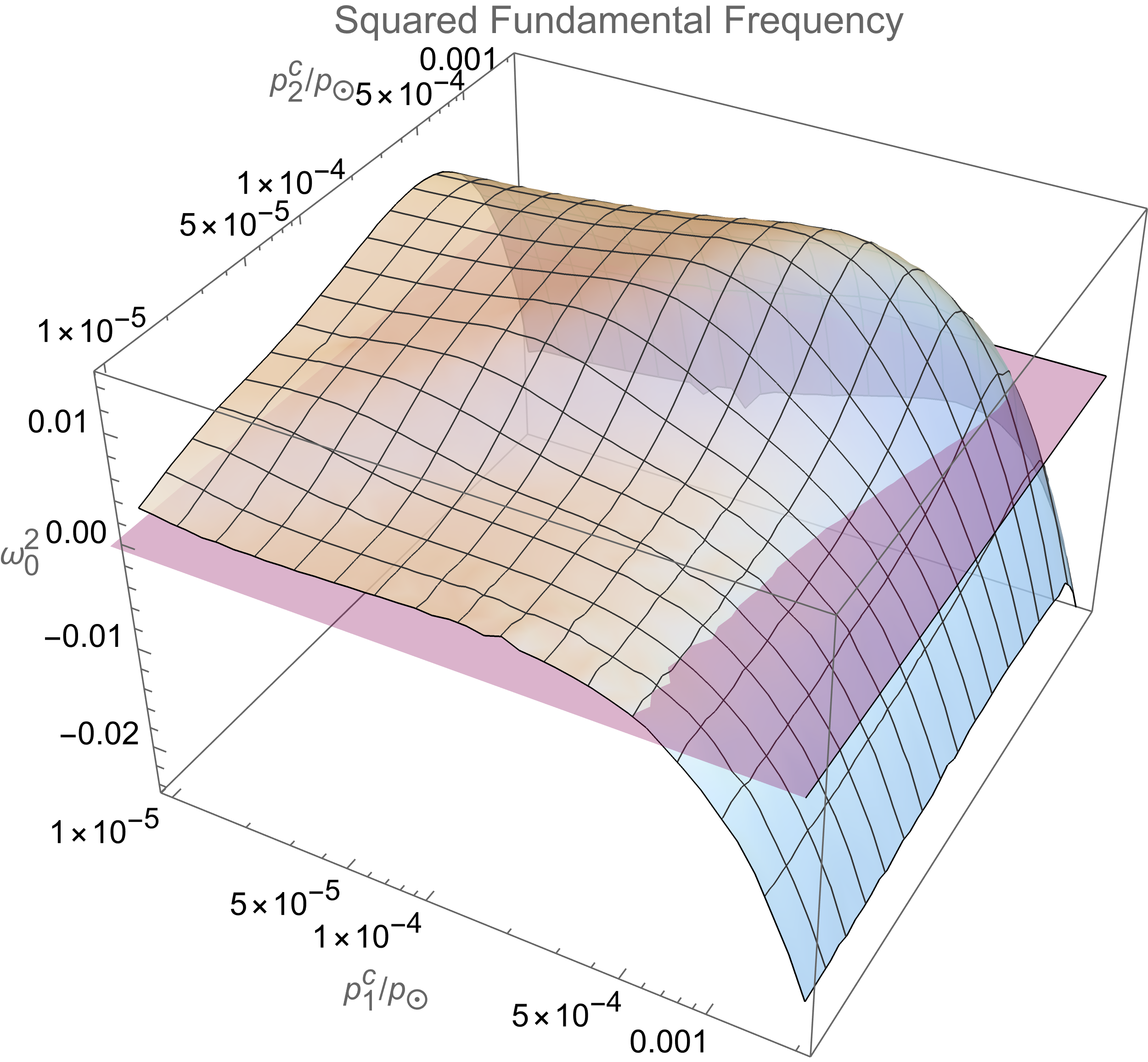} 
\caption{Continuous distribution of the fundamental eigenvalues of radial oscillation modes}
\label{fig:shooting_a}    
\end{subfigure}
\hfill
\begin{subfigure}[b]{0.48\textwidth}
\centering
\includegraphics[width=\linewidth]{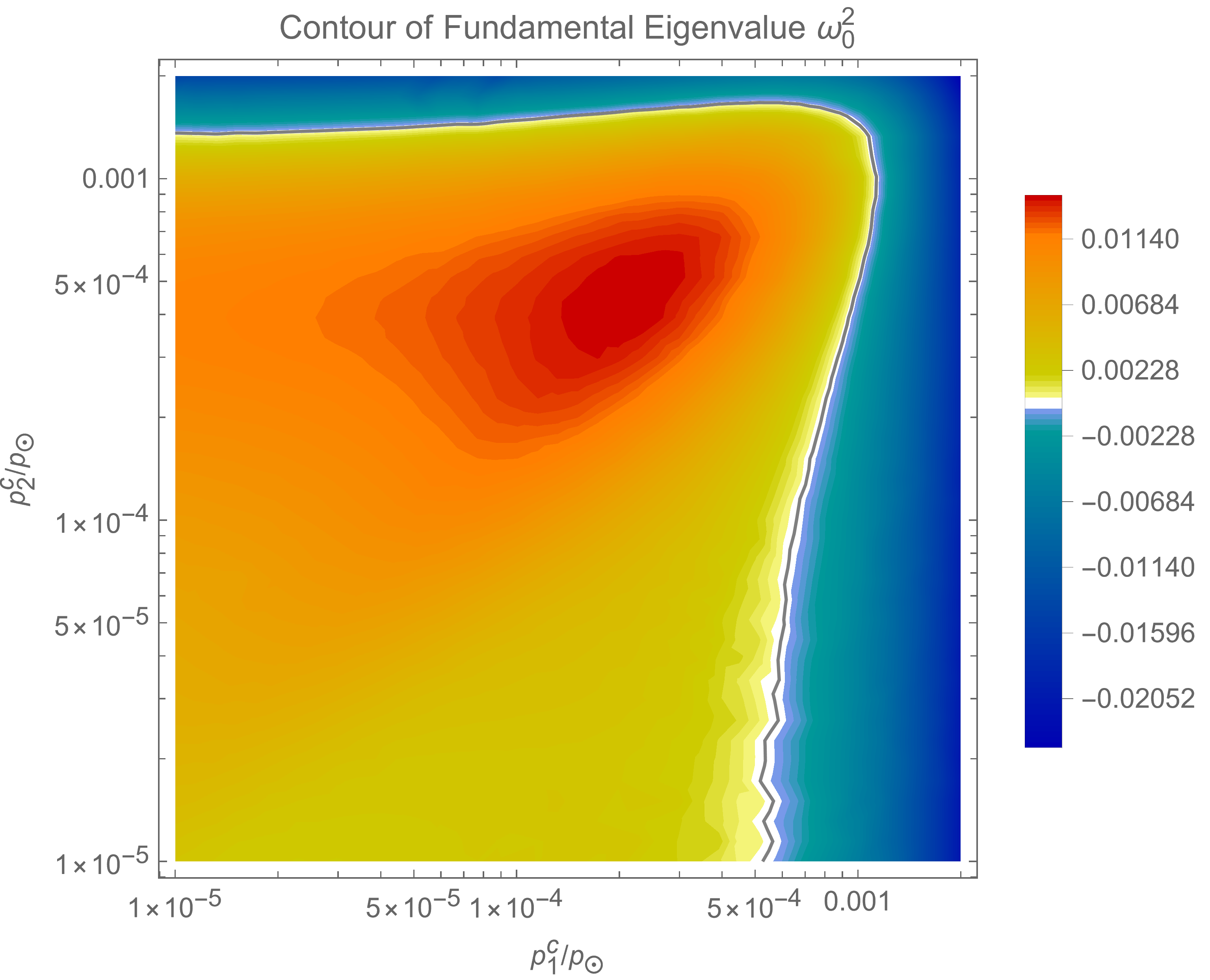} 
\caption{Distribution of stable and unstable configuration points in the central pressure parameter space.}
\label{fig:shooting_b}
\end{subfigure}
\caption{Radial oscillation analysis results for the mixed star model with EoS of Holographic NM ($\ell ^{-7}=10300$) and Bosonic DM ($B_4 = 0.1$). Left panel (Fig.\ref{fig:shooting_a}) shows the continuous distribution of eigenvalues. Right panel (Fig.\ref{fig:shooting_b}) displays the distribution of stable and unstable configuration points in the central pressure parameter space, where the black curve represents configurations where the fundamental radial mode frequency vanishes and its interior indicates the stable region.}
\label{fig:4}
\end{figure*}

This section presents the numerical results of our investigation into the stability properties and macroscopic characteristics of mixed compact stars. We begin with a detailed stability analysis comparing single-fluid and multi-fluid models, followed by an examination of the resulting $M$-$R$ relationships and their implications for astrophysical observations.

\subsection{stability analyze}

\subsubsection*{1. Single Fluid}

As shown in Fig.\ref{fig:single_fluid_eigenvalue}, we show the eigenvalue-central pressure relations for single-fluid models with three different 
EoS: SLy4, Bosonic DM (with $B_4=0.1$) and Holographic NM (with $\ell^{-7}=10300$). The stability distribution and $M$-$R$ relations satisfy the BTM criteria. These numerical results serve as comparisons for the situation of mixed components.

\begin{figure*}[!htb] 
\centering
\begin{subfigure}[b]{0.49\textwidth}
\centering
\includegraphics[width=\linewidth]{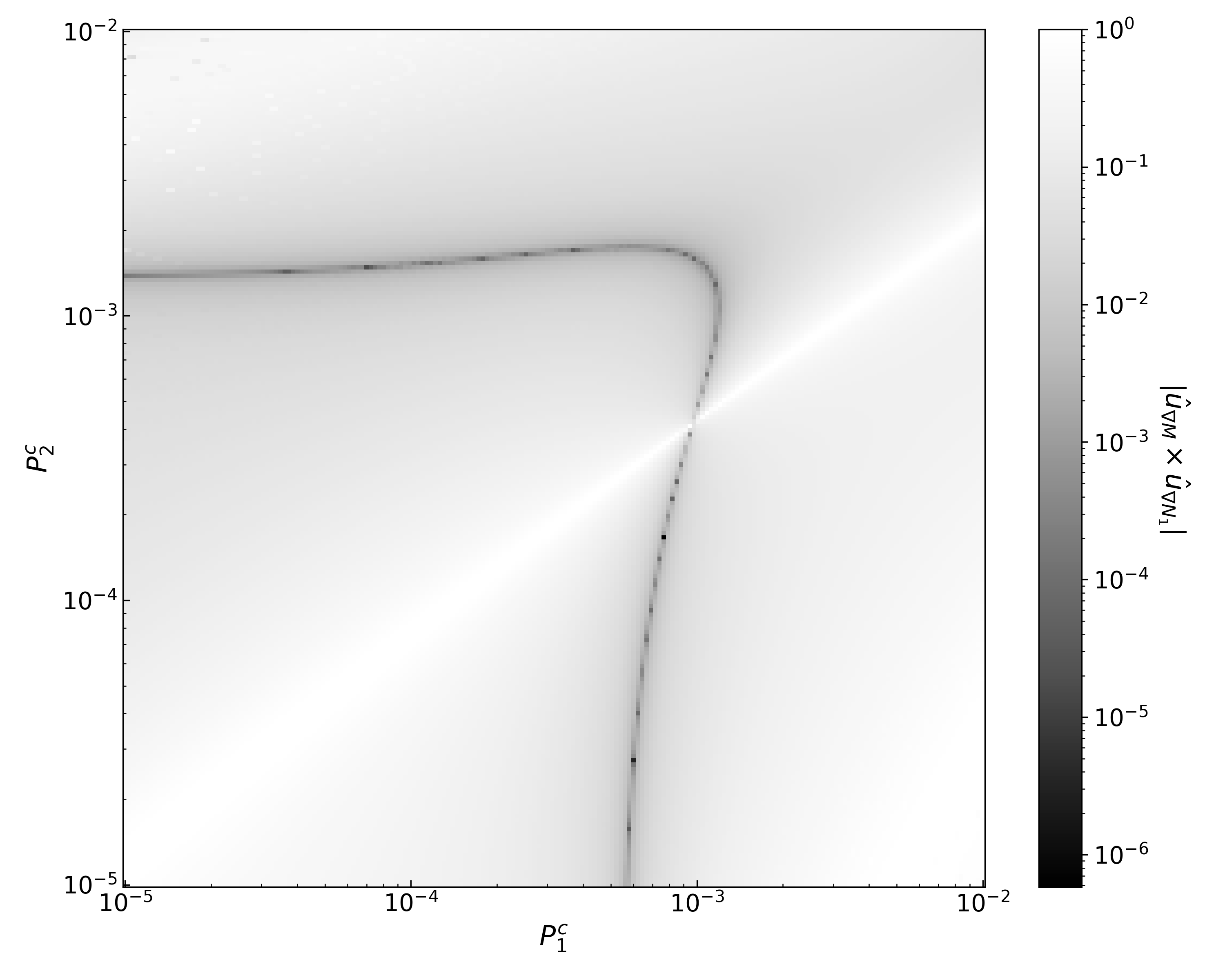} 
\caption{Critical curves for EoS combination 1: \newline
Holographic NM ($\ell ^{-7}=10300$) and Bosonic DM
($B_4 = 0.1$).}
\label{fig:B200_a}
\end{subfigure}
\begin{subfigure}[b]{0.49\textwidth}
\centering
\includegraphics[width=\linewidth]{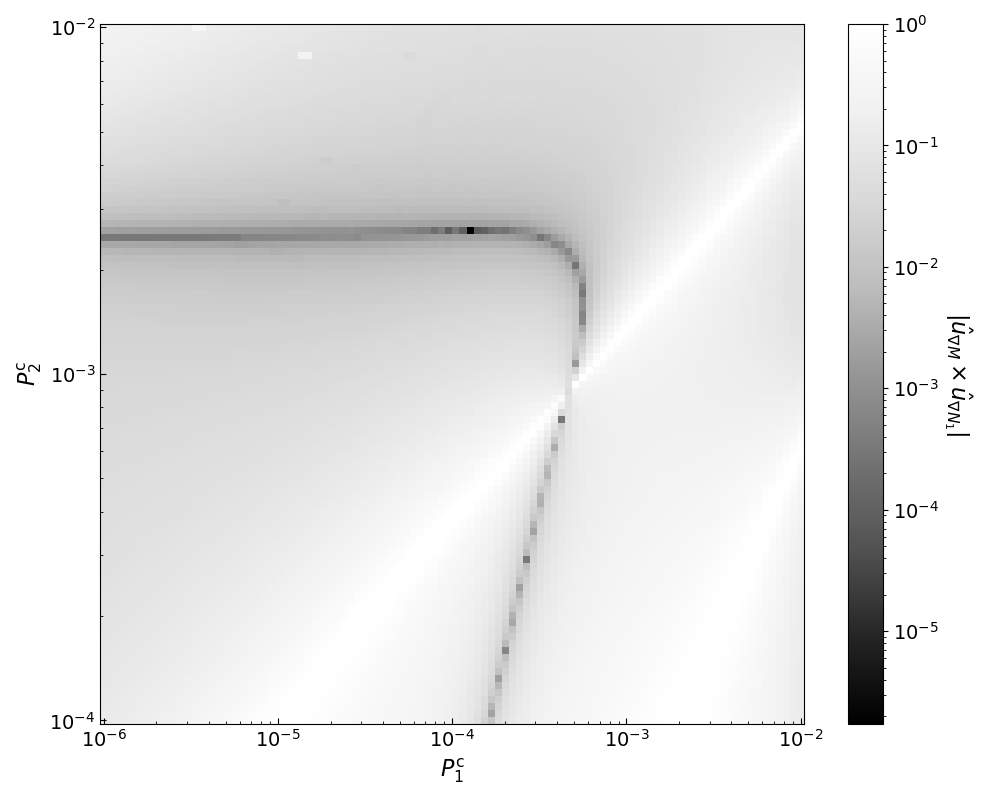} 
\caption{Critical curves for EoS combination 2: \newline
SLy4 and Bosonic DM ($B _4=0.03)$.}
\label{fig:B200_b}
\end{subfigure}
\begin{subfigure}[b]{0.49\textwidth}
\centering
\includegraphics[width=\linewidth]{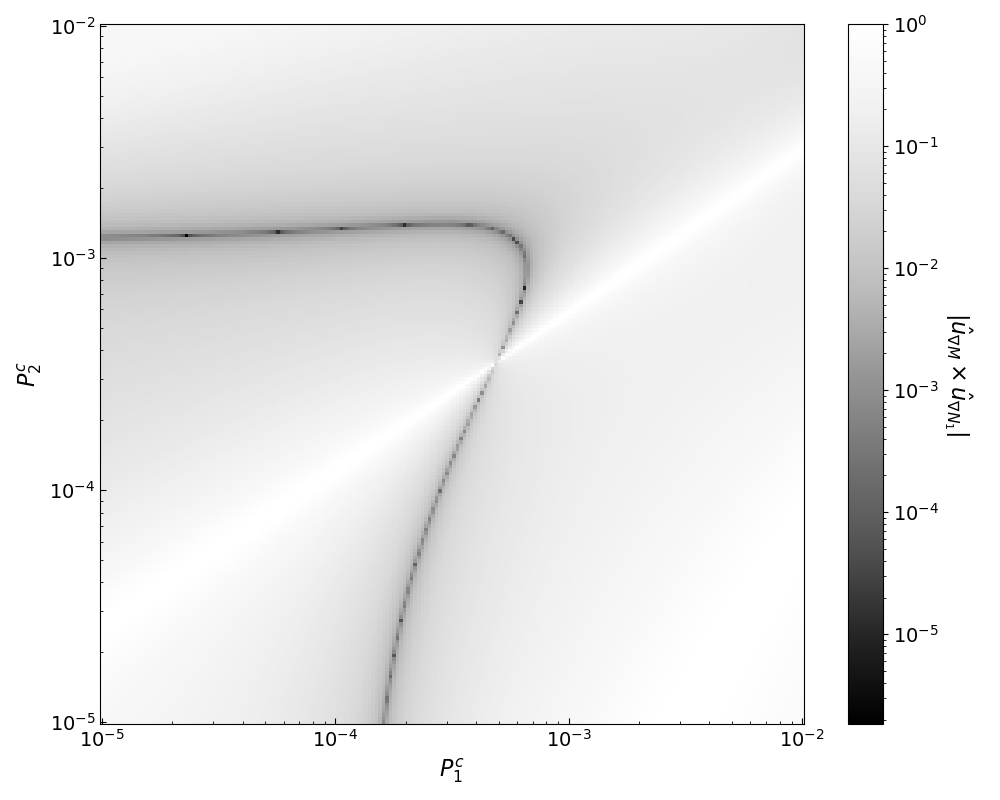} 
\caption{Critical curves for EoS combination 3: \newline
Holographic NM ($\ell ^{-7}=10300$) and Fermionic DM ($m_f=0.5$).}
\label{fig:B200_c}
\end{subfigure}
\begin{subfigure}[b]{0.49\textwidth}
\centering
\includegraphics[width=\linewidth]{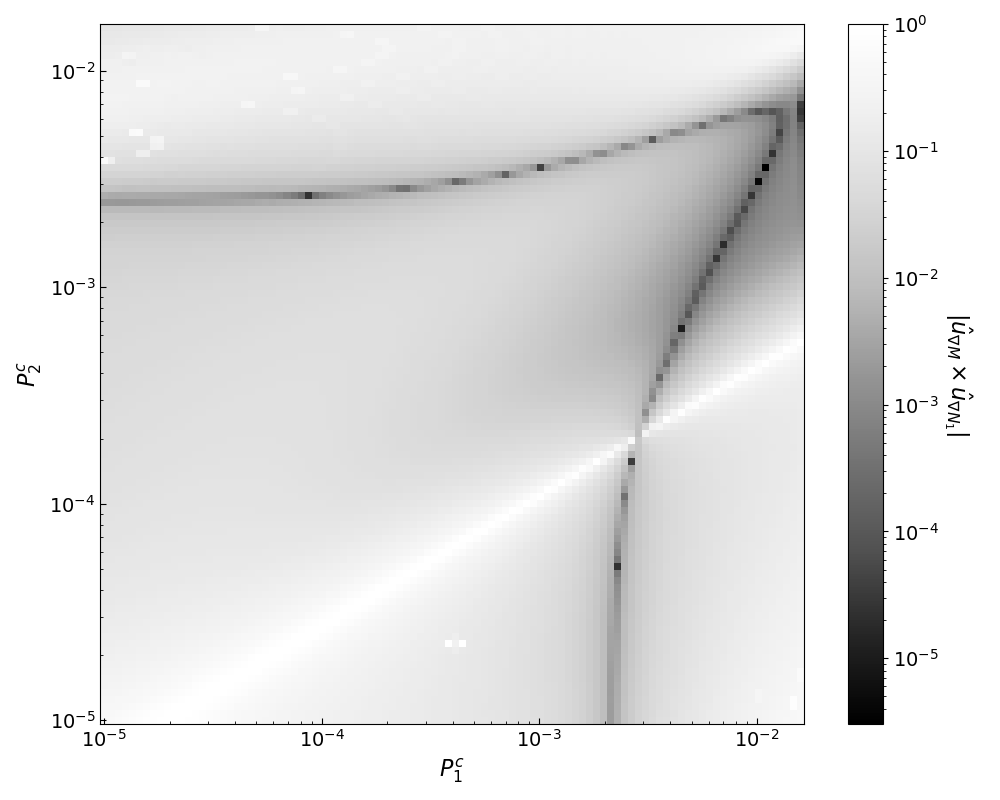} 
\caption{Critical curves for EoS combination 4: \newline
SLy4 and Fermionic DM ($m_f=0.5$).}
\label{fig:B200_d}
\end{subfigure}
\caption{Critical curves for different EoS combinations. The black curves indicate multiple stability boundaries. The units of both the horizontal and vertical axes are in astrophysical units ($p_\odot$), representing the magnitudes of central pressure for the two components. The black-and-white heatmap in the background displays the absolute value of normalized cross product of the gradient directions at each point of the total mass field and the particle number field of the second component; the darker the color, the more the gradient directions of the two fields at that point tend to be parallel. $N_1\equiv N_{DM}$, represents the DM particle number.}
\label{fig:Critical curves}
\end{figure*}

\begin{figure*}[!htb] 
\centering
\begin{subfigure}[b]{0.49\textwidth}
\centering
\includegraphics[width=\linewidth]{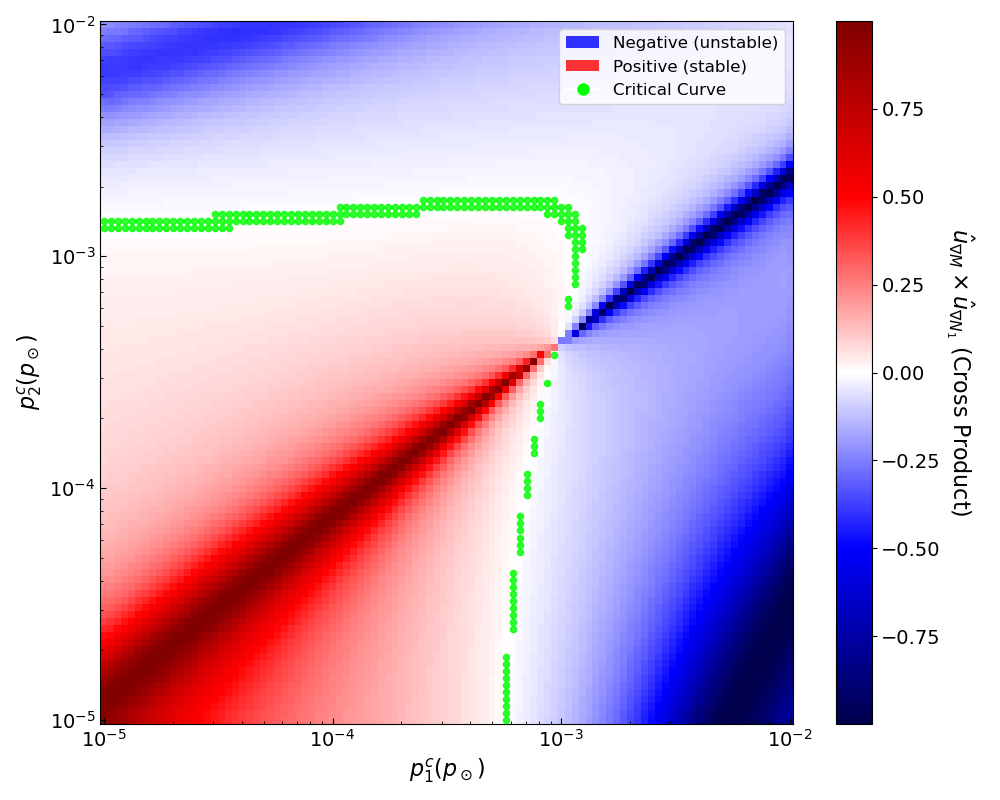} 
\caption{Heatmap of the signed value of cross product $\mathcal{C}_{M,N_1}$. Here the red and blue points refer to stable and unstable configurations. The green points construct the critical curve.}
\label{fig:new_a}    
\end{subfigure}
\hfill
\begin{subfigure}[b]{0.45\textwidth}
\centering
\includegraphics[width=\linewidth]{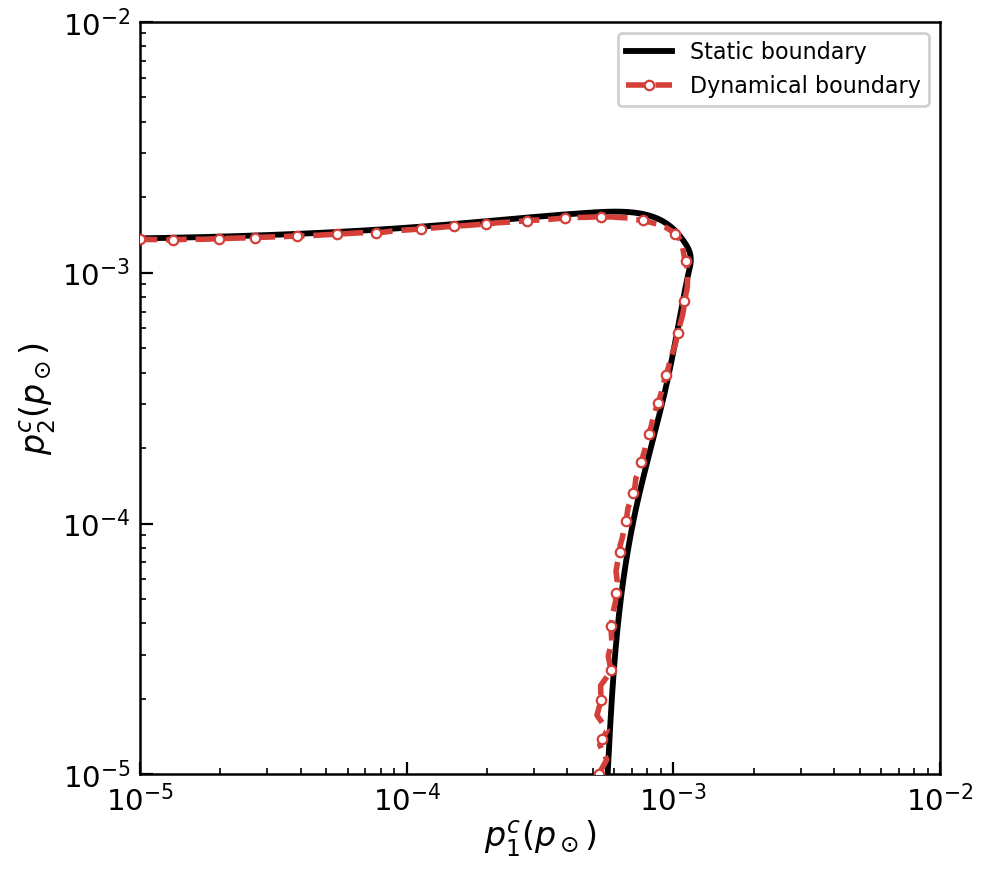} 
\caption{Direct comparison between the static critical curve and the dynamical zero-frequency boundary in the same central-pressure plane.}
\label{fig:new_b}
\end{subfigure}
\caption{Signed-cross-product critical curve and its comparison with the dynamical stability boundary. EoS combination:
Holographic NM ($\ell ^{-7}=10300$) and Bosonic DM
($B_4 = 0.1$).}
\label{fig:new figs_v2}
\end{figure*}

\subsubsection*{2. Multi-Fluid}

The numerical results in Fig.\ref{fig:shooting_a} provides precise squared frequencies of fundamental radial oscillation using Bosonic DM ($B_4 = 0.1$) and
Holographic NM ($\ell ^{-7}=10300$), thus in addition to identifying stable regions, it also reveals the trend of fundamental frequency variation, showing non-monotonic behavior.

To numerically identify the stability boundary in the static approach, we need to locate the points where the relevant gradients become parallel. Throughout this subsection, the gradients are defined in the two-dimensional central-pressure parameter space in Eq.~\ref{eq:parameter-space-gradient} rather than with respect to the radial coordinate inside the star. Thus, the gravitational mass $M$ and the particle numbers $N_I$ are treated as scalar functions on this parameter space.
Previous work usually relied on connecting the tangent points of several contour lines of mass and particle number \cite{PhysRevD.87.084040,kumar2025stabilityanalysistwofluidneutron}. We adopt a more efficient numerical method by computing the normalized signed cross product between gradients. Specifically, for two scalar quantities $X(\sigma)$ and $Y(\sigma)$, chosen from $M$, $N_1$, and $N_2$, we define

\begin{equation}
    \mathcal{C}_{X,Y}
    =
    \hat{u}_{\nabla_\sigma X}
    \times
    \hat{u}_{\nabla_\sigma Y}
    =
    \frac{
    \partial_{p_1^c}X\,\partial_{p_2^c}Y
    -
    \partial_{p_2^c}X\,\partial_{p_1^c}Y
    }{
    |\nabla_\sigma X|\,|\nabla_\sigma Y|
    },
    \label{eq:normalized_cross_product}
\end{equation}
where $\hat{u}_{\nabla_\sigma X}=\nabla_\sigma X/|\nabla_\sigma X|$ is the normalized gradient direction of $X$ in the central-pressure parameter space. The magnitude $|\mathcal{C}_{X,Y}|$ ranges from 0 to 1. Points with $|\mathcal{C}_{X,Y}|=0$ correspond to parallel gradients and form the critical curve, while points with $|\mathcal{C}_{X,Y}|=1$ correspond to orthogonal gradients.
Fig.~\ref{fig:Critical curves} shows the resulting critical curves for different EoS combinations using the method introduced in subsection~\ref{subsec:critical_curve}. In these plots, the black curves mark the zero-contours $|\mathcal{C}_{X,Y}|=0$, namely the locations of the critical curves, while the white curves mark the contours $|\mathcal{C}_{X,Y}|=1$.

Fig.\ref{fig:B200_a} uses Holographic NM with $\ell ^{-7}=10300$ and Bosonic DM
with $B_4 = 0.1$. Fig.\ref{fig:B200_b} uses SLy4 and Bosonic DM with $B _4=0.03$. Fig.\ref{fig:B200_c} uses Holographic NM with $\ell ^{-7}=10300$ and Fermionic DM with $m_f=0.5$. Fig.\ref{fig:B200_d} uses SLy4 and Fermionic DM with $m_f=0.5$. It is clear that Fig.\ref{fig:shooting_b} and Fig.\ref{fig:B200_a} confirm that, for these EoS models, the critical curve obtained from the static parallel‑gradient condition coincides with the boundary where the fundamental radial mode frequency vanishes. This numerically demonstrates that the zero mode identified by the critical curve method is indeed the fundamental mode, and thus the curve constitutes the true stability boundary for these cases.

In Fig.\ref{fig:new figs_v2}, we further illustrate the stability analysis using the signed cross product $\mathcal{C}_{M,N_1}$. By replacing the absolute value used in Fig.\ref{fig:B200_a} with the signed value of the cross product for the same EoS and parameters, we obtain the heatmap shown in Fig.\ref{fig:new_a}. The green points where $\mathcal{C}_{M,N_1}=0$ mark the static critical curve. In Fig.\ref{fig:new_b}, this static boundary is directly compared with the dynamical boundary obtained from the vanishing of the fundamental radial-mode frequency. The agreement between the two curves provides a numerical check of the equivalence between the static criterion and the dynamical stability condition for this EoS combination.

\begin{figure}[!htb]
    \centering
    \includegraphics[width=1\linewidth]{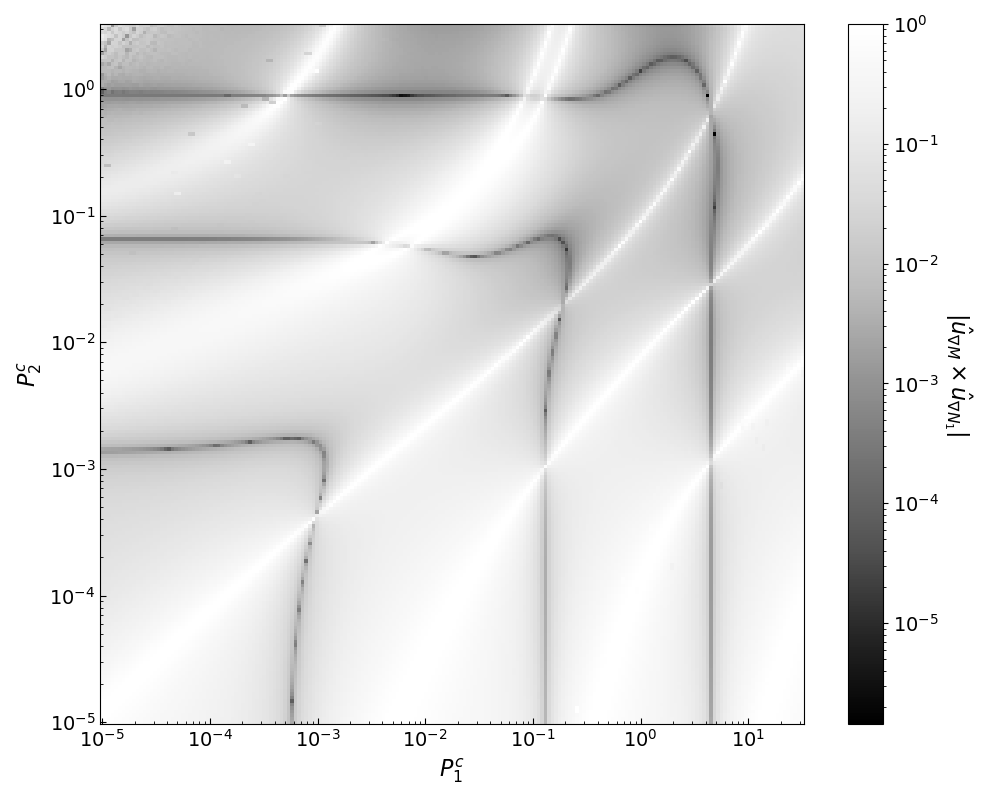}
    \caption{Static gradient-alignment structures in an enlarged central-pressure parameter space. The units of both the horizontal and vertical axes are astrophysical pressure units ($p_\odot$). This figure extends Fig.~\ref{fig:B200_a} using the same EoS combination. The dark regions indicate small values of $|\mathcal{C}_{M,N_{\rm DM}}|$, and the black curves denote the zero-contours $|\mathcal{C}_{M,N_{\rm DM}}|=0$. The innermost black curve corresponds to the stability boundary associated with the fundamental radial mode. The outer nested black curves are interpreted as \textbf{possible} higher-order critical curves suggested by the static gradient-alignment structure.} 
    \label{fig:large}
\end{figure}

Fig.~\ref{fig:large} extends the analysis of Fig.~\ref{fig:B200_a} to a larger parameter space. In this enlarged domain, the static gradient-alignment diagnostic reveals additional nested curve-like structures besides the innermost critical curve. These outer curves are interpreted \textbf{as possible higher-order critical curves} suggested by the macroscopic gradient-alignment structure of the equilibrium-configuration surface.

\begin{figure}[!htb] 
\centering 
\begin{subfigure}[b]{0.45\textwidth}
\centering 
\includegraphics[width=\linewidth]{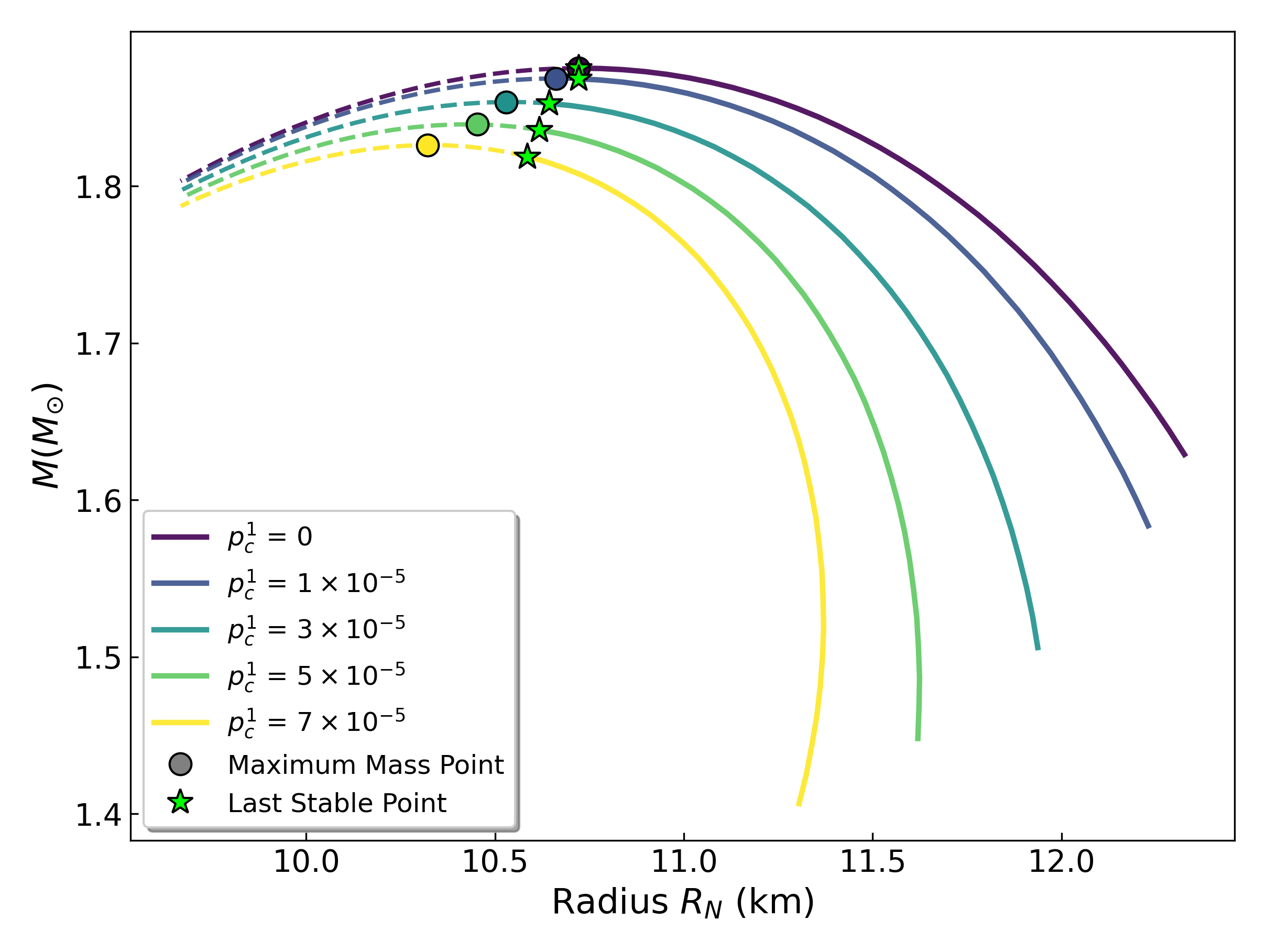} 
\caption{$M$-$R$ curves for mixed stars with a fixed Bosonic DM central pressure $p_1^c$ and varying Holographic NM ($\ell ^{-7}=10300$)
central pressure $p_2^c$. The maximum mass point (circle marker) and the last stable configuration point (star marker), determined by the critical curve method, do not coincide.}
\label{fig:MRdiff1}    
\end{subfigure}

\vspace{10pt}

\begin{subfigure}[b]{0.45\textwidth}
\centering
\includegraphics[width=\linewidth]{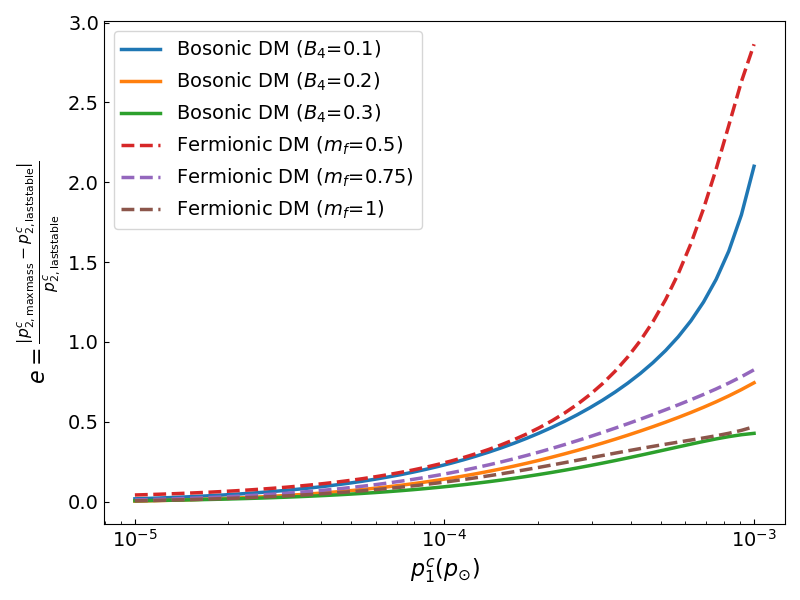} 
\caption{The deviation is described as the relative error between the central pressure of the last stable point and that of the maximum mass point, and it is a function of the DM central pressure $p_1^c$. This explicitly shows that the turning point of stability deviates from the mass peak, invalidating the simple BTM criteria for these sequences.}
\label{fig:MRdiff2}
\end{subfigure}

\caption{Deviation from the BTM criteria in mixed stars. }
\label{fig:7}
\end{figure}

Fig.\ref{fig:MRdiff1} shows the $M$-$R$ curves for mixed stars with a fixed Bosonic DM central pressure $p_1^c$ and varying Holographic NM ($\ell ^{-7}=10300$)
central pressure $p_2^c$. These relations obtained through single-parameter variations in mixed star models no longer satisfy the BTM criteria. Furthermore, the deviation from the BTM criteria can be quantitatively analyzed, as demonstrated in Fig.\ref{fig:MRdiff2}. This panel plots the relative separation between the central pressure of the last stable point, $p_{\text{last-stable}}$, and that of the maximum mass configuration, $p_{\text{max-mass}}$, as a function of the DM properties, showing the trend of the systematic separation as it varies with the central pressure and properties of DM.

Crucially, the magnitude of this deviation is highly sensitive to the EoS properties of the admixed DM. As shown, different DM models (e.g., Bosonic DM versus Fermionic DM) and variations in their microphysical parameters (such as the interaction strength $B_4$ or the fermion mass $m_f$) lead to distinctly different deviation curves. This suggests that the degree to which the BTM criteria is violated is not universal but is instead an indirect probe of the DM's EoS and its gravitational coupling with the NM. Deviations of the neutral stability point from the standard turning-point criterion under various additional physical effects have also been discussed in earlier literature~\cite{Becerra2026, Szewczyk2025, Takami2011}

The deviation of the stability boundary in the multi-fluid configuration space has direct implications for gravitational-wave observables. In particular, the stability boundary directly determines the maximum achievable compactness $M/R$ and the characteristic density scale $M/R^{3}$ of stable configurations. Since the fundamental $f$-mode frequency scales as $f_f \sim \sqrt{M/R^{3}}$ (see, e.g., Refs.~\cite{anderssonkokkotas1998,lau2010universal,bauswein2019postmerger}), precise measurements of $f$-mode frequencies from gravitational-wave observations provide a direct probe of this characteristic scale. Consequently, the deviation of the mixed-star stability boundary from the single-fluid maximum-mass point translates into a distinct upper-limit shift in the predicted $M/R^{3}$ scale, which can be tested against future gravitational-wave data. That is, the configuration that maximizes the $f$-mode frequency is no longer necessarily aligned with the maximum-mass point of single-fluid sequences, but instead follows from the full multi-dimensional stability condition in the central-pressure parameter space of the mixed system.

Therefore, a precise future measurement of a mixed star's mass, radius, and its stability threshold could, in principle, provide a tool to constrain the nature of DM within the star.

\subsubsection*{Resolution Test of the Static Critical Curve Method}
\label{app:resolution}

\begin{figure}[htb]
    \centering    \includegraphics[width=0.9\linewidth]{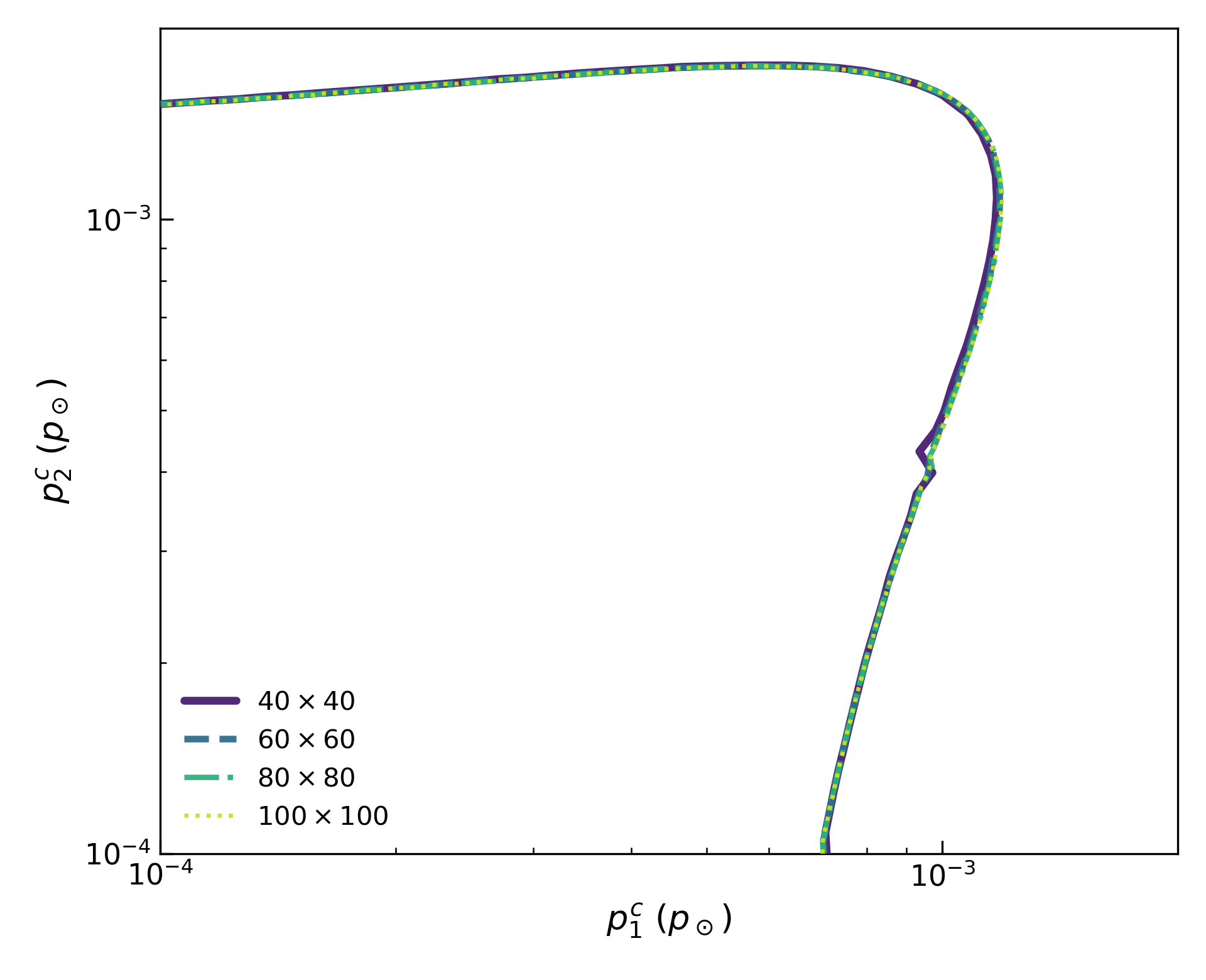}
    \caption{Critical curves computed with four different grid resolutions ($40\times40$, $60\times60$, $80\times80$, $100\times100$) for the same EoS combination and parameters as in Fig.~\ref{fig:B200_a} (Holographic NM $\ell^{-7}=10300$, Bosonic DM $B_4=0.1$). The curves coincide, demonstrating that the static method is resolution‑free.}
    \label{fig:resolution_test}
\end{figure}

To verify that the critical curve obtained from the static gradient‑alignment algorithm is independent of numerical grid resolution, we have recomputed the curve using four different grid sizes in the central‑pressure parameter space: $40\times40$, $60\times60$, $80\times80$, and $100\times100$. The EoS combination and parameters are the same as in Fig.~\ref{fig:B200_a}, i.e., Holographic NM ($\ell^{-7}=10300$) and Bosonic DM ($B_4=0.1$). As shown in Fig.~\ref{fig:resolution_test}, the resulting critical curves are indistinguishable, confirming that the algorithm is resolution‑free.

\subsection{Analysis and application of macroscopic quantities}

\subsubsection*{1. $M$-$R$ Relation}

\begin{figure*}[ht] 
\centering
\begin{subfigure}[b]{0.49\textwidth}
\centering
\includegraphics[width=\linewidth]{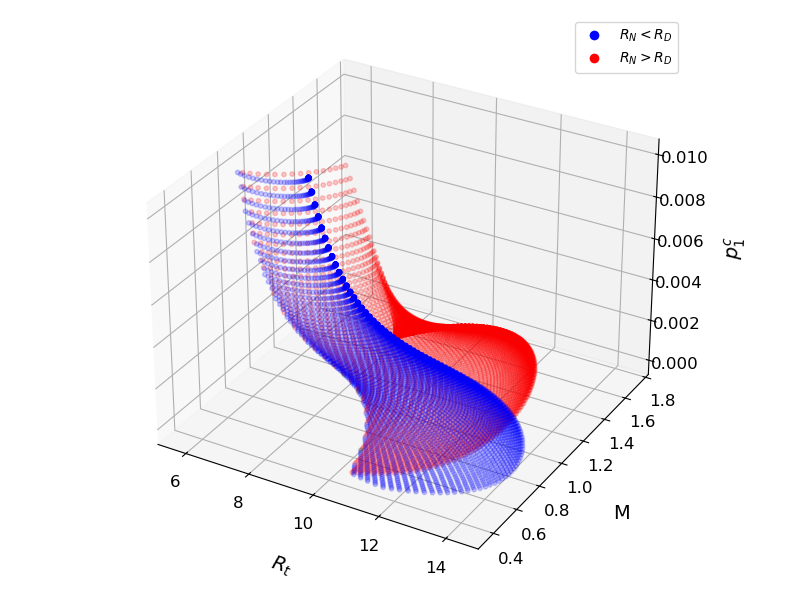} 
\caption{Topological correspondence in $M$-$R$-$p_{1}^{c}$ space.}
\label{fig:3d_mass_radius_a}
\end{subfigure}
\begin{subfigure}[b]{0.49\textwidth}
\centering
\includegraphics[width=\linewidth]{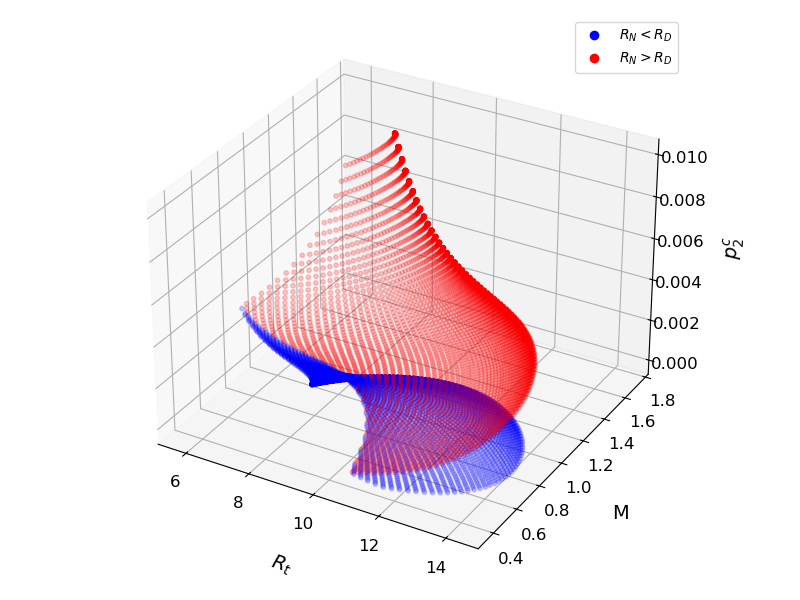} 
\caption{Topological correspondence in $M$-$R$-$p_{2}^{c}$ space.}
\label{fig:3d_mass_radius_b}
\end{subfigure}
\caption{The two-dimensional stable region from the central pressure parameter space is mapped onto a three-dimensional space with axes representing the total mass (in $M_\odot$ unit), the total radius $R_t = max\,(R_N, R_D)$ (in $km$ unit), and one of the central pressures (e.g., $p_{1}^{c}$ and $p_{2}^{c}$ in $p_\odot$ unit ). This unfolds the degeneracies of the two-dimensional $M$-$R$ projection. The red and blue dots distinguishes configurations where the NM radius is larger than the DM radius ($R_N > R_D$, blue) from those where it is smaller ($R_N < R_D$, red). (a) and (b) show the two-dimensional surface of DM and NM pressure for EoS combination in Fig.\ref{fig:B200_a}, Holographic NM ($\ell ^{-7}=10300$) and Bosonic DM ($B_4=0.1$).}
\label{fig:3D M-R}
\end{figure*}

The stable region of equilibrium configurations maps to a continuous topological correspondence region in the $M$-$R_t$-$p_{I}^{c}$ space: a two-dimensional surface in three-dimensional space, as shown in Fig.\ref{fig:3D M-R}. The EoS in Fig.\ref{fig:3D M-R} is Holographic NM ($\ell ^{-7}=10300$) and Bosonic DM ($B_4=0.1$), with vertical axis $p_{1}^{c}$ in Fig.\ref{fig:3d_mass_radius_a} and $p_{2}^{c}$ in Fig.\ref{fig:3d_mass_radius_b}. $R_t$ is the total radius: $R_t=max(R_N,R_D)$. Here $R_N$ and $R_D$ represents the radius of NM and DM. Red and blue dots code the configurations where $R_N>R_D$ and $R_N<R_D$. Any slice along the $p_{I}^{c}$ axis corresponds to an $M$-$R$ curve governed by single parameter variation.

\begin{figure*}[ht] 
\centering
\begin{subfigure}[b]{0.49\textwidth}
\centering
\includegraphics[width=\linewidth]{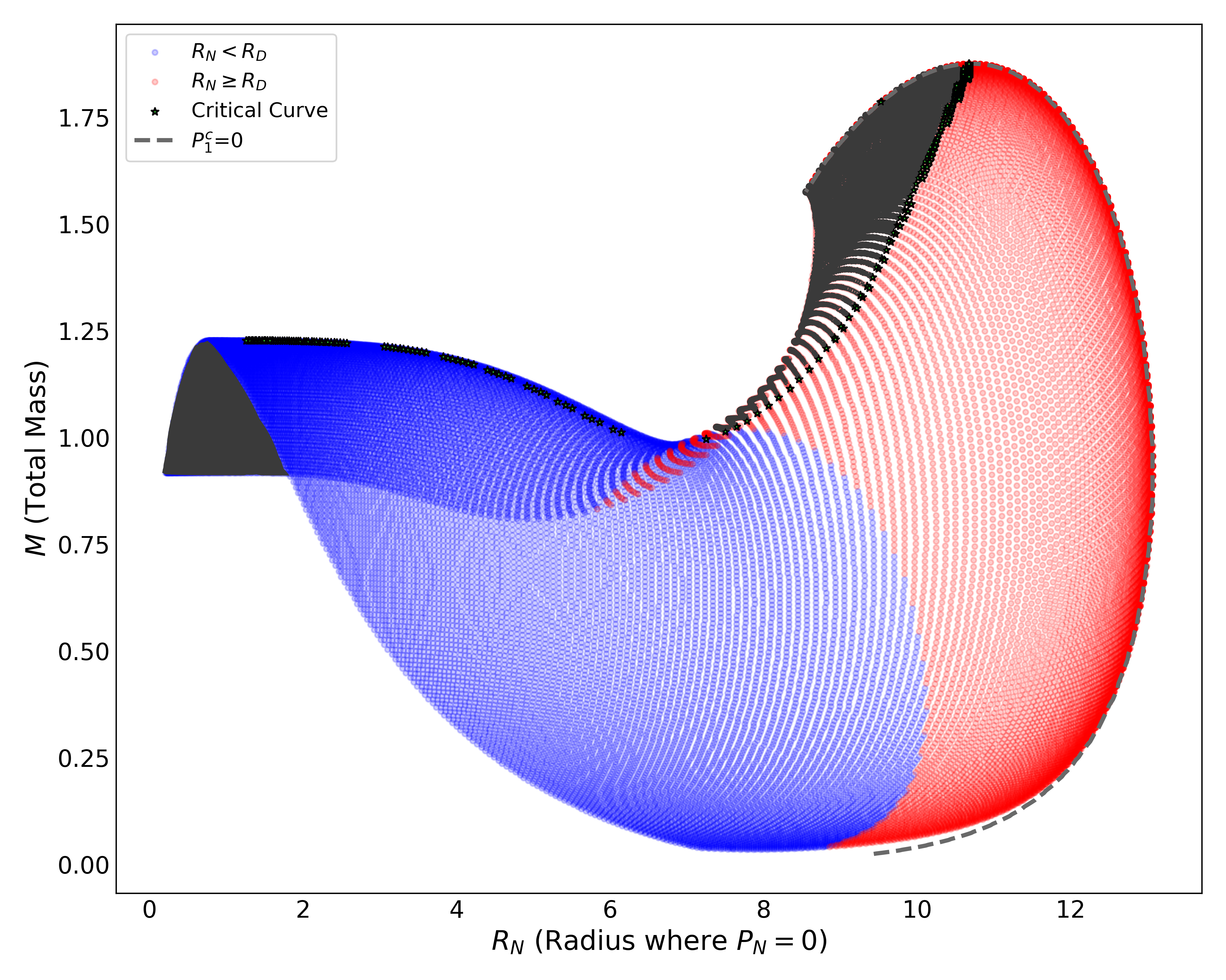} 
\caption{The stable group projected onto the mass versus visible radius plane using Holographic NM and Bosonic DM ($B_4=0.1$).}
\label{fig:2d_mass_visible_radius_projection}
\end{subfigure}
\hfill
\begin{subfigure}[b]{0.49\textwidth}
\centering
\includegraphics[width=\linewidth]{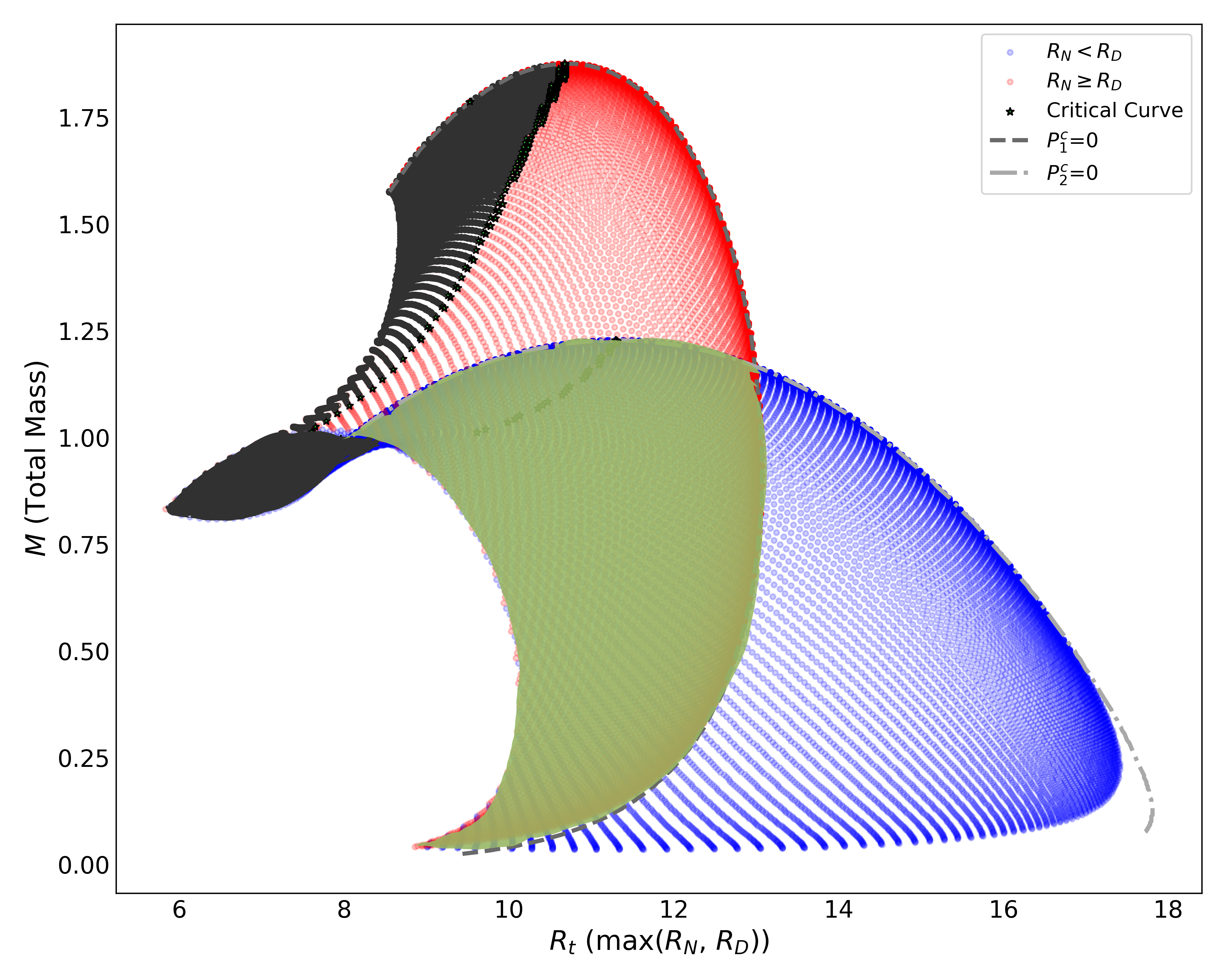} 
\caption{The same stable group projected onto the mass versus maximum radius plane ($M$-$R_t$), where $R_t = \max(R_N, R_D)$.}
\label{fig:2d_mass_radius_projection}
\end{subfigure}
\caption{
In both panels, $R_N$ and $R_D$ are the radius(in $km$ unit)  where the nuclear and DM pressure vanish, respectively. The colored regions represent dynamically stable configurations, while the grey shaded areas are unstable. 
The color coding within the stable group distinguishes between Dark Core configurations ($R_N > R_D$, red) and Dark Halo configurations ($R_N < R_D$, blue). 
The overlapping area between the red and blue regions, highlighted in green, indicates the parameter space where \textbf{twin stars} \cite{PhysRevD.107.115028} can exist: configurations with identical mass and radius but different internal structures (one being a Dark Core, the other a Dark Halo). 
The dashed line represents the boundary of the stable region (the critical curve).
}
\label{fig:2d_projections}
\end{figure*}

\begin{figure*}[!htb] 
\centering
\begin{subfigure}[b]{0.48\textwidth}
\centering
\includegraphics[width=\linewidth]{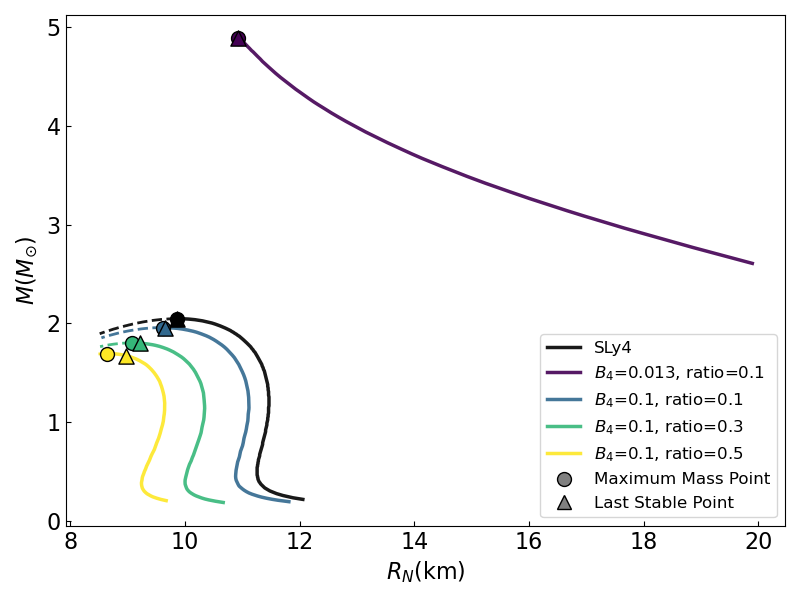} 
\caption{The $M$-$R$ relationship with different parameter settings and different mixing ratios ($ratio\equiv p^c_{DM}/p^c_{NM}$) of DM central pressure, where the mass data is within the Oppenheimer limit and near the mass gap, respectively.}
\label{fig:prospects_a}
\end{subfigure}
\hfill
\begin{subfigure}[b]{0.48\textwidth}
\centering
\includegraphics[width=\linewidth]{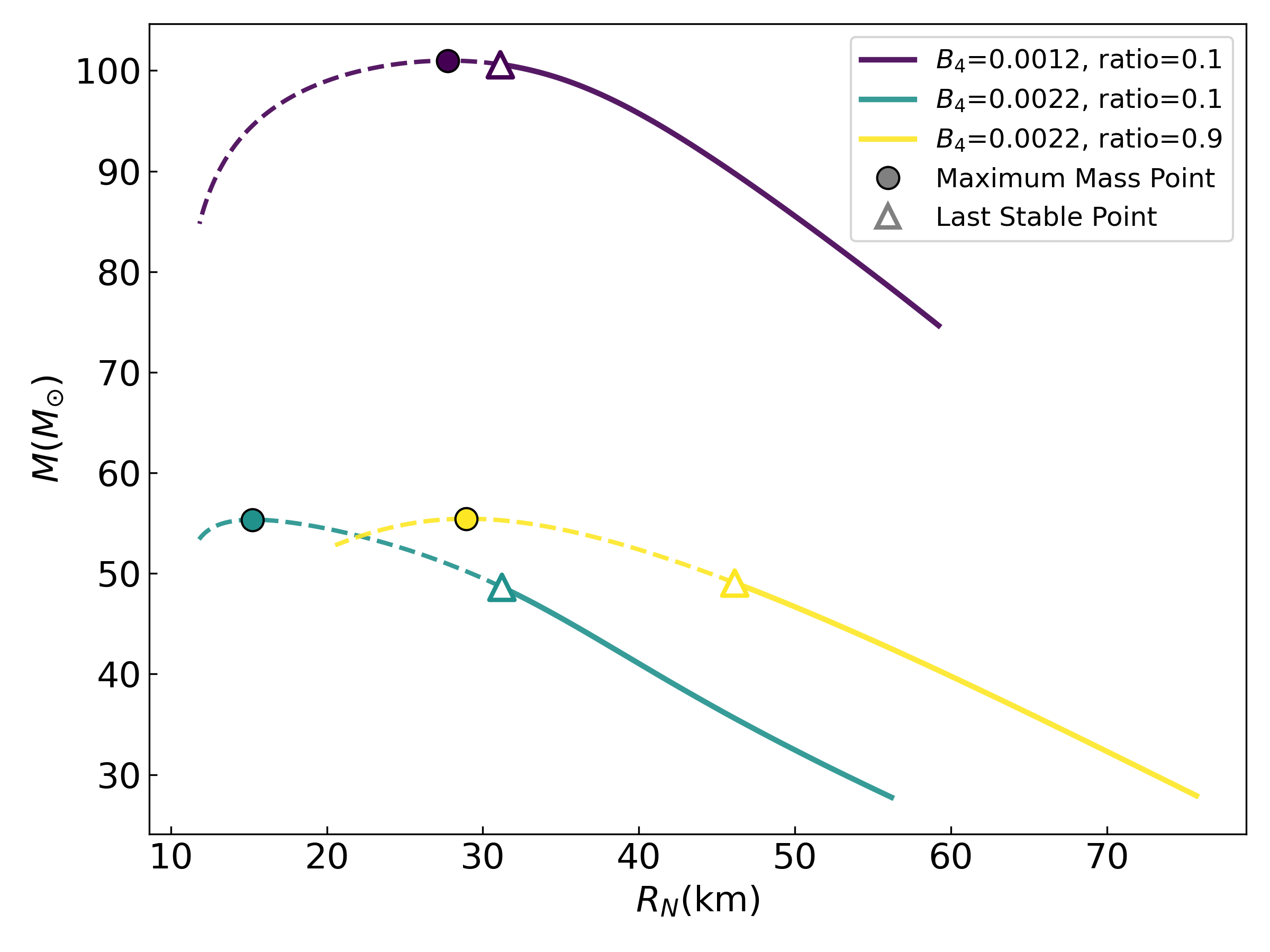} 
\caption{Similar settings to the left panel, but with a larger range of total mass.}
\label{fig:prospects_b}
\end{subfigure}
\caption{
Prediction prospects for admixed star models displayed through the $M$-$R$ relationship, with EoS combination2: Bosonic DM \eqref{eq:self-interaction EoS} and
SLy4.}
\label{fig:prediction_prospects}
\end{figure*}

\begin{table*}[ht]
\centering

\caption{Last stable configurations and corresponding macroscopic quantities for different bosonic DM EoS parameters and DM mixing ratios.}
\label{tab:max_mass}
\begin{tabular}{lccccccc}
\hline
Parameter combinations 
& $M_{\rm LS}$ ($M_\odot$) 
& $R_N$ (km) 
& $R_D$ (km) 
& $R_t$ (km) 
& $M_{\rm LS}/R_N$ 
& $M_{\rm LS}/R_D$ 
& $p_{2,{\rm LS}}^{c}$ ($p_\odot$)\\
\hline
$B_4=0.0012$, ratio=0.1 
& 100.616 
& 31.122 
& 905.64 
& 905.64 
& 3.23 
& 0.111 
& $7.6634 \times 10^{-7}$ \\
$B_4=0.0022$, ratio=0.1 
& 48.649 
& 31.206 
& 594.03 
& 594.03 
& 1.56 
& 0.082 
& $6.9859 \times 10^{-7}$ \\
$B_4=0.0022$, ratio=0.9 
& 49.076 
& 46.133 
& 591.15 
& 591.15 
& 1.06 
& 0.083 
& $7.3168 \times 10^{-8}$ \\
\hline
\end{tabular}

\end{table*}

Fig.\ref{fig:2d_projections} shows the projections of stable configurations onto  both the mass-visible radius plane $M$-$R_N$  and mass-total radius plane $M$-$R_t$ with EoS combination: Holographic NM ($\ell ^{-7}=10300$) and Bosonic DM ($B_4=0.1$). Points within the stability boundary form a ``stable group" in the projection plane, serving as a comparison tool for observational data. Overlapping regions of two colors imply the possible existence of ``twin configurations" with identical mass and radius but distinct halo compositions. The dynamical stability of such twin configurations under non-linear perturbations has recently been investigated in Ref.~\cite{Haque2026}.

This degeneracy between twin configurations can be resolved by combining gravitational-wave and electromagnetic observations. Gravitational-wave measurements constrain the stellar radius through tidal deformability~\cite{Abbott2018}, while electromagnetic observations such as pulse-profile modeling provide independent constraints on the surface emission geometry~\cite{Riley2019}. A systematic discrepancy between the inferred radii, in particular $R_{\mathrm{GW}} > R_{\mathrm{EM}}$, may indicate the presence of an extended low-density component or multi-fluid internal structure, offering a potential observational signature of twin stars~\cite{Chandrakar2026}.

The left boundary line of the ``stable group" in Fig.\ref{fig:2d_mass_radius_projection} coincides with the curve where the radius size relationship between the two fluid components reverses, meaning that for mixed star models with the same mass, the stable configuration that minimizes the total star radius is always the one where the two fluid components have equal radii.

\subsubsection*{2. Prediction prospects}

We employ the strongly self-repulsive bosonic dark matter equation of state introduced in Sec.~\ref{subsec:EoS} to discuss the observational implications of mixed stars. As shown by Colpi, Shapiro, and Wasserman~\cite{colpi1986boson}, a complex scalar field with a quartic self-interaction potential $V(\phi) = \frac{1}{4}\lambda |\phi|^4$ can support compact equilibrium configurations with maximum mass scaling as $M_{\max} \simeq 0.10\ \text{GeV}^2\, M_\odot\, \lambda^{1/2} / m^2$. For $\lambda \gg 1$ (strong self-repulsion), this maximum mass becomes comparable to the Chandrasekhar mass of fermionic stars, making it possible to reach the extreme mass ranges discussed below. The strong self-repulsion (encoded in a small value of $B$ in Eq.~\ref{eq:self-interaction EoS}) is therefore essential for the mixed-star model to access the large masses shown in Fig.~\ref{fig:prediction_prospects}. With this justification, we now examine the predicted mass--radius relations.

Collecting the latest $M$-$R$ credible window from observational data of compact stars on the $M$-$R$ stable groups can impose constraints on DM EoS. Our mixed star model demonstrates the ability to cover a wide range of $M$-$R$ area by adjusting the EoS parameters. For instance, Fig.\ref{fig:2d_mass_visible_radius_projection} covers the mass and radius values of data PSR J0437-4715 from the latest NICER observations ($M = 1.418 \pm 0.037 M_\odot$, $R = 11.36^{+0.95}_{-0.63}$ km) \cite{Choudhury_2024}.

Fig.\ref{fig:prediction_prospects} and Table. \ref{tab:max_mass} show the predicted range of $M$-$R$ relations and compactness analysis of mixed stars under different EoS parameter selections and different DM mixing ratios ($ratio\equiv p^c_{DM}/p^c_{NM}$, the central pressure ratio of DM and NM), indicating that by appropriately adjusting the hardness of the EoS and the ratios of admixed DM, the mixed star models have the potential to become predictive for NS and certain gap events, containing ranges of both 3\~{}5 $M_{\odot}$ and 50\~{}100 $M_{\odot}$ \cite{Vikiaris_2025,zhang2023dark}. While such high-mass objects are conventionally interpreted as black holes formed via hierarchical mergers or primordial mechanisms~\cite{Li2025, zhang2020constraint}, our model suggests an alternative: they could be dark-matter-dominated exotic compact stars~\cite{zhang2020constraint, Lee2021}. The key question is whether these objects might exhibit non-black-hole signatures in gravitational-wave observations. Future third-generation detectors (Einstein Telescope, Cosmic Explorer) might be able to test these predictions by measuring the tidal Love numbers and quasi-normal mode frequencies of candidate events in the upper mass gap~\cite{Pradhan2023, Narikawa2021}. Fig.\ref{fig:prospects_b} shows the situations of massive stars with the observable radius of conventional compact stars and accompanied by a large dark halo.
In addition, the similar values of compactness $M/R$ for the same EoS combinations but  with different parameters and mixing ratios exhibits a slight violation from scaling symmetry for certain kind of single EoS \cite{Maselli:2017vfi, zhang2023dark}. 

\section{Conclusion}\label{sec:Conclusion}
In this work, we have carried out a comprehensive investigation of the stability properties and macroscopic characteristics of DM-NM mixed stars within the framework of general relativity.

We have established a rigorous equivalence between the critical curve method and the emergence of a zero-frequency radial oscillation mode. This foundational link provides a formal justification for using the static critical curve method as a powerful proxy for dynamical stability analysis.

Our numerical calculations, employing four typical EoS combinations—SLy4,
Holographic NM,  Fermionic DM and Bosonic DM—determine the stability regions of mixed stars and validate the equivalence between the two methods numerically, while also revealing complex stability distributions that fundamentally differ fundamentally from those of single-fluid and hybrid stars. Unlike pure NS that obey the classical BTM criteria, stable configurations of mixed stars form continuous two-dimensional surfaces in the three-dimensional parameter space spanned by (one of the) central pressure, mass, and radius. The projection of these stable configurations onto the $M$-$R$ plane reveals a ``stable group" that delineates natural regions where mixed stars should be observationally found. These groups exhibit several noteworthy features, including the existence of twin configurations with identical $M$-$R$ measurements but different internal compositions, and the characteristic property that the most compact stable configurations are those where the two fluid components have equal radii. Moreover, our mixed star model, by adjusting the EoS hardness and the DM mixing ratio, can accommodate regular NS events and also account for certain mass-gap events. It also has the potential to predict massive configurations featuring a compact star radius core accompanied by a large DM halo.

Looking forward, our theoretical framework can be extended to incorporate more sophisticated DM models, such as self-interacting fermionic DM and  axion-like particles. The methodology also provides a foundation for investigating rotating mixed stars, magnetic field effects, and dynamical formation scenarios. As observational capabilities continue to advance, the stable groups and characteristic $M$-$R$ relations identified through our analysis may serve as references for identifying potential mixed star candidates and for constraining the properties of DM through astrophysical observations.

\begin{acknowledgments}
The authors would like to thank  Feng-Li Lin, Dong Lai, Qiyuan Pan, Shao-Feng Ge, Zhoujian Cao, Nobutoshi Yasutake, Chian-Shu Chen, Alessandro Parisi, Chen Zhang, Zihao Zhang, Hao Feng and Jing-Yi Wu for helpful discussions. K.Z. (Hong Zhang) is supported by a classified fund from Shanghai city.
\end{acknowledgments}

\appendix
\section{Review of Static Stability Criteria}
\label{app:static_criteria}

The static stability criteria determine the stability of a star by examining the properties of the entire equilibrium sequence, rather than analyzing perturbations of a single configuration. This approach is computationally efficient as it reuses the data from calculating the equilibrium models.

\subsection*{1. BTM Criteria for Single-Fluid}

The BTM criteria is a method for assessing the radial stability of general-relativistic star models. The key principles of this criterion are (we apply its generalized version in \cite{zhang2020constraint}, that when passing through each extremum of the $M$-$R$ curve, the direction can go along either increasing  or decreasing the core pressure):

\begin{itemize}
    \item Stability changes occur exclusively at the extrema of the $M$-$R$ curve.
    \item A \textbf{counterclockwise bend} at an extremum corresponds to \textbf{gaining one unstable mode}.
    \item A \textbf{clockwise bend} at an extremum corresponds to \textbf{losing one unstable mode}.
    \item The sequence is typically parameterized by central pressure.
\end{itemize}

It is crucial to emphasize that the original BTM criteria was formulated specifically for sequences of \textbf{zero-temperature stars}, representing a "family" of different stars composed of the same cold matter. This framework is directly applicable to the cold NS models considered in this work. In contrast, for sequences of \textbf{hot isentropic stars} with a fixed total baryon number, the stability change at an extremum follows the opposite rule \cite{Bardeen:1966}: a counterclockwise bend leads to the loss of an unstable mode, while a clockwise bend leads to the gain of one.

Due to its intuitive nature, the BTM criteria has been widely used for decades as a primary tool for stability assessment in relativistic astrophysics. The equivalence between the oscillation method and BTM criteria has been proved in \cite{Friedman_Stergioulas_2013,Hadzic2021}.

It is important to note that the BTM criteria is specifically applicable to \textbf{single-fluid models} and becomes invalid for \textbf{multi-fluid models} containing interface or mixed phases. We will provide a more detailed discussion of this fact and its manifestation in subsequent sections.

\subsection*{2. Critical Curve Criteria for Multi-Fluid}
\label{subsec:critical_curve}

The stability analysis of multi-fluid compact stars can be significantly simplified through the application of the critical curve method. This approach, originally developed in \cite{HENRIQUES1990511} for determining the stability of boson-fermion stars, is in fact applicable to any spherically symmetric ideal multi-fluid star model in general relativity, including the DM-NM mixed stars considered in this work. The method provides an efficient criterion for determining star stability by relying solely on equilibrium solutions of the multi-fluid TOV equations, thereby circumventing the computationally intensive task of solving the coupled pulsation equations to identify radial normal modes.

The fundamental premise of this method lies in characterizing a stability boundary in the parameter space of central pressures for the constituent fluids. For a two-fluid system, the critical curve demarcating stable and unstable configurations is formed by a group of points characterizing central conditions. At each point $\boldsymbol{\sigma}=\left( p_{2}^{c}, p_{1}^{c} \right) $, there exists a direction $\delta\boldsymbol{\sigma}$ that fits the condition \cite{HENRIQUES1990511}:
\begin{equation}
\begin{aligned}      \label{eq:critical_curve_general}
    &\nabla_{\boldsymbol{\sigma}} M(\infty) \cdot \delta\boldsymbol{\sigma} = 0 \quad \text{and} \quad \nabla_{\boldsymbol{\sigma}} N_{F} \cdot \delta\boldsymbol{\sigma} = 0 \\
    &\left(\text{or alternatively } \nabla_{\boldsymbol{\sigma}} N_{B}\right),
\end{aligned}
\end{equation}
where \(M(\infty)\) denotes the total gravitational mass of the star (equivalent to the mass within radius \(R\), hereafter denoted simply as \(M\)) while $N_{F}$ and $N_{F}$ are the total particle numbers for fermions and bosons, respectively. For the specific case of DM-NM mixed stars considered in our work, this criterion takes the form:
\begin{equation}
\begin{aligned}      \label{eq:critical_curve_general_NMDM}
    &\nabla_{\boldsymbol{\sigma}} M \cdot \delta\boldsymbol{\sigma} = 0 \quad \text{and} \quad \nabla_{\boldsymbol{\sigma}} N_{NM} \cdot \delta\boldsymbol{\sigma} = 0 \\
    &\left(\text{or alternatively } \nabla_{\boldsymbol{\sigma}} N_{DM}\right),
\end{aligned}
\end{equation}
where \(N_{NM}\) and \(N_{DM}\) represent the particle numbers of DM and NM, respectively.

\section{Review of Radial Oscillation}
\label{app:pulsation_derivation}
\subsection*{1. Single Fluid}
Chandrasekhar, in his 1964 paper, rigorously established the pulsation equation for radial oscillations in single-fluid stars based on radial perturbations and variational principles \cite{Chandrasekhar:1964}, and it is widely applied in many works such as \cite{Alford:2017vca}. For the need of unifying notation, we briefly review the main steps and key equations in the derivation of this equation:
\subsubsection*{Linearization of the Field Equations}
An unperturbed, static, spherically symmetric equilibrium configuration is assumed, satisfying the TOV equations. Its behavior after being subjected to infinitesimal radial perturbations can be used to determine its stability. We define $e^{2\nu \left( r \right)}\equiv \left( 1-\frac{2Gm\left( r \right)}{rc^2} \right) ^{-1}
$, then the metric and fluid variables are written as:
\begin{align*}
    \nu &= \nu_0 + \delta\nu, \quad & \phi &= \phi_0 + \delta\phi, \\
    p &= p_0 + \delta p, \quad & \varepsilon &= \varepsilon_0 + \delta\varepsilon.
\end{align*}
The Einstein field equations and the Einstein-Euler equations are perturbed separately. Neglecting higher-order terms, we obtain the corresponding system of linearized radial perturbation equations:
\begin{align}
	&\frac{\partial}{\partial r}(re^{-2\nu _0}\delta \nu )=\frac{4\pi G}{c^4}r^2\delta \varepsilon ,\label{eq:linearized1}\\
	&\frac{e^{-2\nu _0}}{r}\left( \frac{\partial}{\partial r}\delta \phi -\frac{2 d\phi _0}{dr}\delta \nu \right) =\frac{e^{-2\nu _0}}{r^2}\delta \nu +\frac{4\pi G}{c^4}\delta p,\label{eq:linearized2}\\
	&\frac{e^{-2\nu _0}}{r}\frac{\partial}{\partial t}\delta \nu =-\frac{4\pi G}{c^4}(p_0+\varepsilon _0)\frac{dr}{dt},\label{eq:linearized3}\\
    \begin{split}
        	&e^{2\nu _0-2\phi _0}(p_0+\varepsilon _0)\frac{\partial}{\partial t}\frac{dr}{dt}+\frac{\partial}{\partial r}\delta p+\\
	&(p_0+\varepsilon _0)\frac{\partial}{\partial r}\delta \phi +(\delta p+\delta \varepsilon )\frac{d\phi _0}{dr}=0.
        \end{split}\label{eq:linearized4}
\end{align}

\subsubsection*{Lagrangian Displacement and Reformulating Perturbations}
The \textit{Lagrangian displacement} $\xi$ is defined such that the radial velocity satisfies $v=\frac{dr}{dt}=\frac{\partial \xi}{\partial t}$, i.e., $u^r=e^{-\phi_0}\frac{dr}{dt}=e^{-\phi_0}\frac{\partial \xi}{\partial t}$.
Then, using the first three equations of the linearized perturbation equations \eqref{eq:linearized1}\eqref{eq:linearized2}\eqref{eq:linearized3}, all perturbations of metric and fluid variables except $\delta p$ are expressed in terms of $\xi$ and equilibrium quantities:
\begin{align}
	\delta \nu &=-\frac{4\pi G}{c^4}\xi (p_0+\varepsilon _0)\frac{r}{e^{-2\nu _0}},\label{eq:purtitem1}\\
	\delta \varepsilon &=-\frac{1}{r^2}\frac{\partial}{\partial r}\left[ r^2(p_0+\varepsilon _0)\xi \right] ,\label{eq:purtitem2}\\
	\frac{\partial}{\partial r}\delta \phi &=\frac{4\pi G}{c^4}\left[ \delta p-(p_0+\varepsilon _0)\left( \frac{2d\phi _0}{dr}+\frac{1}{r} \right) \xi \right] \frac{r}{e^{-2\nu _0}}.\label{eq:purtitem3}
\end{align}
Note that the perturbation quantities and the Lagrangian displacement are all functions of $r$ and $t$ (e.g., $\xi(r, t)$, $\delta p(r, t)$, etc.). We now assume that all perturbation quantities $F$ possess a time-harmonic dependence of the form: $F(r, t) = F(r)  e^{i\omega t}$. Based on this separation of variables and in combination of \eqref{eq:linearized4} with \eqref{eq:purtitem1}, \eqref{eq:purtitem2}, and \eqref{eq:purtitem3}, we obtain:
\begin{equation}
\begin{aligned}
&\omega^2 e^{2\nu_0 - 2\phi_0} (p_0 + \varepsilon_0) \xi =  \\&-\frac{d\phi_0}{dr} (p_0 + \varepsilon_0) \frac{d\xi}{dr} \\
& -\left[ \frac{d\phi_0}{dr} \left( \frac{d(\varepsilon_0 + p_0)}{dr} + \frac{(p_0 + \varepsilon_0)}{r} \right) \right. \\
& \left. + (p_0 + \varepsilon_0) \left( \frac{2 d\phi_0}{dr} + \frac{1}{r} \right) \left( \frac{d\nu_0}{dr} + \frac{d\phi_0}{dr} \right) \right] \xi \\
& + \frac{d}{dr} \delta p + \delta p \frac{d}{dr} \left(\nu_0 + 2\phi_0 \right),
\end{aligned}\label{eq.purturbation eq without deltap sub}
\end{equation}
where yet the perturbation equations are thus reduced from complex partial differential equations (PDE) to a relatively simpler eigenvalue problem governed by an ordinary differential equation (ODE), with the perturbation terms now consisting of only the pressure variation.

\subsubsection*{Baryon Number Conservation}
\label{sec:baryon}
An additional relation is needed to connect the pressure perturbation $\delta p$ to the other variables. This is provided by imposing adiabaticity and baryon number conservation, which requires $\nabla_\mu (n u^\mu) = 0$. Defining the perturbation of $n$ as $n = n_0(r) + \delta n(r, t)$ and combining it with $\xi$ defined previously leads to the form of $\delta n$ constituted by the Lagrangian displacement and the unperturbed terms:
\begin{equation}
\delta n =-\frac{e^{\phi _0}}{r^2}\frac{\partial (n_0r^2\xi e^{-\phi _0})}{\partial r}.
\end{equation}
Moreover, the first law of thermodynamics gives $d\varepsilon = \frac{\varepsilon + p}{n} dn$, thus the baryon number density is now related to $p$ or $\varepsilon$. This finally allows the pressure perturbation to be written solely in terms of the Lagrangian displacement:
\begin{equation}
    \delta p = -\xi \frac{d p_0}{d r} - \gamma p_0 \frac{e^{\phi_0}}{r^2} \frac{\partial}{\partial r} \left( r^2 e^{-\phi_0} \xi \right). \label{eq:deltap}
\end{equation}
Here $\gamma$ is defined as \(\gamma =-\frac{n_0}{p_0}\frac{dp_0}{dn_0}\), which is the adiabatic index. This step is crucial as it incorporates the microphysics (EoS) into the dynamical description.

\subsubsection*{The Pulsation Equation and Boundary Conditions}

Substituting the expression for the pressure perturbation $\delta p$ from \eqref{eq:deltap} into the perturbed equation of motion \eqref{eq.purturbation eq without deltap sub}, and after involving the use of the background TOV equations \eqref{eq:tov} to eliminate derivatives of the metric potentials, one arrives at the final pulsation equation:
\begin{equation}
\begin{aligned}
&\omega^2 e^{2(\nu_0 - \phi_0)} (p_0 + \varepsilon_0) \xi = 
 \frac{4}{r} \frac{d p_0}{d r} \xi \\
& - e^{-(\nu_0 + 2\phi_0)} \frac{d}{dr} \left[ e^{(\nu_0 + 3\phi_0)} \frac{\gamma p_0}{r^2} \frac{d}{dr} \left( r^2 e^{-\phi_0} \xi \right) \right] \\
& + \frac{8\pi G}{c^4} e^{2\nu_0} p_0 (p_0 + \varepsilon_0) \xi \\
& - \frac{1}{p_0 + \varepsilon_0} \left( \frac{d p_0}{d r} \right)^2 \xi.
\end{aligned}
\label{eq:pulsation}
\end{equation}

To cast it into the standard Sturm-Liouville form, we need two key simplifications: multiplying the entire equation \eqref{eq:pulsation} by $e^{\nu_0 + 2\phi_0}$ to simplify the differential operator and introducing the new variable $\chi(r) \equiv r^2 e^{-\phi_0(r)} \xi(r)$ to simplify the expression for the Lagrangian perturbation. After manipulation, the equation takes the canonical Sturm-Liouville form:

\begin{equation}
-\frac{d}{dr}\left[ P(r) \frac{d\chi}{dr} \right] + Q(r) \chi = \omega^2 W(r) \chi,
\end{equation}where the coefficient functions are defined as:
\begin{align*}
	P(r)&=\frac{\gamma p_0(r)}{r^2}e^{\nu _0(r)+3\phi _0(r)},\\
	Q(r)&=\left( \frac{4}{r}\frac{dp_0}{dr}+\frac{8\pi G}{c^4}e^{2\nu _0(r)}p_0(r)(p_0(r)+\varepsilon _0(r)) \right.\\
	&\quad \left. -\frac{1}{p_0(r)+\varepsilon _0(r)}\left( \frac{dp_0}{dr} \right) ^2 \right) \frac{e^{\nu _0(r)+3\phi _0(r)}}{r^2},\\
	W(r)&=(p_0(r)+\varepsilon _0(r))\frac{e^{3\nu _0+\phi _0}}{r^2}.
\end{align*}
This formulation demonstrates the self-adjoint nature of the eigenvalue problem, which guarantees real eigenvalues $\omega^2$ as well as orthogonal and complete eigenfunctions.

Solutions to the pulsation equation must satisfy boundary conditions imposed by physical regularity and spherical symmetry:
\begin{itemize}
\item At the center of the star ($r = 0$), all physical quantities must be finite. The radial displacement $\xi$ must vanish at the origin to avoid a directional singularity, i.e., $\xi(0) = 0$. A power series analysis shows that the leading-order of the displacement is linear: $\xi \propto r$. Thus, due to the factor of $r^2$, the scaling variable $\chi(r) = r^2 e^{-\phi_0(r)} \xi(r)$ has an asymptotic behavior $\chi \propto r^3$ near the origin. Therefore, the inner boundary condition is:
\begin{equation}
\chi \to 0 \quad \text{as} \quad r \to 0, \quad \text{with} \quad \frac{\chi}{r^3} \to \text{const.}
\label{eq:bc1}
\end{equation}
In practice, for numerical integration, this is implemented by starting the integration at a small but finite radius $r = r_0$ with the asymptotic form $\chi(r_0) = \chi_0 r_0^3$, where $\chi_0$ is a constant to be determined.
\item At the surface of the star, the appropriate physical boundary condition is that the Lagrangian perturbation of the pressure must vanish:
        \begin{equation}
        \Delta p = 0 \quad \text{at} \quad r = R.\label{bc surface}
        \end{equation}
        And its relation to the Eulerian perturbation is:
        \begin{equation}
        \begin{aligned}
        \Delta p &=
        \delta p + \xi \frac{d p_0}{d r} \\&= - \gamma p_0 \frac{e^{\phi_0}}{r^2} \frac{\partial}{\partial r} \left( r^2 e^{-\phi_0} \xi \right),
        \end{aligned}\label{L perturb of p} \end{equation}\label{eq:Eulerian perturbation}
        Therefore, in numerical calculations, we require that the right side of \ref{eq:Eulerian perturbation} should be zero on the outer boundary.
\end{itemize}

The pulsation equation in Sturm-Liouville form, together with the boundary conditions \eqref{eq:bc1} and \eqref{bc surface}, defines a characteristic value problem for the squared oscillation frequency $\omega^2$.

\subsubsection*{Stability Condition and Variational Principle}
The stability of an equilibrium configuration is determined by the temporal evolution of radial perturbations. From the ansatz $\xi(r, t) = \xi(r) e^{i\omega t}$, real values of $\omega$ (i.e., $\omega^2 > 0$) correspond to stable oscillations, while imaginary $\omega$ ($\omega^2 < 0$) leads to exponentially growing or damping modes, indicating dynamical instability.

The self-adjoint nature of the pulsation equation \eqref{eq:pulsation} permits the formulation of a variational principle for the characteristic frequencies $\omega^2$. Multiplying \eqref{eq:pulsation} by $r^2 \xi e^{(-\phi_0)} $ and integrating over the star volume, followed by integration by parts and application of the boundary conditions, yields the integral expression:

\begin{equation}
\omega^2 \int_0^R e^{(3\nu_0 - \phi_0)} (p_0 + \varepsilon_0) r^2 \xi^2  dr = \text{Numerator},
\label{eq:var_base}
\end{equation}

where the right-hand side ``Numerator'' is given by the integral:
\begin{equation}
\begin{aligned}
\text{Numerator} = &
4 \int_0^R e^{(\nu_0 + \phi_0)} r \frac{d p_0}{d r} \xi^2  dr \\
& + \int_0^R e^{(\nu_0 + 3\phi_0)} \frac{\gamma p_0}{r^2} \left[ \frac{d}{dr} \left( r^2 e^{-\phi_0} \xi \right) \right]^2  dr \\
& - \int_0^R e^{(\nu_0 + \phi_0)} \left( \frac{d p_0}{d r} \right)^2 \frac{r^2 \xi^2}{p_0 + \varepsilon_0}  dr \\
& + \frac{8\pi G}{c^4} \int_0^R e^{(3\nu_0 + \phi_0)} p_0 (p_0 + \varepsilon_0) r^2 \xi^2  dr.
\end{aligned}
\label{eq:variational}
\end{equation}

Equation \eqref{eq:var_base} expresses $\omega^2$ as the ratio of two integrals, known as the \textit{Rayleigh quotient}:
\begin{equation}
\begin{aligned}
    &\omega^2[\xi] = \frac{\text{Numerator}}{\text{Denominator}}, \\
&\text{where} \\ &\text{Denominator} = \int_0^R e^{(3\nu_0 - \phi_0)} (p_0 + \varepsilon_0) r^2 \xi^2  dr.
\end{aligned}
\end{equation}
Here, $\omega^2[\xi]$ is a functional, whose value depends on the choice of the trial function $\xi$. 

The orthogonality of the eigenfunctions, as guaranteed by the self-adjoint Sturm-Liouville form, plays a crucial role in the variational principle. Any admissible trial function $\widetilde{\xi}$ can be expanded as a series in the complete set of orthogonal eigenfunctions $\xi^{(n)}$:
\begin{equation}
\widetilde{\xi} = \sum_{n=0}^{\infty} c_n \xi^{(n)},
\end{equation}
where $c_n$ is the weight coefficients. The functional $\widetilde{\omega }^2\left[ \widetilde{\xi } \right] 
$ corresponding to the trial function is a weighted average of the true eigenvalues $\omega_n^2$:
\begin{equation}
\widetilde{\omega}^2 = \frac{\sum_{n=0}^{\infty} c_n^2 \omega_n^2 \langle \xi^{(n)}, \xi^{(n)} \rangle}{\sum_{n=0}^{\infty} c_n^2 \langle \xi^{(n)}, \xi^{(n)} \rangle},
\end{equation}
where $\langle \cdot, \cdot \rangle$ denotes the inner product with the corresponding weight functions. The Rayleigh quotient is stationary when $\widetilde{\xi}$ is an eigenfunction (see more details in\cite{Caballero_2024}), and attains its minimum value $\omega_0^2$ when $\widetilde{\xi}$ is proportional to the fundamental mode $\xi^{(0)}$.

Therefore, for any trial function $\widetilde{\xi}$ satisfying the boundary conditions, the computed value $\widetilde{\omega}^2$ is an upper bound to the true fundamental mode eigenvalue:
\begin{equation}
\omega_0^2 \leq \widetilde{\omega}^2[\widetilde{\xi}].
\end{equation}

This formulation provides a powerful method for determining dynamical stability. A sufficient condition for dynamical instability is that there exists \textit{any} trial function $\widetilde{\xi}$ for which the ``Numerator'' $\leq 0$. Since the ``Denominator'' is always positive, this implies $\widetilde{\omega}^2 \leq 0$, and consequently, $\omega_0^2 \leq 0$.

\subsection*{2. Multi-Fluid}
\label{subsec:pulsation eqns_Multi Fluid}
For a compact star composed of multiple ideal fluids that are in the absence of any non-gravitational coupling, the pulsation equations for a single fluid can be generalized to the multi-fluid case. The oscillation equations for such a system have been previously derived in the literature and applied to other mixed models (see, e.g., [\cite{Kain:2020oho}\cite{Kain:2021poc}\cite{PhysRevD.110.103038}]). This section briefly outlines the main steps in deriving the equations using the same process as the references.

\subsubsection*{Linearized Field Equations and Energy-Momentum Tensor}

Consider a spherically symmetric system composed of \(\mathcal{N}\) ideal fluids. The total energy-momentum tensor is the sum of the decoupled contributions from each fluid:
\[
T^{\mu\nu}_{\text{total}} = \sum_I T_I^{\mu\nu}, \quad T_I^{\mu\nu} = (\varepsilon_I + p_I) u_I^\mu u_I^\nu + p_I g^{\mu\nu}.
\]
For a spherically symmetric background, assuming the EoS for each fluid depends only on its own energy density (i.e., the fluids only interact through gravity): \(p_I = p_I(\varepsilon_I)\), the radial perturbation variables are set as
\begin{align*}
\phi &= \phi_0 + \delta\phi, \quad \nu = \nu_0 + \delta\nu, \\
\varepsilon_I &= \varepsilon_{I0} + \delta\varepsilon_I, \quad p_I = p_{I0} + \delta p_I,
\end{align*}
and the Lagrangian displacement \(\xi_I\) of the \(I\)-th fluid satisfies $v_I = \dot{\xi}_I$. Similar to the case of a single fluid, linearizing the Einstein field equations and the energy-momentum conservation equations would yield a set of coupled perturbation equations.

\subsubsection*{Lagrangian Displacement and Expression of Perturbations}

Introduce the harmonic time dependence assumption: \(\xi_I(t,r) = \xi_I(r) e^{i\omega t}\), and similarly for other perturbation variables. Similar to the last section, the metric perturbations \(\delta\nu\), \(\delta\phi\) can be expressed in terms of the time-independent \(\xi_I\):
\begin{align}
\delta \nu =&\frac{-8\pi G r}{e^{-2\nu _0}}\sum_I{(}\varepsilon _{I0}+p_{I0})\xi _I,\label{eq:lambda perturb}
\\
\begin{split}
\frac{d\delta \phi}{dr}=&\frac{2(\frac{d\phi _0}{dr}+\frac{d\nu _0}{dr})}{(\varepsilon _{0}^{\mathrm{total}}+p_{0}^{\mathrm{total}})}\times
\\
&\left[ \delta p^{\mathrm{total}}-\left( 2\frac{d\phi _0}{dr}+\frac{1}{r} \right) \sum_I{(}\varepsilon _{I0}+p_{I0})\xi _I \right].
\end{split}
\end{align}
The energy density perturbation of each fluid can be written as:
\begin{equation}
    \begin{aligned}
        \delta \varepsilon _I=&-\frac{1}{r^2}\frac{\partial\left[ r^2(p_{I0}+\varepsilon _{I0})\xi _I \right]}{\partial r} +
\\&
\frac{4\pi G}{e^{-2\nu _0}}(\varepsilon _{I0}+p_{I0})r\sum_J{(}\varepsilon _{J0}+p_{J0})(\xi _J-\xi _I).
    \end{aligned}
\end{equation}\label{eq:varepsilonI_perturb}

\subsubsection*{Particle Number Conservation and Pressure Perturbation}

Similar to the case of a single fluid, through covariant conservation, particle number perturbations are constructed as
\begin{equation}
    \begin{aligned}
        \delta n_I=&-\frac{e^{\phi _0}}{r^2}\frac{\partial\left( r^2e^{-\phi _0}n_{I0}\xi _I \right)}{\partial r} 
\\&
+\frac{4\pi G r}{e^{-2\nu _0}}n_{I0}\sum_J{(}\varepsilon _{J0}+p_{J0})(\xi _J-\xi _I).   \end{aligned}
\end{equation}\label{eq:delta ni}
For each fluid, introduce the adiabatic index \(\gamma_I =-\frac{n_{I0}}{p_{I0}}\frac{dp_{I0}}{dn_{I0}}=\left( \frac{p_{I0}+\varepsilon _{I0}}{p_{I0}} \right) \frac{dp_{I0}}{d\varepsilon _{I0}}
\). The pressure perturbation can then be expressed as:
\begin{equation}
    \begin{split}
        \delta p_I =& -\xi_I \frac{\partial p_{I0}}{\partial r} - \gamma_I p_{I0} \biggl[ \frac{e^{\phi_0}}{r^2} \frac{\partial}{\partial r}\left( r^2 e^{-\phi_0} \xi_I \right) \\&
        - \frac{4\pi G r}{e^{-2\nu _0}} \sum_J (\varepsilon_{J0} + p_{J0}) (\xi_J - \xi_I) \biggr].
    \end{split}
\end{equation}
These perturbation expressions couple the fluid variables and the displacements \(\xi_I\) of all fluids, reflecting the gravitational interaction between the multiple fluids.

\subsubsection*{Multi-fluid Pulsation Equations and Boundary Conditions}

Substituting the above expressions into the linearized equations and introducing the variable \(\chi _I(r)=r^2e^{-\phi _0}\xi _I(r)\)(where \(\phi _c=\phi _0\left( r=0 \right) \)) yield a set of coupled second-order ordinary differential equations:
\begin{equation}
    \partial_r (P_I \chi_I') + (Q_I + \omega^2 W_I) \chi_I + \text{coupling terms} = 0,
\end{equation}
where the coefficients are:
\begin{align*}
    P _I=&\frac{1}{r^2}p_{I0}\gamma _Ie^{2\phi _0},\\
    W_I=&\frac{1}{r^2}(\varepsilon _{I0}+p_{I0})e^{2\nu _0},\\
    \begin{split}
        Q_I=&-\frac{e^{2\phi _0}}{r^2}\left[ \frac{3}{r}\frac{dp_{I0}}{dr}+8\pi e^{2\nu _0}p_{0}^{\mathrm{total}}(\varepsilon _{I0}+p_{I0})+\right.\\
        &\left.e^{2\nu _0}\left( 4\pi r\varepsilon _{0}^{\mathrm{total}}-\frac{m_{0}^{\mathrm{total}}}{r^2} \right) \left( \frac{\varepsilon _{I0}+p_{I0}}{r}-\frac{dp_{I0}}{dr} \right) \right].
    \end{split}
\end{align*}

The coupling terms are explicitly given by:
\begin{align*}
    &\text{coupling terms} =\\&
R\left[ \left( \frac{\varepsilon _{I0}+p_{I0}}{r}-\frac{dp_{I0}}{dr} \right) \sum_J{(}\varepsilon _{J0}+p_{J0})\chi _J+ \right. 
\\&
\left. \frac{r^2(\varepsilon _{I0}+p_{I0})}{e^{2\phi _0}}\sum_J{\left( p_{J0}\gamma _J\frac{e^{2\phi _0}}{r^2}\frac{d\chi _J}{dr} \right)} \right] 
\\&
+S_I\sum_J{(}\varepsilon _{J0}+p_{J0})(\chi _J-\chi _I)
\\&
+\frac{r^2}{e^{2\phi _0}}R^2(\varepsilon _{I0}+p_{I0})
\\&
\sum_J{\sum_K{p_{J0}}}\gamma _J(\varepsilon _{K0}+p_{K0})(\chi _K-\chi _J)
\\&
+R\gamma _Ip_{I0}\sum_J{\left[ (\varepsilon _{J0}\prime+p_{J0}\prime)(\chi _J-\chi _I)+ \right.}
\\&
\left. (\varepsilon _{J0}+p_{J0})(\frac{d\chi _J}{dr}-\frac{d\chi _I}{dr}) \right] ,
\end{align*}
where
\begin{align*}
    &R = 4\pi e^{2\nu_0+2\phi_0}/r,
    \\
    &S_I = R \left[ (\gamma_I - 1) \frac{dp_{I0}}{dr} + \frac{d\gamma_I}{dr} p_{I0} + 
    \right.\\&\left.
    \gamma_I p_{I0} \left( 8\pi r e^{2\nu_0} (\varepsilon_0^{\text{total}} + p_0^{\text{total}}) - \frac{1}{r} \right) \right].
\end{align*}

The boundary conditions are as follows:
\begin{itemize}
\item At the star center (\(r=0\)), \(\chi_I \propto r^3\) for each fluid.
\item At the boundary of each fluid \(r = R_I\), the Lagrangian pressure perturbation must vanish, i.e.,
\begin{equation}
\begin{aligned}
\Delta p_I=&\delta p_I+\xi _I\frac{dp_{I0}}{dr}
\\
=&-\xi _I\frac{dp_{I0}}{dr}+\xi _I\frac{dp_{I0}}{dr}
\\&-\gamma _Ip_{I0}\left[ \frac{e^{\phi _0}}{r^2}\frac{\partial}{\partial r}\left( r^2e^{-\phi _0}\xi_I \right)
    \right.\\&\left.-4\pi Gre^{2\nu _0}\sum_J{(}\varepsilon _{J0}+p_{J0})(\xi _J-\xi _I) \right] 
\\
=&0.
\end{aligned}
\end{equation}
Among this, $p_{I0}$ tends to 0 at the corresponding outer boundary of the fluid. Therefore, since the other quantities ($\xi _I$ and unperturbed quantities) in the last term must be finite, the contribution of this term to the boundary constraint is ignored, and thus the boundary condition reduces to
\begin{equation}
\begin{aligned}
\Delta p_I=&-\gamma _Ip_{I0}\left[ \frac{e^{\phi _0}}{r^2}\frac{\partial}{\partial r}\left( r^2e^{-\phi _0}\xi_I \right)\right] 
\\
=&0.
\end{aligned}
\end{equation}
This outer boundary condition is also consistent with those in \cite{Caballero_2024,kumar2025stabilityanalysistwofluidneutron}.
In \cite{Kain:2020oho} the corresponding term of $e^{\phi _0}$  is $e^{2\phi _0}$ after unifying the convention, which seems to be typo. But this slight discrepancy merely introduces a negligible scaling factor in the numerical results and does not affect the stability boundary.
\end{itemize}

\subsubsection*{Stability Condition and Variational Principle}
The same stability criterion from the single-fluid analysis carries over to the two-fluid case: for perturbations of the form $\xi_I(r, t) = \xi_I(r) e^{i\omega t}$, the configuration is stable for real $\omega$ and dynamically unstable for imaginary $\omega$.

Although the multi-fluid pulsation equations cannot be written in a standard Sturm-Liouville form, one can still give the coupled Rayleigh quotient:
\[
\omega^2[\{\xi_I\}] = \frac{\text{Numerator}[\{\xi_I\}]}{\text{Denominator}[\{\xi_I\}]},
\]
which could be constructed by the integration of the pulsation equation, including various complex coupling terms.
The extreme value of its trial functions are the eigenvalues, and the minimum value is the square of the fundamental frequency. If there exists a set of trial functions \(\{\widetilde{\xi}_I\}\) such that \(\omega^2 < 0\), then the system is unstable to that perturbation mode.
\newpage

\bibliography{ZCWZ}

\begin{thebibliography}{103}%
\makeatletter
\providecommand \@ifxundefined [1]{%
 \@ifx{#1\undefined}
}%
\providecommand \@ifnum [1]{%
 \ifnum #1\expandafter \@firstoftwo
 \else \expandafter \@secondoftwo
 \fi
}%
\providecommand \@ifx [1]{%
 \ifx #1\expandafter \@firstoftwo
 \else \expandafter \@secondoftwo
 \fi
}%
\providecommand \natexlab [1]{#1}%
\providecommand \enquote  [1]{``#1''}%
\providecommand \bibnamefont  [1]{#1}%
\providecommand \bibfnamefont [1]{#1}%
\providecommand \citenamefont [1]{#1}%
\providecommand \href@noop [0]{\@secondoftwo}%
\providecommand \href [0]{\begingroup \@sanitize@url \@href}%
\providecommand \@href[1]{\@@startlink{#1}\@@href}%
\providecommand \@@href[1]{\endgroup#1\@@endlink}%
\providecommand \@sanitize@url [0]{\catcode `\\12\catcode `\$12\catcode
  `\&12\catcode `\#12\catcode `\^12\catcode `\_12\catcode `\%12\relax}%
\providecommand \@@startlink[1]{}%
\providecommand \@@endlink[0]{}%
\providecommand \url  [0]{\begingroup\@sanitize@url \@url }%
\providecommand \@url [1]{\endgroup\@href {#1}{\urlprefix }}%
\providecommand \urlprefix  [0]{URL }%
\providecommand \Eprint [0]{\href }%
\providecommand \doibase [0]{https://doi.org/}%
\providecommand \selectlanguage [0]{\@gobble}%
\providecommand \bibinfo  [0]{\@secondoftwo}%
\providecommand \bibfield  [0]{\@secondoftwo}%
\providecommand \translation [1]{[#1]}%
\providecommand \BibitemOpen [0]{}%
\providecommand \bibitemStop [0]{}%
\providecommand \bibitemNoStop [0]{.\EOS\space}%
\providecommand \EOS [0]{\spacefactor3000\relax}%
\providecommand \BibitemShut  [1]{\csname bibitem#1\endcsname}%
\let\auto@bib@innerbib\@empty
\bibitem [{\citenamefont {Abbott}\ \emph {et~al.}(2017)\citenamefont {Abbott},
  \citenamefont {Abbott},\ and\ \citenamefont
  {Abbott}}]{PhysRevLett.119.161101}%
  \BibitemOpen
  \bibfield  {author} {\bibinfo {author} {\bibfnamefont {B.~P.}\ \bibnamefont
  {Abbott}}, \bibinfo {author} {\bibfnamefont {R.}~\bibnamefont {Abbott}},\
  and\ \bibinfo {author} {\bibfnamefont {e.~a.}\ \bibnamefont {Abbott}}
  (\bibinfo {collaboration} {LIGO Scientific Collaboration and Virgo
  Collaboration}),\ }\href {https://doi.org/10.1103/PhysRevLett.119.161101}
  {\bibfield  {journal} {\bibinfo  {journal} {Phys. Rev. Lett.}\ }\textbf
  {\bibinfo {volume} {119}},\ \bibinfo {pages} {161101} (\bibinfo {year}
  {2017})}\BibitemShut {NoStop}%
\bibitem [{\citenamefont {Chandrasekhar}(1964)}]{Chandrasekhar:1964}%
  \BibitemOpen
  \bibfield  {author} {\bibinfo {author} {\bibfnamefont {S.}~\bibnamefont
  {Chandrasekhar}},\ }\href {https://doi.org/10.1086/147938} {\bibfield
  {journal} {\bibinfo  {journal} {Astrophys. J.}\ }\textbf {\bibinfo {volume}
  {140}},\ \bibinfo {pages} {417} (\bibinfo {year} {1964})}\BibitemShut
  {NoStop}%
\bibitem [{\citenamefont {Bardeen}\ \emph
  {et~al.}(1966{\natexlab{a}})\citenamefont {Bardeen}, \citenamefont {Thorne},\
  and\ \citenamefont {Meltzer}}]{bardeen1966}%
  \BibitemOpen
  \bibfield  {author} {\bibinfo {author} {\bibfnamefont {J.~M.}\ \bibnamefont
  {Bardeen}}, \bibinfo {author} {\bibfnamefont {K.~S.}\ \bibnamefont
  {Thorne}},\ and\ \bibinfo {author} {\bibfnamefont {D.~W.}\ \bibnamefont
  {Meltzer}},\ }\href {https://doi.org/10.1086/148791} {\bibfield  {journal}
  {\bibinfo  {journal} {The Astrophysical Journal}\ }\textbf {\bibinfo {volume}
  {145}},\ \bibinfo {pages} {505} (\bibinfo {year}
  {1966}{\natexlab{a}})}\BibitemShut {NoStop}%
\bibitem [{\citenamefont {MISHRA}\ \emph {et~al.}(1993)\citenamefont {MISHRA},
  \citenamefont {MISRA}, \citenamefont {PANDA},\ and\ \citenamefont
  {PARIDA}}]{MISHRA_1993}%
  \BibitemOpen
  \bibfield  {author} {\bibinfo {author} {\bibfnamefont {H.}~\bibnamefont
  {MISHRA}}, \bibinfo {author} {\bibfnamefont {S.}~\bibnamefont {MISRA}},
  \bibinfo {author} {\bibfnamefont {P.}~\bibnamefont {PANDA}},\ and\ \bibinfo
  {author} {\bibfnamefont {B.}~\bibnamefont {PARIDA}},\ }\href
  {https://doi.org/10.1142/s0218301393000212} {\bibfield  {journal} {\bibinfo
  {journal} {International Journal of Modern Physics E}\ }\textbf {\bibinfo
  {volume} {02}},\ \bibinfo {pages} {547–563} (\bibinfo {year}
  {1993})}\BibitemShut {NoStop}%
\bibitem [{\citenamefont {Ghosh}\ \emph {et~al.}(1995)\citenamefont {Ghosh},
  \citenamefont {Phatak},\ and\ \citenamefont {Sahu}}]{Ghosh_1995}%
  \BibitemOpen
  \bibfield  {author} {\bibinfo {author} {\bibfnamefont {S.~K.}\ \bibnamefont
  {Ghosh}}, \bibinfo {author} {\bibfnamefont {S.~C.}\ \bibnamefont {Phatak}},\
  and\ \bibinfo {author} {\bibfnamefont {P.~K.}\ \bibnamefont {Sahu}},\ }\href
  {https://doi.org/10.1007/bf01299764} {\bibfield  {journal} {\bibinfo
  {journal} {Zeitschrift f\"{u}r Physik A Hadrons and Nuclei}\ }\textbf
  {\bibinfo {volume} {352}},\ \bibinfo {pages} {457–466} (\bibinfo {year}
  {1995})}\BibitemShut {NoStop}%
\bibitem [{\citenamefont {KHADKIKAR}\ \emph {et~al.}(1995)\citenamefont
  {KHADKIKAR}, \citenamefont {MISHRA},\ and\ \citenamefont
  {MISHRA}}]{KHADKIKAR_1995}%
  \BibitemOpen
  \bibfield  {author} {\bibinfo {author} {\bibfnamefont {S.}~\bibnamefont
  {KHADKIKAR}}, \bibinfo {author} {\bibfnamefont {A.}~\bibnamefont {MISHRA}},\
  and\ \bibinfo {author} {\bibfnamefont {H.}~\bibnamefont {MISHRA}},\ }\href
  {https://doi.org/10.1142/s0217732395002787} {\bibfield  {journal} {\bibinfo
  {journal} {Modern Physics Letters A}\ }\textbf {\bibinfo {volume} {10}},\
  \bibinfo {pages} {2651–2663} (\bibinfo {year} {1995})}\BibitemShut
  {NoStop}%
\bibitem [{\citenamefont {Di~Clemente}\ \emph {et~al.}(2020)\citenamefont
  {Di~Clemente}, \citenamefont {Mannarelli},\ and\ \citenamefont
  {Tonelli}}]{Di_Clemente_2020}%
  \BibitemOpen
  \bibfield  {author} {\bibinfo {author} {\bibfnamefont {F.}~\bibnamefont
  {Di~Clemente}}, \bibinfo {author} {\bibfnamefont {M.}~\bibnamefont
  {Mannarelli}},\ and\ \bibinfo {author} {\bibfnamefont {F.}~\bibnamefont
  {Tonelli}},\ }\bibfield  {journal} {\bibinfo  {journal} {Physical Review D}\
  }\textbf {\bibinfo {volume} {101}},\ \href
  {https://doi.org/10.1103/physrevd.101.103003} {10.1103/physrevd.101.103003}
  (\bibinfo {year} {2020})\BibitemShut {NoStop}%
\bibitem [{\citenamefont {Henriques}\ \emph {et~al.}(1990)\citenamefont
  {Henriques}, \citenamefont {Liddle},\ and\ \citenamefont
  {Moorhouse}}]{HENRIQUES1990511}%
  \BibitemOpen
  \bibfield  {author} {\bibinfo {author} {\bibfnamefont {A.}~\bibnamefont
  {Henriques}}, \bibinfo {author} {\bibfnamefont {A.~R.}\ \bibnamefont
  {Liddle}},\ and\ \bibinfo {author} {\bibfnamefont {R.}~\bibnamefont
  {Moorhouse}},\ }\href
  {https://doi.org/https://doi.org/10.1016/0370-2693(90)90789-9} {\bibfield
  {journal} {\bibinfo  {journal} {Physics Letters B}\ }\textbf {\bibinfo
  {volume} {251}},\ \bibinfo {pages} {511} (\bibinfo {year}
  {1990})}\BibitemShut {NoStop}%
\bibitem [{\citenamefont {Zhang}\ \emph {et~al.}(2025)\citenamefont {Zhang},
  \citenamefont {Li}, \citenamefont {Zhang}, \citenamefont {Shen},\ and\
  \citenamefont {Zhang}}]{Zhang_2025}%
  \BibitemOpen
  \bibfield  {author} {\bibinfo {author} {\bibfnamefont {N.}~\bibnamefont
  {Zhang}}, \bibinfo {author} {\bibfnamefont {B.-A.}\ \bibnamefont {Li}},
  \bibinfo {author} {\bibfnamefont {J.}~\bibnamefont {Zhang}}, \bibinfo
  {author} {\bibfnamefont {W.}~\bibnamefont {Shen}},\ and\ \bibinfo {author}
  {\bibfnamefont {H.}~\bibnamefont {Zhang}},\ }\href
  {https://doi.org/10.3390/sym17101669} {\bibfield  {journal} {\bibinfo
  {journal} {Symmetry}\ }\textbf {\bibinfo {volume} {17}},\ \bibinfo {pages}
  {1669} (\bibinfo {year} {2025})}\BibitemShut {NoStop}%
\bibitem [{\citenamefont
  {Panotopoulos}(2019)}]{panotopoulos2019compactstarschallengeview}%
  \BibitemOpen
  \bibfield  {author} {\bibinfo {author} {\bibfnamefont {G.}~\bibnamefont
  {Panotopoulos}},\ }\href@noop {} {} (\bibinfo {year} {2019}),\ \bibinfo
  {note} {arXiv:1910.02279}\BibitemShut {NoStop}%
\bibitem [{\citenamefont {Rezaei}(2018)}]{Rezaei_2018}%
  \BibitemOpen
  \bibfield  {author} {\bibinfo {author} {\bibfnamefont {Z.}~\bibnamefont
  {Rezaei}},\ }\href {https://doi.org/10.1142/s0218271819500020} {\bibfield
  {journal} {\bibinfo  {journal} {International Journal of Modern Physics D}\
  }\textbf {\bibinfo {volume} {27}},\ \bibinfo {pages} {1950002} (\bibinfo
  {year} {2018})}\BibitemShut {NoStop}%
\bibitem [{\citenamefont {Wang}\ \emph {et~al.}(2019)\citenamefont {Wang},
  \citenamefont {Qi}, \citenamefont {Yang}, \citenamefont {Zhang},\ and\
  \citenamefont {Wang}}]{Wang_2019}%
  \BibitemOpen
  \bibfield  {author} {\bibinfo {author} {\bibfnamefont {X.~D.}\ \bibnamefont
  {Wang}}, \bibinfo {author} {\bibfnamefont {B.}~\bibnamefont {Qi}}, \bibinfo
  {author} {\bibfnamefont {G.~L.}\ \bibnamefont {Yang}}, \bibinfo {author}
  {\bibfnamefont {N.~B.}\ \bibnamefont {Zhang}},\ and\ \bibinfo {author}
  {\bibfnamefont {S.~Y.}\ \bibnamefont {Wang}},\ }\href
  {https://doi.org/10.1142/s0218271819501487} {\bibfield  {journal} {\bibinfo
  {journal} {International Journal of Modern Physics D}\ }\textbf {\bibinfo
  {volume} {28}},\ \bibinfo {pages} {1950148} (\bibinfo {year}
  {2019})}\BibitemShut {NoStop}%
\bibitem [{\citenamefont {Mukhopadhyay}\ and\ \citenamefont
  {Schaffner-Bielich}(2016)}]{Mukhopadhyay_2016}%
  \BibitemOpen
  \bibfield  {author} {\bibinfo {author} {\bibfnamefont {P.}~\bibnamefont
  {Mukhopadhyay}}\ and\ \bibinfo {author} {\bibfnamefont {J.}~\bibnamefont
  {Schaffner-Bielich}},\ }\bibfield  {journal} {\bibinfo  {journal} {Physical
  Review D}\ }\textbf {\bibinfo {volume} {93}},\ \href
  {https://doi.org/10.1103/physrevd.93.083009} {10.1103/physrevd.93.083009}
  (\bibinfo {year} {2016})\BibitemShut {NoStop}%
\bibitem [{\citenamefont {Xiang}\ \emph {et~al.}(2014)\citenamefont {Xiang},
  \citenamefont {Jiang}, \citenamefont {Zhang},\ and\ \citenamefont
  {Yang}}]{Xiang_2014}%
  \BibitemOpen
  \bibfield  {author} {\bibinfo {author} {\bibfnamefont {Q.-F.}\ \bibnamefont
  {Xiang}}, \bibinfo {author} {\bibfnamefont {W.-Z.}\ \bibnamefont {Jiang}},
  \bibinfo {author} {\bibfnamefont {D.-R.}\ \bibnamefont {Zhang}},\ and\
  \bibinfo {author} {\bibfnamefont {R.-Y.}\ \bibnamefont {Yang}},\ }\bibfield
  {journal} {\bibinfo  {journal} {Physical Review C}\ }\textbf {\bibinfo
  {volume} {89}},\ \href {https://doi.org/10.1103/physrevc.89.025803}
  {10.1103/physrevc.89.025803} (\bibinfo {year} {2014})\BibitemShut {NoStop}%
\bibitem [{\citenamefont {Leung}\ \emph {et~al.}(2011)\citenamefont {Leung},
  \citenamefont {Chu},\ and\ \citenamefont {Lin}}]{Leung_2011}%
  \BibitemOpen
  \bibfield  {author} {\bibinfo {author} {\bibfnamefont {S.-C.}\ \bibnamefont
  {Leung}}, \bibinfo {author} {\bibfnamefont {M.-C.}\ \bibnamefont {Chu}},\
  and\ \bibinfo {author} {\bibfnamefont {L.-M.}\ \bibnamefont {Lin}},\ }\href
  {https://doi.org/10.1103/PhysRevD.84.107301} {\bibfield  {journal} {\bibinfo
  {journal} {Phys. Rev. D}\ }\textbf {\bibinfo {volume} {84}},\ \bibinfo
  {pages} {107301} (\bibinfo {year} {2011})}\BibitemShut {NoStop}%
\bibitem [{\citenamefont {Leung}\ \emph {et~al.}(2012)\citenamefont {Leung},
  \citenamefont {Chu},\ and\ \citenamefont {Lin}}]{Leung_2012}%
  \BibitemOpen
  \bibfield  {author} {\bibinfo {author} {\bibfnamefont {S.-C.}\ \bibnamefont
  {Leung}}, \bibinfo {author} {\bibfnamefont {M.-C.}\ \bibnamefont {Chu}},\
  and\ \bibinfo {author} {\bibfnamefont {L.-M.}\ \bibnamefont {Lin}},\ }\href
  {https://doi.org/10.1103/PhysRevD.85.103528} {\bibfield  {journal} {\bibinfo
  {journal} {Phys. Rev. D}\ }\textbf {\bibinfo {volume} {85}},\ \bibinfo
  {pages} {103528} (\bibinfo {year} {2012})}\BibitemShut {NoStop}%
\bibitem [{\citenamefont {Zhang}\ \emph {et~al.}(2022)\citenamefont {Zhang},
  \citenamefont {Huang}, \citenamefont {Tsao},\ and\ \citenamefont
  {Lin}}]{zhang2022gw170817}%
  \BibitemOpen
  \bibfield  {author} {\bibinfo {author} {\bibfnamefont {K.}~\bibnamefont
  {Zhang}}, \bibinfo {author} {\bibfnamefont {G.-Z.}\ \bibnamefont {Huang}},
  \bibinfo {author} {\bibfnamefont {J.-S.}\ \bibnamefont {Tsao}},\ and\
  \bibinfo {author} {\bibfnamefont {F.-L.}\ \bibnamefont {Lin}},\ }\href@noop
  {} {\bibfield  {journal} {\bibinfo  {journal} {The European Physical Journal
  C}\ }\textbf {\bibinfo {volume} {82}},\ \bibinfo {pages} {1} (\bibinfo {year}
  {2022})}\BibitemShut {NoStop}%
\bibitem [{\citenamefont {Kain}(2020)}]{Kain:2020oho}%
  \BibitemOpen
  \bibfield  {author} {\bibinfo {author} {\bibfnamefont {B.}~\bibnamefont
  {Kain}},\ }\href {https://doi.org/10.1103/PhysRevD.102.023001} {\bibfield
  {journal} {\bibinfo  {journal} {Phys. Rev. D}\ }\textbf {\bibinfo {volume}
  {102}},\ \bibinfo {pages} {023001} (\bibinfo {year} {2020})}\BibitemShut
  {NoStop}%
\bibitem [{\citenamefont {Kain}(2021)}]{Kain:2021poc}%
  \BibitemOpen
  \bibfield  {author} {\bibinfo {author} {\bibfnamefont {B.}~\bibnamefont
  {Kain}},\ }\href {https://doi.org/10.1103/PhysRevD.103.043009} {\bibfield
  {journal} {\bibinfo  {journal} {Phys. Rev. D}\ }\textbf {\bibinfo {volume}
  {103}},\ \bibinfo {pages} {043009} (\bibinfo {year} {2021})}\BibitemShut
  {NoStop}%
\bibitem [{\citenamefont {Chen}\ \emph {et~al.}()\citenamefont {Chen},
  \citenamefont {Zhou}, \citenamefont {Feng},\ and\ \citenamefont
  {Zhang}}]{Chen2025Equivalence}%
  \BibitemOpen
  \bibfield  {author} {\bibinfo {author} {\bibfnamefont {T.-S.}\ \bibnamefont
  {Chen}}, \bibinfo {author} {\bibfnamefont {X.-D.}\ \bibnamefont {Zhou}},
  \bibinfo {author} {\bibfnamefont {H.}~\bibnamefont {Feng}},\ and\ \bibinfo
  {author} {\bibfnamefont {K.}~\bibnamefont {Zhang}},\ }\href@noop {} {\
  }\Eprint {https://arxiv.org/abs/2512.01640v2} {arXiv:2512.01640v2}
  \BibitemShut {NoStop}%
\bibitem [{\citenamefont {Tolman}(1939)}]{PhysRev.55.364}%
  \BibitemOpen
  \bibfield  {author} {\bibinfo {author} {\bibfnamefont {R.~C.}\ \bibnamefont
  {Tolman}},\ }\href {https://doi.org/10.1103/PhysRev.55.364} {\bibfield
  {journal} {\bibinfo  {journal} {Phys. Rev.}\ }\textbf {\bibinfo {volume}
  {55}},\ \bibinfo {pages} {364} (\bibinfo {year} {1939})}\BibitemShut
  {NoStop}%
\bibitem [{\citenamefont {Oppenheimer}\ and\ \citenamefont
  {Volkoff}(1939)}]{oppenheimer1939massive}%
  \BibitemOpen
  \bibfield  {author} {\bibinfo {author} {\bibfnamefont {J.~R.}\ \bibnamefont
  {Oppenheimer}}\ and\ \bibinfo {author} {\bibfnamefont {G.~M.}\ \bibnamefont
  {Volkoff}},\ }\href@noop {} {\bibfield  {journal} {\bibinfo  {journal}
  {Physical Review}\ }\textbf {\bibinfo {volume} {55}},\ \bibinfo {pages} {374}
  (\bibinfo {year} {1939})}\BibitemShut {NoStop}%
\bibitem [{\citenamefont {Panotopoulos}\ and\ \citenamefont
  {Lopes}(2017)}]{panotopoulos2017dark}%
  \BibitemOpen
  \bibfield  {author} {\bibinfo {author} {\bibfnamefont {G.}~\bibnamefont
  {Panotopoulos}}\ and\ \bibinfo {author} {\bibfnamefont {I.}~\bibnamefont
  {Lopes}},\ }\href@noop {} {\bibfield  {journal} {\bibinfo  {journal}
  {Physical Review D}\ }\textbf {\bibinfo {volume} {96}},\ \bibinfo {pages}
  {083004} (\bibinfo {year} {2017})}\BibitemShut {NoStop}%
\bibitem [{\citenamefont {Kolb}(2018)}]{kolb2018early}%
  \BibitemOpen
  \bibfield  {author} {\bibinfo {author} {\bibfnamefont {E.}~\bibnamefont
  {Kolb}},\ }\href@noop {} {\emph {\bibinfo {title} {The early universe}}}\
  (\bibinfo  {publisher} {CRC press},\ \bibinfo {year} {2018})\BibitemShut
  {NoStop}%
\bibitem [{\citenamefont {Steffen}(2009)}]{steffen2009dark}%
  \BibitemOpen
  \bibfield  {author} {\bibinfo {author} {\bibfnamefont {F.~D.}\ \bibnamefont
  {Steffen}},\ }\href@noop {} {\bibfield  {journal} {\bibinfo  {journal} {The
  European Physical Journal C}\ }\textbf {\bibinfo {volume} {59}},\ \bibinfo
  {pages} {557} (\bibinfo {year} {2009})}\BibitemShut {NoStop}%
\bibitem [{\citenamefont {G{\"u}ver}\ \emph {et~al.}(2014)\citenamefont
  {G{\"u}ver}, \citenamefont {Erkoca}, \citenamefont {Reno},\ and\
  \citenamefont {Sarcevic}}]{guver2014capture}%
  \BibitemOpen
  \bibfield  {author} {\bibinfo {author} {\bibfnamefont {T.}~\bibnamefont
  {G{\"u}ver}}, \bibinfo {author} {\bibfnamefont {A.~E.}\ \bibnamefont
  {Erkoca}}, \bibinfo {author} {\bibfnamefont {M.~H.}\ \bibnamefont {Reno}},\
  and\ \bibinfo {author} {\bibfnamefont {I.}~\bibnamefont {Sarcevic}},\
  }\href@noop {} {\bibfield  {journal} {\bibinfo  {journal} {Journal of
  Cosmology and Astroparticle Physics}\ }\textbf {\bibinfo {volume}
  {2014}}\bibinfo  {number} { (05)},\ \bibinfo {pages} {013}}\BibitemShut
  {NoStop}%
\bibitem [{\citenamefont {Davoudiasl}(2013)}]{PhysRevD.88.095004}%
  \BibitemOpen
\bibfield  {number} {  }\bibfield  {author} {\bibinfo {author} {\bibfnamefont
  {H.}~\bibnamefont {Davoudiasl}},\ }\href
  {https://doi.org/10.1103/PhysRevD.88.095004} {\bibfield  {journal} {\bibinfo
  {journal} {Phys. Rev. D}\ }\textbf {\bibinfo {volume} {88}},\ \bibinfo
  {pages} {095004} (\bibinfo {year} {2013})}\BibitemShut {NoStop}%
\bibitem [{\citenamefont {Zentner}\ and\ \citenamefont
  {Hearin}(2011)}]{PhysRevD.84.101302}%
  \BibitemOpen
  \bibfield  {author} {\bibinfo {author} {\bibfnamefont {A.~R.}\ \bibnamefont
  {Zentner}}\ and\ \bibinfo {author} {\bibfnamefont {A.~P.}\ \bibnamefont
  {Hearin}},\ }\href {https://doi.org/10.1103/PhysRevD.84.101302} {\bibfield
  {journal} {\bibinfo  {journal} {Phys. Rev. D}\ }\textbf {\bibinfo {volume}
  {84}},\ \bibinfo {pages} {101302} (\bibinfo {year} {2011})}\BibitemShut
  {NoStop}%
\bibitem [{\citenamefont {Tulin}\ and\ \citenamefont {Yu}(2018)}]{Tulin_2018}%
  \BibitemOpen
  \bibfield  {author} {\bibinfo {author} {\bibfnamefont {S.}~\bibnamefont
  {Tulin}}\ and\ \bibinfo {author} {\bibfnamefont {H.-B.}\ \bibnamefont {Yu}},\
  }\href {https://doi.org/10.1016/j.physrep.2017.11.004} {\bibfield  {journal}
  {\bibinfo  {journal} {Physics Reports}\ }\textbf {\bibinfo {volume} {730}},\
  \bibinfo {pages} {1–57} (\bibinfo {year} {2018})}\BibitemShut {NoStop}%
\bibitem [{\citenamefont {van~den Aarssen}\ \emph {et~al.}(2012)\citenamefont
  {van~den Aarssen}, \citenamefont {Bringmann},\ and\ \citenamefont
  {Pfrommer}}]{van_den_Aarssen_2012}%
  \BibitemOpen
  \bibfield  {author} {\bibinfo {author} {\bibfnamefont {L.~G.}\ \bibnamefont
  {van~den Aarssen}}, \bibinfo {author} {\bibfnamefont {T.}~\bibnamefont
  {Bringmann}},\ and\ \bibinfo {author} {\bibfnamefont {C.}~\bibnamefont
  {Pfrommer}},\ }\bibfield  {journal} {\bibinfo  {journal} {Physical Review
  Letters}\ }\textbf {\bibinfo {volume} {109}},\ \href
  {https://doi.org/10.1103/physrevlett.109.231301}
  {10.1103/physrevlett.109.231301} (\bibinfo {year} {2012})\BibitemShut
  {NoStop}%
\bibitem [{\citenamefont {Tulin}\ \emph {et~al.}(2013)\citenamefont {Tulin},
  \citenamefont {Yu},\ and\ \citenamefont {Zurek}}]{PhysRevLett.110.111301}%
  \BibitemOpen
  \bibfield  {author} {\bibinfo {author} {\bibfnamefont {S.}~\bibnamefont
  {Tulin}}, \bibinfo {author} {\bibfnamefont {H.-B.}\ \bibnamefont {Yu}},\ and\
  \bibinfo {author} {\bibfnamefont {K.~M.}\ \bibnamefont {Zurek}},\ }\href
  {https://doi.org/10.1103/PhysRevLett.110.111301} {\bibfield  {journal}
  {\bibinfo  {journal} {Phys. Rev. Lett.}\ }\textbf {\bibinfo {volume} {110}},\
  \bibinfo {pages} {111301} (\bibinfo {year} {2013})}\BibitemShut {NoStop}%
\bibitem [{\citenamefont {Feng}\ \emph {et~al.}(2010)\citenamefont {Feng},
  \citenamefont {Kaplinghat},\ and\ \citenamefont {Yu}}]{Feng_2010}%
  \BibitemOpen
  \bibfield  {author} {\bibinfo {author} {\bibfnamefont {J.~L.}\ \bibnamefont
  {Feng}}, \bibinfo {author} {\bibfnamefont {M.}~\bibnamefont {Kaplinghat}},\
  and\ \bibinfo {author} {\bibfnamefont {H.-B.}\ \bibnamefont {Yu}},\
  }\bibfield  {journal} {\bibinfo  {journal} {Physical Review Letters}\
  }\textbf {\bibinfo {volume} {104}},\ \href
  {https://doi.org/10.1103/physrevlett.104.151301}
  {10.1103/physrevlett.104.151301} (\bibinfo {year} {2010})\BibitemShut
  {NoStop}%
\bibitem [{\citenamefont {Zhang}\ \emph {et~al.}(2023)\citenamefont {Zhang},
  \citenamefont {Luo}, \citenamefont {Tsao}, \citenamefont {Chen},\ and\
  \citenamefont {Lin}}]{zhang2023dark}%
  \BibitemOpen
  \bibfield  {author} {\bibinfo {author} {\bibfnamefont {K.}~\bibnamefont
  {Zhang}}, \bibinfo {author} {\bibfnamefont {L.-W.}\ \bibnamefont {Luo}},
  \bibinfo {author} {\bibfnamefont {J.-S.}\ \bibnamefont {Tsao}}, \bibinfo
  {author} {\bibfnamefont {C.-S.}\ \bibnamefont {Chen}},\ and\ \bibinfo
  {author} {\bibfnamefont {F.-L.}\ \bibnamefont {Lin}},\ }\href@noop {}
  {\bibfield  {journal} {\bibinfo  {journal} {Results in Physics}\ ,\ \bibinfo
  {pages} {106967}} (\bibinfo {year} {2023})}\BibitemShut {NoStop}%
\bibitem [{\citenamefont {Aprile}\ \emph {et~al.}(2010)\citenamefont {Aprile},
  \citenamefont {Arisaka}, \citenamefont {Arneodo} \emph
  {et~al.}}]{PhysRevLett.105.131302}%
  \BibitemOpen
  \bibfield  {author} {\bibinfo {author} {\bibfnamefont {E.}~\bibnamefont
  {Aprile}}, \bibinfo {author} {\bibfnamefont {K.}~\bibnamefont {Arisaka}},
  \bibinfo {author} {\bibfnamefont {F.}~\bibnamefont {Arneodo}}, \emph {et~al.}
  (\bibinfo {collaboration} {XENON100 Collaboration}),\ }\href
  {https://doi.org/10.1103/PhysRevLett.105.131302} {\bibfield  {journal}
  {\bibinfo  {journal} {Phys. Rev. Lett.}\ }\textbf {\bibinfo {volume} {105}},\
  \bibinfo {pages} {131302} (\bibinfo {year} {2010})}\BibitemShut {NoStop}%
\bibitem [{\citenamefont {Ahmed}\ \emph {et~al.}(2011)\citenamefont {Ahmed},
  \citenamefont {Akerib}, \citenamefont {Arrenberg} \emph
  {et~al.}}]{PhysRevLett.106.131302}%
  \BibitemOpen
  \bibfield  {author} {\bibinfo {author} {\bibfnamefont {Z.}~\bibnamefont
  {Ahmed}}, \bibinfo {author} {\bibfnamefont {D.~S.}\ \bibnamefont {Akerib}},
  \bibinfo {author} {\bibfnamefont {S.}~\bibnamefont {Arrenberg}}, \emph
  {et~al.} (\bibinfo {collaboration} {CDMS Collaboration}),\ }\href
  {https://doi.org/10.1103/PhysRevLett.106.131302} {\bibfield  {journal}
  {\bibinfo  {journal} {Phys. Rev. Lett.}\ }\textbf {\bibinfo {volume} {106}},\
  \bibinfo {pages} {131302} (\bibinfo {year} {2011})}\BibitemShut {NoStop}%
\bibitem [{\citenamefont {Aalseth}\ \emph {et~al.}(2011)\citenamefont
  {Aalseth}, \citenamefont {Barbeau}, \citenamefont {Bowden}, \citenamefont
  {Cabrera-Palmer}, \citenamefont {Colaresi}, \citenamefont {Collar},
  \citenamefont {Dazeley}, \citenamefont {De~Lurgio}, \citenamefont {Fast},
  \citenamefont {Fields} \emph {et~al.}}]{aalseth2011results}%
  \BibitemOpen
  \bibfield  {author} {\bibinfo {author} {\bibfnamefont {C.~E.}\ \bibnamefont
  {Aalseth}}, \bibinfo {author} {\bibfnamefont {P.}~\bibnamefont {Barbeau}},
  \bibinfo {author} {\bibfnamefont {N.}~\bibnamefont {Bowden}}, \bibinfo
  {author} {\bibfnamefont {B.}~\bibnamefont {Cabrera-Palmer}}, \bibinfo
  {author} {\bibfnamefont {J.}~\bibnamefont {Colaresi}}, \bibinfo {author}
  {\bibfnamefont {J.}~\bibnamefont {Collar}}, \bibinfo {author} {\bibfnamefont
  {S.}~\bibnamefont {Dazeley}}, \bibinfo {author} {\bibfnamefont
  {P.}~\bibnamefont {De~Lurgio}}, \bibinfo {author} {\bibfnamefont {J.~E.}\
  \bibnamefont {Fast}}, \bibinfo {author} {\bibfnamefont {N.}~\bibnamefont
  {Fields}}, \emph {et~al.},\ }\href@noop {} {\bibfield  {journal} {\bibinfo
  {journal} {Physical Review Letters}\ }\textbf {\bibinfo {volume} {106}},\
  \bibinfo {pages} {131301} (\bibinfo {year} {2011})}\BibitemShut {NoStop}%
\bibitem [{\citenamefont {Bernabei}\ \emph {et~al.}(2008)\citenamefont
  {Bernabei}, \citenamefont {Belli}, \citenamefont {Cappella}, \citenamefont
  {Cerulli}, \citenamefont {Dai}, \citenamefont {d’Angelo}, \citenamefont
  {He}, \citenamefont {Incicchitti}, \citenamefont {Kuang}, \citenamefont {Ma}
  \emph {et~al.}}]{bernabei2008first}%
  \BibitemOpen
  \bibfield  {author} {\bibinfo {author} {\bibfnamefont {R.}~\bibnamefont
  {Bernabei}}, \bibinfo {author} {\bibfnamefont {P.}~\bibnamefont {Belli}},
  \bibinfo {author} {\bibfnamefont {F.}~\bibnamefont {Cappella}}, \bibinfo
  {author} {\bibfnamefont {R.}~\bibnamefont {Cerulli}}, \bibinfo {author}
  {\bibfnamefont {C.}~\bibnamefont {Dai}}, \bibinfo {author} {\bibfnamefont
  {A.}~\bibnamefont {d’Angelo}}, \bibinfo {author} {\bibfnamefont
  {H.}~\bibnamefont {He}}, \bibinfo {author} {\bibfnamefont {A.}~\bibnamefont
  {Incicchitti}}, \bibinfo {author} {\bibfnamefont {H.}~\bibnamefont {Kuang}},
  \bibinfo {author} {\bibfnamefont {J.}~\bibnamefont {Ma}}, \emph {et~al.},\
  }\href@noop {} {\bibfield  {journal} {\bibinfo  {journal} {The European
  Physical Journal C}\ }\textbf {\bibinfo {volume} {56}},\ \bibinfo {pages}
  {333} (\bibinfo {year} {2008})}\BibitemShut {NoStop}%
\bibitem [{\citenamefont {Goldman}\ and\ \citenamefont
  {Nussinov}(1989)}]{PhysRevD.40.3221}%
  \BibitemOpen
  \bibfield  {author} {\bibinfo {author} {\bibfnamefont {I.}~\bibnamefont
  {Goldman}}\ and\ \bibinfo {author} {\bibfnamefont {S.}~\bibnamefont
  {Nussinov}},\ }\href {https://doi.org/10.1103/PhysRevD.40.3221} {\bibfield
  {journal} {\bibinfo  {journal} {Phys. Rev. D}\ }\textbf {\bibinfo {volume}
  {40}},\ \bibinfo {pages} {3221} (\bibinfo {year} {1989})}\BibitemShut
  {NoStop}%
\bibitem [{\citenamefont {Kouvaris}(2008)}]{PhysRevD.77.023006}%
  \BibitemOpen
  \bibfield  {author} {\bibinfo {author} {\bibfnamefont {C.}~\bibnamefont
  {Kouvaris}},\ }\href {https://doi.org/10.1103/PhysRevD.77.023006} {\bibfield
  {journal} {\bibinfo  {journal} {Phys. Rev. D}\ }\textbf {\bibinfo {volume}
  {77}},\ \bibinfo {pages} {023006} (\bibinfo {year} {2008})}\BibitemShut
  {NoStop}%
\bibitem [{\citenamefont {Bertone}\ and\ \citenamefont
  {Fairbairn}(2008)}]{PhysRevD.77.043515}%
  \BibitemOpen
  \bibfield  {author} {\bibinfo {author} {\bibfnamefont {G.}~\bibnamefont
  {Bertone}}\ and\ \bibinfo {author} {\bibfnamefont {M.}~\bibnamefont
  {Fairbairn}},\ }\href {https://doi.org/10.1103/PhysRevD.77.043515} {\bibfield
   {journal} {\bibinfo  {journal} {Phys. Rev. D}\ }\textbf {\bibinfo {volume}
  {77}},\ \bibinfo {pages} {043515} (\bibinfo {year} {2008})}\BibitemShut
  {NoStop}%
\bibitem [{\citenamefont {de~Lavallaz}\ and\ \citenamefont
  {Fairbairn}(2010)}]{PhysRevD.81.123521}%
  \BibitemOpen
  \bibfield  {author} {\bibinfo {author} {\bibfnamefont {A.}~\bibnamefont
  {de~Lavallaz}}\ and\ \bibinfo {author} {\bibfnamefont {M.}~\bibnamefont
  {Fairbairn}},\ }\href {https://doi.org/10.1103/PhysRevD.81.123521} {\bibfield
   {journal} {\bibinfo  {journal} {Phys. Rev. D}\ }\textbf {\bibinfo {volume}
  {81}},\ \bibinfo {pages} {123521} (\bibinfo {year} {2010})}\BibitemShut
  {NoStop}%
\bibitem [{\citenamefont {Kouvaris}\ and\ \citenamefont
  {Tinyakov}(2010)}]{PhysRevD.82.063531}%
  \BibitemOpen
  \bibfield  {author} {\bibinfo {author} {\bibfnamefont {C.}~\bibnamefont
  {Kouvaris}}\ and\ \bibinfo {author} {\bibfnamefont {P.}~\bibnamefont
  {Tinyakov}},\ }\href {https://doi.org/10.1103/PhysRevD.82.063531} {\bibfield
  {journal} {\bibinfo  {journal} {Phys. Rev. D}\ }\textbf {\bibinfo {volume}
  {82}},\ \bibinfo {pages} {063531} (\bibinfo {year} {2010})}\BibitemShut
  {NoStop}%
\bibitem [{\citenamefont {Brito}\ \emph {et~al.}(2015)\citenamefont {Brito},
  \citenamefont {Cardoso},\ and\ \citenamefont
  {Okawa}}]{PhysRevLett.115.111301}%
  \BibitemOpen
  \bibfield  {author} {\bibinfo {author} {\bibfnamefont {R.}~\bibnamefont
  {Brito}}, \bibinfo {author} {\bibfnamefont {V.}~\bibnamefont {Cardoso}},\
  and\ \bibinfo {author} {\bibfnamefont {H.}~\bibnamefont {Okawa}},\ }\href
  {https://doi.org/10.1103/PhysRevLett.115.111301} {\bibfield  {journal}
  {\bibinfo  {journal} {Phys. Rev. Lett.}\ }\textbf {\bibinfo {volume} {115}},\
  \bibinfo {pages} {111301} (\bibinfo {year} {2015})}\BibitemShut {NoStop}%
\bibitem [{\citenamefont {Cermeno}\ \emph {et~al.}(2017)\citenamefont
  {Cermeno}, \citenamefont {P{\'e}rez-Garc{\'i}a},\ and\ \citenamefont
  {Silk}}]{Cermeno_Perez_Garcia_Silk_2017}%
  \BibitemOpen
  \bibfield  {author} {\bibinfo {author} {\bibfnamefont {M.}~\bibnamefont
  {Cermeno}}, \bibinfo {author} {\bibfnamefont {M.~{\'A}.}\ \bibnamefont
  {P{\'e}rez-Garc{\'i}a}},\ and\ \bibinfo {author} {\bibfnamefont
  {J.}~\bibnamefont {Silk}},\ }\href {https://doi.org/10.1017/pasa.2017.38}
  {\bibfield  {journal} {\bibinfo  {journal} {Publications of the Astronomical
  Society of Australia}\ }\textbf {\bibinfo {volume} {34}},\ \bibinfo {pages}
  {e043} (\bibinfo {year} {2017})}\BibitemShut {NoStop}%
\bibitem [{\citenamefont {Gresham}\ and\ \citenamefont
  {Zurek}(2019{\natexlab{a}})}]{gresham2019asymmetric}%
  \BibitemOpen
  \bibfield  {author} {\bibinfo {author} {\bibfnamefont {M.~I.}\ \bibnamefont
  {Gresham}}\ and\ \bibinfo {author} {\bibfnamefont {K.~M.}\ \bibnamefont
  {Zurek}},\ }\href@noop {} {\bibfield  {journal} {\bibinfo  {journal}
  {Physical Review D}\ }\textbf {\bibinfo {volume} {99}},\ \bibinfo {pages}
  {083008} (\bibinfo {year} {2019}{\natexlab{a}})}\BibitemShut {NoStop}%
\bibitem [{\citenamefont {Ellis}\ \emph {et~al.}(2018)\citenamefont {Ellis},
  \citenamefont {H\"utsi}, \citenamefont {Kannike}, \citenamefont {Marzola},
  \citenamefont {Raidal},\ and\ \citenamefont {Vaskonen}}]{PhysRevD.97.123007}%
  \BibitemOpen
  \bibfield  {author} {\bibinfo {author} {\bibfnamefont {J.}~\bibnamefont
  {Ellis}}, \bibinfo {author} {\bibfnamefont {G.}~\bibnamefont {H\"utsi}},
  \bibinfo {author} {\bibfnamefont {K.}~\bibnamefont {Kannike}}, \bibinfo
  {author} {\bibfnamefont {L.}~\bibnamefont {Marzola}}, \bibinfo {author}
  {\bibfnamefont {M.}~\bibnamefont {Raidal}},\ and\ \bibinfo {author}
  {\bibfnamefont {V.}~\bibnamefont {Vaskonen}},\ }\href
  {https://doi.org/10.1103/PhysRevD.97.123007} {\bibfield  {journal} {\bibinfo
  {journal} {Phys. Rev. D}\ }\textbf {\bibinfo {volume} {97}},\ \bibinfo
  {pages} {123007} (\bibinfo {year} {2018})}\BibitemShut {NoStop}%
\bibitem [{\citenamefont {Deliyergiyev}\ \emph {et~al.}(2019)\citenamefont
  {Deliyergiyev}, \citenamefont {Del~Popolo}, \citenamefont {Tolos},
  \citenamefont {Le~Delliou}, \citenamefont {Lee},\ and\ \citenamefont
  {Burgio}}]{PhysRevD.99.063015}%
  \BibitemOpen
  \bibfield  {author} {\bibinfo {author} {\bibfnamefont {M.}~\bibnamefont
  {Deliyergiyev}}, \bibinfo {author} {\bibfnamefont {A.}~\bibnamefont
  {Del~Popolo}}, \bibinfo {author} {\bibfnamefont {L.}~\bibnamefont {Tolos}},
  \bibinfo {author} {\bibfnamefont {M.}~\bibnamefont {Le~Delliou}}, \bibinfo
  {author} {\bibfnamefont {X.}~\bibnamefont {Lee}},\ and\ \bibinfo {author}
  {\bibfnamefont {F.}~\bibnamefont {Burgio}},\ }\href
  {https://doi.org/10.1103/PhysRevD.99.063015} {\bibfield  {journal} {\bibinfo
  {journal} {Phys. Rev. D}\ }\textbf {\bibinfo {volume} {99}},\ \bibinfo
  {pages} {063015} (\bibinfo {year} {2019})}\BibitemShut {NoStop}%
\bibitem [{\citenamefont {Nelson}\ \emph {et~al.}(2019)\citenamefont {Nelson},
  \citenamefont {Reddy},\ and\ \citenamefont {Zhou}}]{nelson2019dark}%
  \BibitemOpen
  \bibfield  {author} {\bibinfo {author} {\bibfnamefont {A.~E.}\ \bibnamefont
  {Nelson}}, \bibinfo {author} {\bibfnamefont {S.}~\bibnamefont {Reddy}},\ and\
  \bibinfo {author} {\bibfnamefont {D.}~\bibnamefont {Zhou}},\ }\href@noop {}
  {\bibfield  {journal} {\bibinfo  {journal} {Journal of Cosmology and
  Astroparticle Physics}\ }\textbf {\bibinfo {volume} {2019}}\bibinfo  {number}
  { (07)},\ \bibinfo {pages} {012}}\BibitemShut {NoStop}%
\bibitem [{\citenamefont {Bramante}\ and\ \citenamefont
  {Raj}(2024)}]{bramante2025darkmattercompactstars}%
  \BibitemOpen
\bibfield  {number} {  }\bibfield  {author} {\bibinfo {author} {\bibfnamefont
  {J.}~\bibnamefont {Bramante}}\ and\ \bibinfo {author} {\bibfnamefont
  {N.}~\bibnamefont {Raj}},\ }\href@noop {} {} (\bibinfo {year}
  {2024})\BibitemShut {NoStop}%
\bibitem [{\citenamefont {Alford}\ \emph {et~al.}(2005)\citenamefont {Alford},
  \citenamefont {Braby}, \citenamefont {Paris},\ and\ \citenamefont
  {Reddy}}]{Alford_2005}%
  \BibitemOpen
  \bibfield  {author} {\bibinfo {author} {\bibfnamefont {M.}~\bibnamefont
  {Alford}}, \bibinfo {author} {\bibfnamefont {M.}~\bibnamefont {Braby}},
  \bibinfo {author} {\bibfnamefont {M.}~\bibnamefont {Paris}},\ and\ \bibinfo
  {author} {\bibfnamefont {S.}~\bibnamefont {Reddy}},\ }\href@noop {}
  {\bibfield  {journal} {\bibinfo  {journal} {The Astrophysical Journal}\
  }\textbf {\bibinfo {volume} {629}},\ \bibinfo {pages} {969} (\bibinfo {year}
  {2005})}\BibitemShut {NoStop}%
\bibitem [{\citenamefont {Issifu}\ \emph {et~al.}(2025)\citenamefont {Issifu},
  \citenamefont {Konstantinou}, \citenamefont {Thakur},\ and\ \citenamefont
  {Frederico}}]{issifu2025rotatingprotoneutronstarsadmixed}%
  \BibitemOpen
  \bibfield  {author} {\bibinfo {author} {\bibfnamefont {A.}~\bibnamefont
  {Issifu}}, \bibinfo {author} {\bibfnamefont {A.}~\bibnamefont
  {Konstantinou}}, \bibinfo {author} {\bibfnamefont {P.}~\bibnamefont
  {Thakur}},\ and\ \bibinfo {author} {\bibfnamefont {T.}~\bibnamefont
  {Frederico}},\ }\href@noop {} {} (\bibinfo {year} {2025})\BibitemShut
  {NoStop}%
\bibitem [{\citenamefont {Zhang}\ and\ \citenamefont
  {Lin}(2020)}]{zhang2020constraint}%
  \BibitemOpen
  \bibfield  {author} {\bibinfo {author} {\bibfnamefont {K.}~\bibnamefont
  {Zhang}}\ and\ \bibinfo {author} {\bibfnamefont {F.-L.}\ \bibnamefont
  {Lin}},\ }\href@noop {} {\bibfield  {journal} {\bibinfo  {journal}
  {Universe}\ }\textbf {\bibinfo {volume} {6}},\ \bibinfo {pages} {231}
  (\bibinfo {year} {2020})}\BibitemShut {NoStop}%
\bibitem [{\citenamefont {Pitz}\ and\ \citenamefont
  {Schaffner-Bielich}(2025)}]{pitz2024generatingultracompactneutronstars}%
  \BibitemOpen
  \bibfield  {author} {\bibinfo {author} {\bibfnamefont {S.~L.}\ \bibnamefont
  {Pitz}}\ and\ \bibinfo {author} {\bibfnamefont {J.}~\bibnamefont
  {Schaffner-Bielich}},\ }\href@noop {} {} (\bibinfo {year} {2025})\BibitemShut
  {NoStop}%
\bibitem [{\citenamefont {Kumar}\ and\ \citenamefont
  {Sotani}(2025)}]{kumar2025stabilityanalysistwofluidneutron}%
  \BibitemOpen
  \bibfield  {author} {\bibinfo {author} {\bibfnamefont {A.}~\bibnamefont
  {Kumar}}\ and\ \bibinfo {author} {\bibfnamefont {H.}~\bibnamefont {Sotani}},\
  }\href@noop {} {} (\bibinfo {year} {2025})\BibitemShut {NoStop}%
\bibitem [{\citenamefont {Ivanytskyi}\ \emph {et~al.}(2020)\citenamefont
  {Ivanytskyi}, \citenamefont {Sagun},\ and\ \citenamefont
  {Lopes}}]{Ivanytskyi_2020}%
  \BibitemOpen
  \bibfield  {author} {\bibinfo {author} {\bibfnamefont {O.}~\bibnamefont
  {Ivanytskyi}}, \bibinfo {author} {\bibfnamefont {V.}~\bibnamefont {Sagun}},\
  and\ \bibinfo {author} {\bibfnamefont {I.}~\bibnamefont {Lopes}},\ }\bibfield
   {journal} {\bibinfo  {journal} {Physical Review D}\ }\textbf {\bibinfo
  {volume} {102}},\ \href {https://doi.org/10.1103/physrevd.102.063028}
  {10.1103/physrevd.102.063028} (\bibinfo {year} {2020})\BibitemShut {NoStop}%
\bibitem [{\citenamefont {Gresham}\ and\ \citenamefont
  {Zurek}(2019{\natexlab{b}})}]{Gresham_2019}%
  \BibitemOpen
  \bibfield  {author} {\bibinfo {author} {\bibfnamefont {M.~I.}\ \bibnamefont
  {Gresham}}\ and\ \bibinfo {author} {\bibfnamefont {K.~M.}\ \bibnamefont
  {Zurek}},\ }\bibfield  {journal} {\bibinfo  {journal} {Physical Review D}\
  }\textbf {\bibinfo {volume} {99}},\ \href
  {https://doi.org/10.1103/physrevd.99.083008} {10.1103/physrevd.99.083008}
  (\bibinfo {year} {2019}{\natexlab{b}})\BibitemShut {NoStop}%
\bibitem [{\citenamefont {Bardeen}\ \emph
  {et~al.}(1966{\natexlab{b}})\citenamefont {Bardeen}, \citenamefont {Thorne},\
  and\ \citenamefont {Meltzer}}]{Bardeen:1966}%
  \BibitemOpen
  \bibfield  {author} {\bibinfo {author} {\bibfnamefont {J.~M.}\ \bibnamefont
  {Bardeen}}, \bibinfo {author} {\bibfnamefont {K.~S.}\ \bibnamefont
  {Thorne}},\ and\ \bibinfo {author} {\bibfnamefont {D.~W.}\ \bibnamefont
  {Meltzer}},\ }\href {https://doi.org/10.1086/148796} {\bibfield  {journal}
  {\bibinfo  {journal} {Astrophys. J.}\ }\textbf {\bibinfo {volume} {145}},\
  \bibinfo {pages} {505} (\bibinfo {year} {1966}{\natexlab{b}})}\BibitemShut
  {NoStop}%
\bibitem [{\citenamefont {Shahrbaf}\ \emph
  {et~al.}(2026{\natexlab{a}})\citenamefont {Shahrbaf}, \citenamefont
  {Rafiei~Karkevandi}, \citenamefont {Ayriyan},\ and\ \citenamefont
  {Typel}}]{shahrbaf2024observational}%
  \BibitemOpen
  \bibfield  {author} {\bibinfo {author} {\bibfnamefont {M.}~\bibnamefont
  {Shahrbaf}}, \bibinfo {author} {\bibfnamefont {D.}~\bibnamefont
  {Rafiei~Karkevandi}}, \bibinfo {author} {\bibfnamefont {A.}~\bibnamefont
  {Ayriyan}},\ and\ \bibinfo {author} {\bibfnamefont {S.}~\bibnamefont
  {Typel}},\ }\href@noop {} {\bibfield  {journal} {\bibinfo  {journal}
  {Astronomy \& Astrophysics}\ }\textbf {\bibinfo {volume} {706}},\ \bibinfo
  {pages} {A203} (\bibinfo {year} {2026}{\natexlab{a}})}\BibitemShut {NoStop}%
\bibitem [{\citenamefont {Shahrbaf}\ \emph
  {et~al.}(2026{\natexlab{b}})\citenamefont {Shahrbaf}, \citenamefont
  {Thakur},\ and\ \citenamefont
  {Rafiei~Karkevandi}}]{shahrbaf2025probingstrangedarkmatter}%
  \BibitemOpen
  \bibfield  {author} {\bibinfo {author} {\bibfnamefont {M.}~\bibnamefont
  {Shahrbaf}}, \bibinfo {author} {\bibfnamefont {P.}~\bibnamefont {Thakur}},\
  and\ \bibinfo {author} {\bibfnamefont {D.}~\bibnamefont
  {Rafiei~Karkevandi}},\ }\href@noop {} {} (\bibinfo {year}
  {2026}{\natexlab{b}})\BibitemShut {NoStop}%
\bibitem [{\citenamefont {Mukhopadhyay}\ \emph {et~al.}(2017)\citenamefont
  {Mukhopadhyay}, \citenamefont {Atta}, \citenamefont {Imam}, \citenamefont
  {Basu},\ and\ \citenamefont {Samanta}}]{Mukhopadhyay_2017}%
  \BibitemOpen
  \bibfield  {author} {\bibinfo {author} {\bibfnamefont {S.}~\bibnamefont
  {Mukhopadhyay}}, \bibinfo {author} {\bibfnamefont {D.}~\bibnamefont {Atta}},
  \bibinfo {author} {\bibfnamefont {K.}~\bibnamefont {Imam}}, \bibinfo {author}
  {\bibfnamefont {D.~N.}\ \bibnamefont {Basu}},\ and\ \bibinfo {author}
  {\bibfnamefont {C.}~\bibnamefont {Samanta}},\ }\bibfield  {journal} {\bibinfo
   {journal} {The European Physical Journal C}\ }\textbf {\bibinfo {volume}
  {77}},\ \href {https://doi.org/10.1140/epjc/s10052-017-5006-3}
  {10.1140/epjc/s10052-017-5006-3} (\bibinfo {year} {2017})\BibitemShut
  {NoStop}%
\bibitem [{\citenamefont {Pandharipande}\ \emph {et~al.}(1989)\citenamefont
  {Pandharipande}, \citenamefont {Pethick},\ and\ \citenamefont
  {Ravenhall}}]{Pandharipande1989}%
  \BibitemOpen
  \bibfield  {author} {\bibinfo {author} {\bibfnamefont {V.~R.}\ \bibnamefont
  {Pandharipande}}, \bibinfo {author} {\bibfnamefont {C.~J.}\ \bibnamefont
  {Pethick}},\ and\ \bibinfo {author} {\bibfnamefont {D.~G.}\ \bibnamefont
  {Ravenhall}},\ }\href@noop {} {\bibfield  {journal} {\bibinfo  {journal}
  {Nuclear Physics A}\ }\textbf {\bibinfo {volume} {A498}},\ \bibinfo {pages}
  {313c} (\bibinfo {year} {1989})}\BibitemShut {NoStop}%
\bibitem [{\citenamefont {Akmal}\ \emph {et~al.}(1998)\citenamefont {Akmal},
  \citenamefont {Pandharipande},\ and\ \citenamefont
  {Ravenhall}}]{akmal1998equation}%
  \BibitemOpen
  \bibfield  {author} {\bibinfo {author} {\bibfnamefont {A.}~\bibnamefont
  {Akmal}}, \bibinfo {author} {\bibfnamefont {V.}~\bibnamefont
  {Pandharipande}},\ and\ \bibinfo {author} {\bibfnamefont {D.}~\bibnamefont
  {Ravenhall}},\ }\href@noop {} {\bibfield  {journal} {\bibinfo  {journal}
  {Physical Review C}\ }\textbf {\bibinfo {volume} {58}},\ \bibinfo {pages}
  {1804} (\bibinfo {year} {1998})}\BibitemShut {NoStop}%
\bibitem [{\citenamefont {Wiringa}\ \emph {et~al.}(1988)\citenamefont
  {Wiringa}, \citenamefont {Fiks},\ and\ \citenamefont
  {Fabrocini}}]{wiringa1988equation}%
  \BibitemOpen
  \bibfield  {author} {\bibinfo {author} {\bibfnamefont {R.~B.}\ \bibnamefont
  {Wiringa}}, \bibinfo {author} {\bibfnamefont {V.}~\bibnamefont {Fiks}},\ and\
  \bibinfo {author} {\bibfnamefont {A.}~\bibnamefont {Fabrocini}},\ }\href@noop
  {} {\bibfield  {journal} {\bibinfo  {journal} {Physical Review C}\ }\textbf
  {\bibinfo {volume} {38}},\ \bibinfo {pages} {1010} (\bibinfo {year}
  {1988})}\BibitemShut {NoStop}%
\bibitem [{\citenamefont {M{\"u}ther}\ \emph {et~al.}(1987)\citenamefont
  {M{\"u}ther}, \citenamefont {Prakash},\ and\ \citenamefont
  {Ainsworth}}]{muther1987nuclear}%
  \BibitemOpen
  \bibfield  {author} {\bibinfo {author} {\bibfnamefont {H.}~\bibnamefont
  {M{\"u}ther}}, \bibinfo {author} {\bibfnamefont {M.}~\bibnamefont
  {Prakash}},\ and\ \bibinfo {author} {\bibfnamefont {T.}~\bibnamefont
  {Ainsworth}},\ }\href@noop {} {\bibfield  {journal} {\bibinfo  {journal}
  {Physics Letters B}\ }\textbf {\bibinfo {volume} {199}},\ \bibinfo {pages}
  {469} (\bibinfo {year} {1987})}\BibitemShut {NoStop}%
\bibitem [{\citenamefont {Mueller}\ and\ \citenamefont
  {Serot}(1996)}]{mueller1996relativistic}%
  \BibitemOpen
  \bibfield  {author} {\bibinfo {author} {\bibfnamefont {H.}~\bibnamefont
  {Mueller}}\ and\ \bibinfo {author} {\bibfnamefont {B.~D.}\ \bibnamefont
  {Serot}},\ }\href@noop {} {\bibfield  {journal} {\bibinfo  {journal} {Nuclear
  Physics A}\ }\textbf {\bibinfo {volume} {606}},\ \bibinfo {pages} {508}
  (\bibinfo {year} {1996})}\BibitemShut {NoStop}%
\bibitem [{\citenamefont {Lackey}\ \emph {et~al.}(2006)\citenamefont {Lackey},
  \citenamefont {Nayyar},\ and\ \citenamefont {Owen}}]{PhysRevD.73.024021}%
  \BibitemOpen
  \bibfield  {author} {\bibinfo {author} {\bibfnamefont {B.~D.}\ \bibnamefont
  {Lackey}}, \bibinfo {author} {\bibfnamefont {M.}~\bibnamefont {Nayyar}},\
  and\ \bibinfo {author} {\bibfnamefont {B.~J.}\ \bibnamefont {Owen}},\ }\href
  {https://doi.org/10.1103/PhysRevD.73.024021} {\bibfield  {journal} {\bibinfo
  {journal} {Phys. Rev. D}\ }\textbf {\bibinfo {volume} {73}},\ \bibinfo
  {pages} {024021} (\bibinfo {year} {2006})}\BibitemShut {NoStop}%
\bibitem [{\citenamefont {Goriely}\ \emph {et~al.}(2009)\citenamefont
  {Goriely}, \citenamefont {Chamel},\ and\ \citenamefont
  {Pearson}}]{goriely2009skyrme}%
  \BibitemOpen
  \bibfield  {author} {\bibinfo {author} {\bibfnamefont {S.}~\bibnamefont
  {Goriely}}, \bibinfo {author} {\bibfnamefont {N.}~\bibnamefont {Chamel}},\
  and\ \bibinfo {author} {\bibfnamefont {J.}~\bibnamefont {Pearson}},\
  }\href@noop {} {\bibfield  {journal} {\bibinfo  {journal} {Physical review
  letters}\ }\textbf {\bibinfo {volume} {102}},\ \bibinfo {pages} {152503}
  (\bibinfo {year} {2009})}\BibitemShut {NoStop}%
\bibitem [{\citenamefont {Chabanat}\ \emph {et~al.}(1998)\citenamefont
  {Chabanat}, \citenamefont {Bonche}, \citenamefont {Haensel}, \citenamefont
  {Meyer},\ and\ \citenamefont {Schaeffer}}]{CHABANAT1998231}%
  \BibitemOpen
  \bibfield  {author} {\bibinfo {author} {\bibfnamefont {E.}~\bibnamefont
  {Chabanat}}, \bibinfo {author} {\bibfnamefont {P.}~\bibnamefont {Bonche}},
  \bibinfo {author} {\bibfnamefont {P.}~\bibnamefont {Haensel}}, \bibinfo
  {author} {\bibfnamefont {J.}~\bibnamefont {Meyer}},\ and\ \bibinfo {author}
  {\bibfnamefont {R.}~\bibnamefont {Schaeffer}},\ }\href
  {https://doi.org/https://doi.org/10.1016/S0375-9474(98)00180-8} {\bibfield
  {journal} {\bibinfo  {journal} {Nuclear Physics A}\ }\textbf {\bibinfo
  {volume} {635}},\ \bibinfo {pages} {231} (\bibinfo {year}
  {1998})}\BibitemShut {NoStop}%
\bibitem [{\citenamefont {Chabanat}\ \emph {et~al.}(1997)\citenamefont
  {Chabanat}, \citenamefont {Bonche}, \citenamefont {Haensel}, \citenamefont
  {Meyer},\ and\ \citenamefont {Schaeffer}}]{CHABANAT1997710}%
  \BibitemOpen
  \bibfield  {author} {\bibinfo {author} {\bibfnamefont {E.}~\bibnamefont
  {Chabanat}}, \bibinfo {author} {\bibfnamefont {P.}~\bibnamefont {Bonche}},
  \bibinfo {author} {\bibfnamefont {P.}~\bibnamefont {Haensel}}, \bibinfo
  {author} {\bibfnamefont {J.}~\bibnamefont {Meyer}},\ and\ \bibinfo {author}
  {\bibfnamefont {R.}~\bibnamefont {Schaeffer}},\ }\href
  {https://doi.org/https://doi.org/10.1016/S0375-9474(97)00596-4} {\bibfield
  {journal} {\bibinfo  {journal} {Nuclear Physics A}\ }\textbf {\bibinfo
  {volume} {627}},\ \bibinfo {pages} {710} (\bibinfo {year}
  {1997})}\BibitemShut {NoStop}%
\bibitem [{\citenamefont {Douchin}\ and\ \citenamefont
  {Haensel}(2001)}]{Douchin2001}%
  \BibitemOpen
  \bibfield  {author} {\bibinfo {author} {\bibfnamefont {F.}~\bibnamefont
  {Douchin}}\ and\ \bibinfo {author} {\bibfnamefont {P.}~\bibnamefont
  {Haensel}},\ }\href {https://doi.org/10.1051/0004-6361:20011402} {\bibfield
  {journal} {\bibinfo  {journal} {Astronomy \& Astrophysics}\ }\textbf
  {\bibinfo {volume} {380}},\ \bibinfo {pages} {151} (\bibinfo {year}
  {2001})}\BibitemShut {NoStop}%
\bibitem [{\citenamefont {Choudhury}\ \emph {et~al.}(2024)\citenamefont
  {Choudhury}, \citenamefont {Salmi}, \citenamefont {Vinciguerra},
  \citenamefont {Riley}, \citenamefont {Kini}, \citenamefont {Watts},
  \citenamefont {Dorsman}, \citenamefont {Bogdanov}, \citenamefont {Guillot},
  \citenamefont {Ray}, \citenamefont {Reardon}, \citenamefont {Remillard},
  \citenamefont {Bilous}, \citenamefont {Huppenkothen}, \citenamefont
  {Lattimer}, \citenamefont {Rutherford}, \citenamefont {Arzoumanian},
  \citenamefont {Gendreau}, \citenamefont {Morsink},\ and\ \citenamefont
  {Ho}}]{Choudhury_2024}%
  \BibitemOpen
  \bibfield  {author} {\bibinfo {author} {\bibfnamefont {D.}~\bibnamefont
  {Choudhury}}, \bibinfo {author} {\bibfnamefont {T.}~\bibnamefont {Salmi}},
  \bibinfo {author} {\bibfnamefont {S.}~\bibnamefont {Vinciguerra}}, \bibinfo
  {author} {\bibfnamefont {T.~E.}\ \bibnamefont {Riley}}, \bibinfo {author}
  {\bibfnamefont {Y.}~\bibnamefont {Kini}}, \bibinfo {author} {\bibfnamefont
  {A.~L.}\ \bibnamefont {Watts}}, \bibinfo {author} {\bibfnamefont
  {B.}~\bibnamefont {Dorsman}}, \bibinfo {author} {\bibfnamefont
  {S.}~\bibnamefont {Bogdanov}}, \bibinfo {author} {\bibfnamefont
  {S.}~\bibnamefont {Guillot}}, \bibinfo {author} {\bibfnamefont {P.~S.}\
  \bibnamefont {Ray}}, \bibinfo {author} {\bibfnamefont {D.~J.}\ \bibnamefont
  {Reardon}}, \bibinfo {author} {\bibfnamefont {R.~A.}\ \bibnamefont
  {Remillard}}, \bibinfo {author} {\bibfnamefont {A.~V.}\ \bibnamefont
  {Bilous}}, \bibinfo {author} {\bibfnamefont {D.}~\bibnamefont
  {Huppenkothen}}, \bibinfo {author} {\bibfnamefont {J.~M.}\ \bibnamefont
  {Lattimer}}, \bibinfo {author} {\bibfnamefont {N.}~\bibnamefont
  {Rutherford}}, \bibinfo {author} {\bibfnamefont {Z.}~\bibnamefont
  {Arzoumanian}}, \bibinfo {author} {\bibfnamefont {K.~C.}\ \bibnamefont
  {Gendreau}}, \bibinfo {author} {\bibfnamefont {S.~M.}\ \bibnamefont
  {Morsink}},\ and\ \bibinfo {author} {\bibfnamefont {W.~C.~G.}\ \bibnamefont
  {Ho}},\ }\href {https://doi.org/10.3847/2041-8213/ad5a6f} {\bibfield
  {journal} {\bibinfo  {journal} {The Astrophysical Journal Letters}\ }\textbf
  {\bibinfo {volume} {971}},\ \bibinfo {pages} {L20} (\bibinfo {year}
  {2024})}\BibitemShut {NoStop}%
\bibitem [{\citenamefont {Li}\ \emph {et~al.}(2024)\citenamefont {Li},
  \citenamefont {Wu},\ and\ \citenamefont {Zhang}}]{Li_2024}%
  \BibitemOpen
  \bibfield  {author} {\bibinfo {author} {\bibfnamefont {W.}~\bibnamefont
  {Li}}, \bibinfo {author} {\bibfnamefont {J.-Y.}\ \bibnamefont {Wu}},\ and\
  \bibinfo {author} {\bibfnamefont {K.}~\bibnamefont {Zhang}},\ }\href
  {https://doi.org/10.1016/j.rinp.2024.107893} {\bibfield  {journal} {\bibinfo
  {journal} {Results in Physics}\ }\textbf {\bibinfo {volume} {64}},\ \bibinfo
  {pages} {107893} (\bibinfo {year} {2024})}\BibitemShut {NoStop}%
\bibitem [{\citenamefont {Sakai}\ and\ \citenamefont
  {Sugimoto}(2005)}]{Sakai_2005}%
  \BibitemOpen
  \bibfield  {author} {\bibinfo {author} {\bibfnamefont {T.}~\bibnamefont
  {Sakai}}\ and\ \bibinfo {author} {\bibfnamefont {S.}~\bibnamefont
  {Sugimoto}},\ }\href {https://doi.org/10.1143/ptp.113.843} {\bibfield
  {journal} {\bibinfo  {journal} {Progress of Theoretical Physics}\ }\textbf
  {\bibinfo {volume} {113}},\ \bibinfo {pages} {843–882} (\bibinfo {year}
  {2005})}\BibitemShut {NoStop}%
\bibitem [{\citenamefont {Witten}(1984)}]{witten1984cosmic}%
  \BibitemOpen
  \bibfield  {author} {\bibinfo {author} {\bibfnamefont {E.}~\bibnamefont
  {Witten}},\ }\href@noop {} {\bibfield  {journal} {\bibinfo  {journal}
  {Physical Review D}\ }\textbf {\bibinfo {volume} {30}},\ \bibinfo {pages}
  {272} (\bibinfo {year} {1984})}\BibitemShut {NoStop}%
\bibitem [{\citenamefont {Shapiro}\ and\ \citenamefont
  {Teukolsky}(1983)}]{shapiro1983black}%
  \BibitemOpen
  \bibfield  {author} {\bibinfo {author} {\bibfnamefont {S.~L.}\ \bibnamefont
  {Shapiro}}\ and\ \bibinfo {author} {\bibfnamefont {S.~A.}\ \bibnamefont
  {Teukolsky}},\ }\href@noop {} {\emph {\bibinfo {title} {Black Holes, White
  Dwarfs and Neutron Stars: The Physics of Compact Objects}}}\ (\bibinfo
  {publisher} {John Wiley \& Sons},\ \bibinfo {address} {New York},\ \bibinfo
  {year} {1983})\BibitemShut {NoStop}%
\bibitem [{\citenamefont {Glendenning}(2000)}]{glendenning2000compact}%
  \BibitemOpen
  \bibfield  {author} {\bibinfo {author} {\bibfnamefont {N.~K.}\ \bibnamefont
  {Glendenning}},\ }\href@noop {} {\emph {\bibinfo {title} {Compact Stars:
  Nuclear Physics, Particle Physics, and General Relativity}}},\ \bibinfo
  {edition} {2nd}\ ed.\ (\bibinfo  {publisher} {Springer},\ \bibinfo {address}
  {New York},\ \bibinfo {year} {2000})\BibitemShut {NoStop}%
\bibitem [{\citenamefont {Colpi}\ \emph {et~al.}(1986)\citenamefont {Colpi},
  \citenamefont {Shapiro},\ and\ \citenamefont {Wasserman}}]{colpi1986boson}%
  \BibitemOpen
  \bibfield  {author} {\bibinfo {author} {\bibfnamefont {M.}~\bibnamefont
  {Colpi}}, \bibinfo {author} {\bibfnamefont {S.~L.}\ \bibnamefont {Shapiro}},\
  and\ \bibinfo {author} {\bibfnamefont {I.}~\bibnamefont {Wasserman}},\
  }\href@noop {} {\bibfield  {journal} {\bibinfo  {journal} {Physical review
  letters}\ }\textbf {\bibinfo {volume} {57}},\ \bibinfo {pages} {2485}
  (\bibinfo {year} {1986})}\BibitemShut {NoStop}%
\bibitem [{\citenamefont {Caballero}\ \emph
  {et~al.}(2024{\natexlab{a}})\citenamefont {Caballero}, \citenamefont
  {Ripley},\ and\ \citenamefont {Yunes}}]{Caballero_2024}%
  \BibitemOpen
  \bibfield  {author} {\bibinfo {author} {\bibfnamefont {D.~A.}\ \bibnamefont
  {Caballero}}, \bibinfo {author} {\bibfnamefont {J.}~\bibnamefont {Ripley}},\
  and\ \bibinfo {author} {\bibfnamefont {N.}~\bibnamefont {Yunes}},\ }\bibfield
   {journal} {\bibinfo  {journal} {Physical Review D}\ }\textbf {\bibinfo
  {volume} {110}},\ \href {https://doi.org/10.1103/physrevd.110.103038}
  {10.1103/physrevd.110.103038} (\bibinfo {year}
  {2024}{\natexlab{a}})\BibitemShut {NoStop}%
\bibitem [{\citenamefont {Weinberg}(2013)}]{weinberg2013gravitation}%
  \BibitemOpen
  \bibfield  {author} {\bibinfo {author} {\bibfnamefont {S.}~\bibnamefont
  {Weinberg}},\ }\href@noop {} {\emph {\bibinfo {title} {Gravitation and
  cosmology: principles and applications of the general theory of
  relativity}}}\ (\bibinfo  {publisher} {John Wiley \& Sons},\ \bibinfo {year}
  {2013})\BibitemShut {NoStop}%
\bibitem [{\citenamefont {Goldman}\ \emph {et~al.}(2013)\citenamefont
  {Goldman}, \citenamefont {Mohapatra}, \citenamefont {Nussinov}, \citenamefont
  {Rosenbaum},\ and\ \citenamefont {Teplitz}}]{Goldman2013}%
  \BibitemOpen
  \bibfield  {author} {\bibinfo {author} {\bibfnamefont {I.}~\bibnamefont
  {Goldman}}, \bibinfo {author} {\bibfnamefont {R.~N.}\ \bibnamefont
  {Mohapatra}}, \bibinfo {author} {\bibfnamefont {S.}~\bibnamefont {Nussinov}},
  \bibinfo {author} {\bibfnamefont {D.}~\bibnamefont {Rosenbaum}},\ and\
  \bibinfo {author} {\bibfnamefont {V.}~\bibnamefont {Teplitz}},\ }\href@noop
  {} {\bibfield  {journal} {\bibinfo  {journal} {Phys. Lett. B}\ }\textbf
  {\bibinfo {volume} {725}},\ \bibinfo {pages} {200} (\bibinfo {year}
  {2013})}\BibitemShut {NoStop}%
\bibitem [{\citenamefont {Friedman}(1978)}]{Friedman:1978}%
  \BibitemOpen
  \bibfield  {author} {\bibinfo {author} {\bibfnamefont {J.~L.}\ \bibnamefont
  {Friedman}},\ }\href {https://doi.org/10.1007/BF01196933} {\bibfield
  {journal} {\bibinfo  {journal} {Communications in Mathematical Physics}\
  }\textbf {\bibinfo {volume} {63}},\ \bibinfo {pages} {243} (\bibinfo {year}
  {1978})}\BibitemShut {NoStop}%
\bibitem [{\citenamefont {Valdez-Alvarado}\ \emph {et~al.}(2013)\citenamefont
  {Valdez-Alvarado}, \citenamefont {Palenzuela}, \citenamefont {Alic},\ and\
  \citenamefont {Ure\~na L\'opez}}]{PhysRevD.87.084040}%
  \BibitemOpen
  \bibfield  {author} {\bibinfo {author} {\bibfnamefont {S.}~\bibnamefont
  {Valdez-Alvarado}}, \bibinfo {author} {\bibfnamefont {C.}~\bibnamefont
  {Palenzuela}}, \bibinfo {author} {\bibfnamefont {D.}~\bibnamefont {Alic}},\
  and\ \bibinfo {author} {\bibfnamefont {L.~A.}\ \bibnamefont {Ure\~na
  L\'opez}},\ }\href {https://doi.org/10.1103/PhysRevD.87.084040} {\bibfield
  {journal} {\bibinfo  {journal} {Phys. Rev. D}\ }\textbf {\bibinfo {volume}
  {87}},\ \bibinfo {pages} {084040} (\bibinfo {year} {2013})}\BibitemShut
  {NoStop}%
\bibitem [{\citenamefont {Becerra}\ \emph {et~al.}(2026)\citenamefont
  {Becerra}, \citenamefont {Becerra-Vergara} \emph {et~al.}}]{Becerra2026}%
  \BibitemOpen
  \bibfield  {author} {\bibinfo {author} {\bibfnamefont {L.~M.}\ \bibnamefont
  {Becerra}}, \bibinfo {author} {\bibfnamefont {E.~A.}\ \bibnamefont
  {Becerra-Vergara}}, \emph {et~al.},\ }\href@noop {} {\bibfield  {journal}
  {\bibinfo  {journal} {Phys. Rev. D}\ }\textbf {\bibinfo {volume} {113}},\
  \bibinfo {pages} {023009} (\bibinfo {year} {2026})}\BibitemShut {NoStop}%
\bibitem [{\citenamefont {Szewczyk}\ \emph {et~al.}(2025)\citenamefont
  {Szewczyk}, \citenamefont {Gondek-Rosińska},\ and\ \citenamefont
  {Cerdá-Durán}}]{Szewczyk2025}%
  \BibitemOpen
  \bibfield  {author} {\bibinfo {author} {\bibfnamefont {P.}~\bibnamefont
  {Szewczyk}}, \bibinfo {author} {\bibfnamefont {D.}~\bibnamefont
  {Gondek-Rosińska}},\ and\ \bibinfo {author} {\bibfnamefont {P.}~\bibnamefont
  {Cerdá-Durán}},\ }\href@noop {} {\bibfield  {journal} {\bibinfo  {journal}
  {Astrophys. J.}\ }\textbf {\bibinfo {volume} {990}},\ \bibinfo {pages} {199}
  (\bibinfo {year} {2025})}\BibitemShut {NoStop}%
\bibitem [{\citenamefont {Takami}\ \emph {et~al.}(2011)\citenamefont {Takami},
  \citenamefont {Rezzolla},\ and\ \citenamefont {Yoshida}}]{Takami2011}%
  \BibitemOpen
  \bibfield  {author} {\bibinfo {author} {\bibfnamefont {K.}~\bibnamefont
  {Takami}}, \bibinfo {author} {\bibfnamefont {L.}~\bibnamefont {Rezzolla}},\
  and\ \bibinfo {author} {\bibfnamefont {S.}~\bibnamefont {Yoshida}},\
  }\href@noop {} {\bibfield  {journal} {\bibinfo  {journal} {Mon. Not. Roy.
  Astron. Soc.}\ }\textbf {\bibinfo {volume} {416}},\ \bibinfo {pages} {L1}
  (\bibinfo {year} {2011})}\BibitemShut {NoStop}%
\bibitem [{\citenamefont {Andersson}\ and\ \citenamefont
  {Kokkotas}(1998)}]{anderssonkokkotas1998}%
  \BibitemOpen
  \bibfield  {author} {\bibinfo {author} {\bibfnamefont {N.}~\bibnamefont
  {Andersson}}\ and\ \bibinfo {author} {\bibfnamefont {K.~D.}\ \bibnamefont
  {Kokkotas}},\ }\href@noop {} {\bibfield  {journal} {\bibinfo  {journal}
  {Monthly Notices of the Royal Astronomical Society}\ }\textbf {\bibinfo
  {volume} {299}},\ \bibinfo {pages} {1059} (\bibinfo {year}
  {1998})}\BibitemShut {NoStop}%
\bibitem [{\citenamefont {Lau}\ \emph {et~al.}(2010)\citenamefont {Lau},
  \citenamefont {Leung},\ and\ \citenamefont {Lin}}]{lau2010universal}%
  \BibitemOpen
  \bibfield  {author} {\bibinfo {author} {\bibfnamefont {H.~K.}\ \bibnamefont
  {Lau}}, \bibinfo {author} {\bibfnamefont {P.~T.}\ \bibnamefont {Leung}},\
  and\ \bibinfo {author} {\bibfnamefont {L.~M.}\ \bibnamefont {Lin}},\
  }\href@noop {} {\bibfield  {journal} {\bibinfo  {journal} {The Astrophysical
  Journal}\ }\textbf {\bibinfo {volume} {714}},\ \bibinfo {pages} {1234}
  (\bibinfo {year} {2010})}\BibitemShut {NoStop}%
\bibitem [{\citenamefont {Bauswein}\ \emph {et~al.}(2019)\citenamefont
  {Bauswein}, \citenamefont {Blacker},\ and\ \citenamefont
  {et~al.}}]{bauswein2019postmerger}%
  \BibitemOpen
  \bibfield  {author} {\bibinfo {author} {\bibfnamefont {A.}~\bibnamefont
  {Bauswein}}, \bibinfo {author} {\bibfnamefont {S.}~\bibnamefont {Blacker}},\
  and\ \bibinfo {author} {\bibnamefont {et~al.}},\ }\href@noop {} {\bibfield
  {journal} {\bibinfo  {journal} {Physical Review Letters}\ }\textbf {\bibinfo
  {volume} {122}},\ \bibinfo {pages} {061102} (\bibinfo {year}
  {2019})}\BibitemShut {NoStop}%
\bibitem [{\citenamefont {Hippert}\ \emph {et~al.}(2023)\citenamefont
  {Hippert}, \citenamefont {Dillingham}, \citenamefont {Tan}, \citenamefont
  {Curtin}, \citenamefont {Noronha-Hostler},\ and\ \citenamefont
  {Yunes}}]{PhysRevD.107.115028}%
  \BibitemOpen
  \bibfield  {author} {\bibinfo {author} {\bibfnamefont {M.}~\bibnamefont
  {Hippert}}, \bibinfo {author} {\bibfnamefont {E.}~\bibnamefont {Dillingham}},
  \bibinfo {author} {\bibfnamefont {H.}~\bibnamefont {Tan}}, \bibinfo {author}
  {\bibfnamefont {D.}~\bibnamefont {Curtin}}, \bibinfo {author} {\bibfnamefont
  {J.}~\bibnamefont {Noronha-Hostler}},\ and\ \bibinfo {author} {\bibfnamefont
  {N.}~\bibnamefont {Yunes}},\ }\href
  {https://doi.org/10.1103/PhysRevD.107.115028} {\bibfield  {journal} {\bibinfo
   {journal} {Phys. Rev. D}\ }\textbf {\bibinfo {volume} {107}},\ \bibinfo
  {pages} {115028} (\bibinfo {year} {2023})}\BibitemShut {NoStop}%
\bibitem [{\citenamefont {Haque}\ \emph {et~al.}(2026)\citenamefont {Haque},
  \citenamefont {Rezzolla},\ and\ \citenamefont {Mallick}}]{Haque2026}%
  \BibitemOpen
  \bibfield  {author} {\bibinfo {author} {\bibfnamefont {S.}~\bibnamefont
  {Haque}}, \bibinfo {author} {\bibfnamefont {L.}~\bibnamefont {Rezzolla}},\
  and\ \bibinfo {author} {\bibfnamefont {R.}~\bibnamefont {Mallick}},\
  }\href@noop {} {\bibfield  {journal} {\bibinfo  {journal} {Phys. Rev. D}\
  }\textbf {\bibinfo {volume} {113}},\ \bibinfo {pages} {103044} (\bibinfo
  {year} {2026})}\BibitemShut {NoStop}%
\bibitem [{\citenamefont {{Abbott}}\ \emph {et~al.}(2018)\citenamefont
  {{Abbott}}, \citenamefont {{Abbott}},\ and\ \citenamefont
  {{Abbott}}}]{Abbott2018}%
  \BibitemOpen
  \bibfield  {author} {\bibinfo {author} {\bibfnamefont {B.~P.}\ \bibnamefont
  {{Abbott}}}, \bibinfo {author} {\bibfnamefont {R.}~\bibnamefont {{Abbott}}},\
  and\ \bibinfo {author} {\bibfnamefont {T.~D. e.~a.}\ \bibnamefont {{Abbott}}}
  (\bibinfo {collaboration} {The LIGO Scientific Collaboration and the Virgo
  Collaboration}),\ }\href {https://doi.org/10.1103/PhysRevLett.121.161101}
  {\bibfield  {journal} {\bibinfo  {journal} {Phys. Rev. Lett.}\ }\textbf
  {\bibinfo {volume} {121}},\ \bibinfo {pages} {161101} (\bibinfo {year}
  {2018})}\BibitemShut {NoStop}%
\bibitem [{\citenamefont {Riley}\ \emph {et~al.}(2019)\citenamefont {Riley},
  \citenamefont {Watts}, \citenamefont {Bogdanov}, \citenamefont {Ray},
  \citenamefont {Ludlam}, \citenamefont {Guillot}, \citenamefont {Arzoumanian},
  \citenamefont {Baker}, \citenamefont {Bilous}, \citenamefont {Chakrabarty},
  \citenamefont {Gendreau}, \citenamefont {Harding}, \citenamefont {Ho},
  \citenamefont {Lattimer}, \citenamefont {Morsink},\ and\ \citenamefont
  {Strohmayer}}]{Riley2019}%
  \BibitemOpen
  \bibfield  {author} {\bibinfo {author} {\bibfnamefont {T.~E.}\ \bibnamefont
  {Riley}}, \bibinfo {author} {\bibfnamefont {A.~L.}\ \bibnamefont {Watts}},
  \bibinfo {author} {\bibfnamefont {S.}~\bibnamefont {Bogdanov}}, \bibinfo
  {author} {\bibfnamefont {P.~S.}\ \bibnamefont {Ray}}, \bibinfo {author}
  {\bibfnamefont {R.~M.}\ \bibnamefont {Ludlam}}, \bibinfo {author}
  {\bibfnamefont {S.}~\bibnamefont {Guillot}}, \bibinfo {author} {\bibfnamefont
  {Z.}~\bibnamefont {Arzoumanian}}, \bibinfo {author} {\bibfnamefont {C.~L.}\
  \bibnamefont {Baker}}, \bibinfo {author} {\bibfnamefont {A.~V.}\ \bibnamefont
  {Bilous}}, \bibinfo {author} {\bibfnamefont {D.}~\bibnamefont {Chakrabarty}},
  \bibinfo {author} {\bibfnamefont {K.~C.}\ \bibnamefont {Gendreau}}, \bibinfo
  {author} {\bibfnamefont {A.~K.}\ \bibnamefont {Harding}}, \bibinfo {author}
  {\bibfnamefont {W.~C.~G.}\ \bibnamefont {Ho}}, \bibinfo {author}
  {\bibfnamefont {J.~M.}\ \bibnamefont {Lattimer}}, \bibinfo {author}
  {\bibfnamefont {S.~M.}\ \bibnamefont {Morsink}},\ and\ \bibinfo {author}
  {\bibfnamefont {T.~E.}\ \bibnamefont {Strohmayer}},\ }\href
  {https://doi.org/10.3847/2041-8213/ab481c} {\bibfield  {journal} {\bibinfo
  {journal} {The Astrophysical Journal Letters}\ }\textbf {\bibinfo {volume}
  {887}},\ \bibinfo {pages} {L21} (\bibinfo {year} {2019})}\BibitemShut
  {NoStop}%
\bibitem [{\citenamefont {{Chandrakar}}\ \emph {et~al.}(2026)\citenamefont
  {{Chandrakar}}, \citenamefont {{Aravind}}, \citenamefont {{Thakur}},
  \citenamefont {{Malik}},\ and\ \citenamefont {{Jha}}}]{Chandrakar2026}%
  \BibitemOpen
  \bibfield  {author} {\bibinfo {author} {\bibfnamefont {H.}~\bibnamefont
  {{Chandrakar}}}, \bibinfo {author} {\bibnamefont {{Aravind}}}, \bibinfo
  {author} {\bibfnamefont {P.}~\bibnamefont {{Thakur}}}, \bibinfo {author}
  {\bibfnamefont {T.}~\bibnamefont {{Malik}}},\ and\ \bibinfo {author}
  {\bibfnamefont {T.~K.}\ \bibnamefont {{Jha}}},\ }\href
  {https://doi.org/10.1016/j.jspc.2026.100304} {\bibfield  {journal} {\bibinfo
  {journal} {Journal of Subatomic Particles and Cosmology}\ }\textbf {\bibinfo
  {volume} {5}},\ \bibinfo {pages} {100304} (\bibinfo {year}
  {2026})}\BibitemShut {NoStop}%
\bibitem [{\citenamefont {Vikiaris}\ \emph {et~al.}(2025)\citenamefont
  {Vikiaris}, \citenamefont {Petousis}, \citenamefont {Veselský},\ and\
  \citenamefont {Moustakidis}}]{Vikiaris_2025}%
  \BibitemOpen
  \bibfield  {author} {\bibinfo {author} {\bibfnamefont {M.}~\bibnamefont
  {Vikiaris}}, \bibinfo {author} {\bibfnamefont {V.}~\bibnamefont {Petousis}},
  \bibinfo {author} {\bibfnamefont {M.}~\bibnamefont {Veselský}},\ and\
  \bibinfo {author} {\bibfnamefont {C.~C.}\ \bibnamefont {Moustakidis}},\
  }\bibfield  {journal} {\bibinfo  {journal} {International Journal of Modern
  Physics D}\ }\textbf {\bibinfo {volume} {34}},\ \href
  {https://doi.org/10.1142/s0218271825500646} {10.1142/s0218271825500646}
  (\bibinfo {year} {2025})\BibitemShut {NoStop}%
\bibitem [{\citenamefont {Li}\ \emph {et~al.}(2026)\citenamefont {Li},
  \citenamefont {Tang}, \citenamefont {Xue},\ and\ \citenamefont
  {Fan}}]{Li2025}%
  \BibitemOpen
  \bibfield  {author} {\bibinfo {author} {\bibfnamefont {Y.-J.}\ \bibnamefont
  {Li}}, \bibinfo {author} {\bibfnamefont {S.-P.}\ \bibnamefont {Tang}},
  \bibinfo {author} {\bibfnamefont {L.-Q.}\ \bibnamefont {Xue}},\ and\ \bibinfo
  {author} {\bibfnamefont {Y.-Z.}\ \bibnamefont {Fan}},\ }\href@noop {}
  {\bibfield  {journal} {\bibinfo  {journal} {The Astrophysical Journal}\
  }\textbf {\bibinfo {volume} {999}},\ \bibinfo {pages} {127} (\bibinfo {year}
  {2026})}\BibitemShut {NoStop}%
\bibitem [{\citenamefont {Lee}\ \emph {et~al.}(2021)\citenamefont {Lee},
  \citenamefont {Chu},\ and\ \citenamefont {Lin}}]{Lee2021}%
  \BibitemOpen
  \bibfield  {author} {\bibinfo {author} {\bibfnamefont {B.~K.}\ \bibnamefont
  {Lee}}, \bibinfo {author} {\bibfnamefont {M.-c.}\ \bibnamefont {Chu}},\ and\
  \bibinfo {author} {\bibfnamefont {L.-M.}\ \bibnamefont {Lin}},\ }\href@noop
  {} {\bibfield  {journal} {\bibinfo  {journal} {The Astrophysical Journal}\
  }\textbf {\bibinfo {volume} {922}},\ \bibinfo {pages} {242} (\bibinfo {year}
  {2021})}\BibitemShut {NoStop}%
\bibitem [{\citenamefont {Pradhan}\ \emph {et~al.}(2023)\citenamefont
  {Pradhan}, \citenamefont {Vijaykumar},\ and\ \citenamefont
  {Chatterjee}}]{Pradhan2023}%
  \BibitemOpen
  \bibfield  {author} {\bibinfo {author} {\bibfnamefont {B.~K.}\ \bibnamefont
  {Pradhan}}, \bibinfo {author} {\bibfnamefont {A.}~\bibnamefont
  {Vijaykumar}},\ and\ \bibinfo {author} {\bibfnamefont {D.}~\bibnamefont
  {Chatterjee}},\ }\href {https://doi.org/10.1103/PhysRevD.107.023010}
  {\bibfield  {journal} {\bibinfo  {journal} {Phys. Rev. D}\ }\textbf {\bibinfo
  {volume} {107}},\ \bibinfo {pages} {023010} (\bibinfo {year}
  {2023})}\BibitemShut {NoStop}%
\bibitem [{\citenamefont {Narikawa}\ \emph {et~al.}(2021)\citenamefont
  {Narikawa}, \citenamefont {Uchikata},\ and\ \citenamefont
  {Tanaka}}]{Narikawa2021}%
  \BibitemOpen
  \bibfield  {author} {\bibinfo {author} {\bibfnamefont {T.}~\bibnamefont
  {Narikawa}}, \bibinfo {author} {\bibfnamefont {N.}~\bibnamefont {Uchikata}},\
  and\ \bibinfo {author} {\bibfnamefont {T.}~\bibnamefont {Tanaka}},\
  }\href@noop {} {\bibfield  {journal} {\bibinfo  {journal} {Physical Review
  D}\ }\textbf {\bibinfo {volume} {104}},\ \bibinfo {pages} {084056} (\bibinfo
  {year} {2021})}\BibitemShut {NoStop}%
\bibitem [{\citenamefont {Maselli}\ \emph {et~al.}(2017)\citenamefont
  {Maselli}, \citenamefont {Pnigouras}, \citenamefont {Nielsen}, \citenamefont
  {Kouvaris},\ and\ \citenamefont {Kokkotas}}]{Maselli:2017vfi}%
  \BibitemOpen
  \bibfield  {author} {\bibinfo {author} {\bibfnamefont {A.}~\bibnamefont
  {Maselli}}, \bibinfo {author} {\bibfnamefont {P.}~\bibnamefont {Pnigouras}},
  \bibinfo {author} {\bibfnamefont {N.~G.}\ \bibnamefont {Nielsen}}, \bibinfo
  {author} {\bibfnamefont {C.}~\bibnamefont {Kouvaris}},\ and\ \bibinfo
  {author} {\bibfnamefont {K.~D.}\ \bibnamefont {Kokkotas}},\ }\href
  {https://doi.org/10.1103/PhysRevD.96.023005} {\bibfield  {journal} {\bibinfo
  {journal} {Phys. Rev. D}\ }\textbf {\bibinfo {volume} {96}},\ \bibinfo
  {pages} {023005} (\bibinfo {year} {2017})}\BibitemShut {NoStop}%
\bibitem [{\citenamefont {Friedman}\ and\ \citenamefont
  {Stergioulas}(2013)}]{Friedman_Stergioulas_2013}%
  \BibitemOpen
  \bibfield  {author} {\bibinfo {author} {\bibfnamefont {J.~L.}\ \bibnamefont
  {Friedman}}\ and\ \bibinfo {author} {\bibfnamefont {N.}~\bibnamefont
  {Stergioulas}},\ }\href@noop {} {\emph {\bibinfo {title} {Rotating
  Relativistic Stars}}},\ Cambridge Monographs on Mathematical Physics\
  (\bibinfo  {publisher} {Cambridge University Press},\ \bibinfo {year}
  {2013})\BibitemShut {NoStop}%
\bibitem [{\citenamefont {Had\v{z}i\'{c}}\ and\ \citenamefont
  {Lin}(2021)}]{Hadzic2021}%
  \BibitemOpen
  \bibfield  {author} {\bibinfo {author} {\bibfnamefont {M.}~\bibnamefont
  {Had\v{z}i\'{c}}}\ and\ \bibinfo {author} {\bibfnamefont {Z.}~\bibnamefont
  {Lin}},\ }\href {https://doi.org/10.1007/s00220-021-04197-6} {\bibfield
  {journal} {\bibinfo  {journal} {Communications in Mathematical Physics}\
  }\textbf {\bibinfo {volume} {387}},\ \bibinfo {pages} {729} (\bibinfo {year}
  {2021})}\BibitemShut {NoStop}%
\bibitem [{\citenamefont {Alford}\ \emph {et~al.}(2017)\citenamefont {Alford},
  \citenamefont {Harris},\ and\ \citenamefont {Sachdeva}}]{Alford:2017vca}%
  \BibitemOpen
  \bibfield  {author} {\bibinfo {author} {\bibfnamefont {M.~G.}\ \bibnamefont
  {Alford}}, \bibinfo {author} {\bibfnamefont {S.~P.}\ \bibnamefont {Harris}},\
  and\ \bibinfo {author} {\bibfnamefont {P.~S.}\ \bibnamefont {Sachdeva}},\
  }\href {https://doi.org/10.3847/1538-4357/aa8509} {\bibfield  {journal}
  {\bibinfo  {journal} {The Astrophysical Journal}\ }\textbf {\bibinfo {volume}
  {847}},\ \bibinfo {pages} {109} (\bibinfo {year} {2017})}\BibitemShut
  {NoStop}%
\bibitem [{\citenamefont {Caballero}\ \emph
  {et~al.}(2024{\natexlab{b}})\citenamefont {Caballero}, \citenamefont
  {Ripley},\ and\ \citenamefont {Yunes}}]{PhysRevD.110.103038}%
  \BibitemOpen
  \bibfield  {author} {\bibinfo {author} {\bibfnamefont {D.~A.}\ \bibnamefont
  {Caballero}}, \bibinfo {author} {\bibfnamefont {J.}~\bibnamefont {Ripley}},\
  and\ \bibinfo {author} {\bibfnamefont {N.}~\bibnamefont {Yunes}},\ }\href
  {https://doi.org/10.1103/PhysRevD.110.103038} {\bibfield  {journal} {\bibinfo
   {journal} {Phys. Rev. D}\ }\textbf {\bibinfo {volume} {110}},\ \bibinfo
  {pages} {103038} (\bibinfo {year} {2024}{\natexlab{b}})}\BibitemShut
  {NoStop}%
\end{thebibliography}%
\end{document}